\documentclass[preprintnumbers, superscriptaddress, showpacs, nofootinbib, amsfonts, amsmath, amssymb, aps, prd, notitlepage, floatfix]{revtex4-2}

\usepackage[colorlinks=true,citecolor=blue,linkcolor=blue,urlcolor=blue]{hyperref}
\usepackage[offset=1.5em]{simpler-wick}
\usepackage{xcolor, graphicx, slashed,  accents,  multirow,caption, subcaption, bm}
\usepackage[most]{tcolorbox}
\usepackage[normalem]{ulem}
\usepackage{empheq}
\usepackage[export]{adjustbox}
\usepackage{physics}

\usepackage{ragged2e} 
\DeclareCaptionJustification{justified}{\justifying}

\allowdisplaybreaks

\newcommand{\as}{\alpha_s}

\newcommand{\xo}{x_{10}}

\newcommand{\xti}{x_{21}}
\newcommand{\dprime}{\prime \prime}

\def\eq#1{{Eq.~(\ref{#1})}}
\def\eqs#1{{Eqs.~(\ref{#1})}}
\def\fig#1{{Fig.~\ref{#1}}}
\def\figs#1{{Figs.~\ref{#1}}}
\newcommand{\ben}{\begin{eqnarray*}}
\newcommand{\een}{\end{eqnarray*}}
\newcommand{\un}[1]{\bm{#1}}

\newcommand{\thalf}{\tfrac{1}{2}}

\newcommand{\cc}{\mbox{c.c.}}

\newcommand{\dhd}{{\textstyle d}
\lower.03ex\hbox{\kern-0.38em$^{\scriptstyle-}$}\kern-0.05em{}}
\newcommand{\dbar}{{\textstyle \delta}
\lower.03ex\hbox{\kern-0.38em$^{\scriptstyle-}$}\kern-0.05em{}}
\newcommand{\half}{{1\over 2}}

\newcommand{\rvec}[1]{\overrightarrow{#1}}
\newcommand{\lvec}[1]{\overleftarrow{#1}}

\makeatletter
\DeclareRobustCommand{\cev}[1]{%
  {\mathpalette\do@cev{#1}}%
}
\newcommand{\do@cev}[2]{%
  \vbox{\offinterlineskip
    \sbox\z@{$\m@th#1 x$}%
    \ialign{##\cr
      \hidewidth\reflectbox{$\m@th#1\vec{}\mkern4mu$}\hidewidth\cr
      \noalign{\kern-\ht\z@}
      $\m@th#1#2$\cr
    }%
  }%
}
\makeatother

\newcommand{\iUV}{
    \int \displaylimits^z_{\frac{1}{s \xo^2}} \frac{\dd z^\prime }{z^\prime} \int \displaylimits^{\xo^2}_{\frac{1}{z^\prime s}} \frac{\dd\xti^2}{\xti^2}
}
\newcommand{\iIR}{\int \displaylimits^{z}_{\frac{\Lambda^2}{s}} \frac{\dd z^\prime}{z^\prime} \int \displaylimits^{\mathrm{min}[\frac{z}{z^\prime} \xo^2, \frac{1}{\Lambda^2}]}_{\mathrm{max}[\xo^2, \frac{1}{z^{\prime}s}]} \frac{\dd \xti^2}{\xti^2}}

\newcommand{\inUV}{\int \displaylimits^{z^\prime}_{\frac{1}{sx_{10}^2}} \frac{\dd z^{\dprime}}{z^{\dprime}} \int \displaylimits^{\mathrm{min}[\xo^2, \xti^2 \frac{z^\prime}{z^{\dprime}}]}_{\frac{1}{z^{\dprime} s}} \frac{\dd x_{32}^2}{x_{32}^2}}

\newcommand{\inIR}{\int \displaylimits^{z^\prime \frac{\xti^2}{\xo^2}}_{\frac{\Lambda^2}{s}} \frac{\dd z^{\dprime}}{z^{\dprime}} 
    \int \displaylimits^{\mathrm{min}[\frac{z^\prime}{z^{\dprime}} \xti^2, \frac{1}{\Lambda^2}]}_{\mathrm{max}[\xo^2, \frac{1}{z^{\dprime}s}]} \frac{\dd x_{32}^2}{x_{32}^2} }

\newcommand{\tI}{\widetilde{I}}
\newcommand{\tG}{\widetilde{G}}
\newcommand{\tGm}{\widetilde{\Gamma}}

\newcommand{\Sec}[1]{Sec.~\ref{#1}}

\begin{document}

\title{Feasibility study of extracting orbital angular momentum from elastic dijet \\ and dihadron production at the EIC}

    \makeatletter  
\def\@fnsymbol#1{\ensuremath{\ifcase#1\or *\or \dagger\or \ddagger\or
   \mathsection\or \mathparagraph\or \|\or **\or \dagger\dagger
   \or \ddagger\ddagger \or \mathsection\mathsection \else\@ctrerr\fi}}
    \makeatother

\author{Brandon Manley}
\email[Corresponding author: ]{brandon.manley@uconn.edu}
\affiliation{Department of Physics, The Ohio State University, Columbus, Ohio 43210, USA}
\affiliation{Department of Physics, University of Connecticut, Storrs, Connecticut 06269, USA}

\author{Daniel~Adamiak }
\affiliation{Michigan State University, East Lansing, Michigan 48824, USA}

\author{Nicholas~Baldonado}
\affiliation{Department of Physics, Kennesaw State University, Marietta, Georgia 30062, USA}

\author{G. Zardo Becker}
\affiliation{Physics Institute, Federal University of Santa Catarina, Florianópolis, SC 88049-900, Brazil}

\author{\\ Benjamin Kegerreis}
\affiliation{Department of Physics, Lebanon Valley College, Annville, Pennsylvania 17003, USA}

\author{Yuri~V.~Kovchegov} 
\affiliation{Department of Physics, The Ohio State University, Columbus, Ohio 43210, USA}

\author{Ming~Li}
\affiliation{Department of Physics, Hampton University, Hampton, Virginia 23668, USA}

\author{W.~Melnitchouk}
\affiliation{Theory Center, Jefferson Lab, Newport News, Virginia 23606, USA}

\author{Daniel~Pitonyak }
\affiliation{Department of Physics, Lebanon Valley College, Annville, Pennsylvania 17003, USA}
\author{Nobuo~Sato}
\affiliation{Theory Center, Jefferson Lab, Newport News, Virginia 23606, USA}
\author{Matthew~D.~Sievert}
\affiliation{Department of Physics, New Mexico State University, Las Cruces, New Mexico 88003, USA}
\author{Andrey~Tarasov}
\affiliation{Department of Physics and Astronomy, North Carolina State University, Raleigh, North Carolina 27695, USA}
\author{Yossathorn~Tawabutr}
         \affiliation{Center of Excellence in High Energy Physics, Faculty of Science, Chulalongkorn University, 
         Bangkok 10330, Thailand
\\ 
\vspace*{0.2cm}
{\bf
JAM Collaboration\\
\vspace*{0.1cm}(Small-$x$ Analysis Group)
\vspace*{0.2cm}
}
}

\date{\today}

\preprint{JLAB-THY-26-5042}

\begin{abstract}
We perform the first feasibility study of extracting quark and gluon orbital angular momentum (OAM) distributions in the proton at small parton momentum fractions $x$ from elastic dijet and dihadron production in longitudinally polarized $ep$ collisions at the  Electron-Ion Collider (EIC). In the small-$x$ formalism, the OAM distributions are related to the impact parameter moments of polarized dipole amplitudes, which enter the elastic dijet double-spin asymmetry (DSA) through azimuthal angular correlations among the jets, outgoing proton, and outgoing electron. We generalize earlier results for elastic dijet cross sections to dihadron production and employ the small-$x$ helicity formalism to generate EIC pseudodata for proton, deuteron, and $^3$He DSAs at Bjorken $x < 0.01$. Fitting the elastic dijet and dihadron pseudodata, together with existing world data for polarized deep-inelastic scattering (DIS) and semi-inclusive DIS (SIDIS) at $x<0.1$, EIC pseudodata for polarized DIS and SIDIS, and jet production data in polarized $pp$ collisions, we find that EIC DSA measurements would allow direct extraction of quark and gluon OAM distributions at small $x$, and substantially reduce the uncertainties on small-$x$ helicity PDFs. 
\end{abstract}

\maketitle

\section{Introduction}
\label{sec:intro}

Determining the orbital angular momentum (OAM) carried by quarks and gluons (or partons) in the proton is essential for understanding the proton's spin structure~\cite{Jaffe:1989jz, Ji:1996ek}. While significant progress has been made in extracting the quark and gluon helicity parton distribution functions (hPDFs) from polarized scattering data~\cite{Gluck:2000dy, Jimenez-Delgado:2013boa, Nocera:2014gqa, Leader:2014uua, Sato:2016tuz, Ethier:2017zbq, Zhou:2022wzm, Cocuzza:2022jye, Borsa:2024mss, Dumitru:2024pcv, Bertone:2024taw, Hunt-Smith:2024khs, Cruz-Martinez:2025ahf, Adamiak:2025mdy, Cocuzza:2025qvf}, the corresponding OAM distributions remain largely unknown. In particular, the region of small parton momentum fractions $x$, where the uncertainties on the helicity distributions are substantial, offers an important opportunity for studying the contributions of partonic orbital motion to the proton spin.

Recent theoretical developments have established a connection between the quark and gluon OAM distributions at small $x$ and observables in elastic particle production in polarized electron-proton collisions~\cite{Hatta:2016aoc, Bhattacharya:2022vvo, Bhattacharya:2024sck, Kovchegov:2024wjs, Kovchegov:2026lik}.
In particular, the longitudinal double-spin asymmetry (DSA) in elastic dijet production provides sensitivity to the impact parameter moments of polarized dipole amplitudes, which are related to the OAM distributions, as was shown in Refs.~\cite{Kovchegov:2019rrz, Kovchegov:2023yzd} using 
the small-$x$ light-cone operator treatment (LCOT) \cite{Altinoluk:2014oxa, Balitsky:2015qba, Balitsky:2016dgz, Kovchegov:2017lsr, Kovchegov:2018znm, Chirilli:2018kkw, Jalilian-Marian:2018iui, Jalilian-Marian:2019kaf, Altinoluk:2020oyd, Boussarie:2020vzf, Boussarie:2020fpb, Kovchegov:2021iyc, Kovchegov:2022kyy, Li:2023tlw}. Unlike the corresponding single-spin asymmetry, the DSA does not require knowledge of the imaginary part of the unpolarized dipole amplitude, which is poorly constrained. The theoretical framework for the OAM distributions at small $x$, their evolution in the large-$N_c\&N_f$ limit, and the elastic dijet DSA has been developed in Refs.~\cite{Boussarie:2019icw, Kovchegov:2019rrz, Kovchegov:2023yzd, Kovchegov:2024wjs}, including in the companion theory paper~\cite{Becker:2026xcp}.

An important question, however, is whether the predicted asymmetries can be measured with sufficient precision at the planned Electron-Ion Collider (EIC) to allow a meaningful extraction of the quark and gluon OAM distributions. Addressing this question requires going beyond the theoretical sensitivity of the observables to assess the projected experimental uncertainties, the available kinematic coverage, and the constraints that such measurements would provide on global QCD analysis.
Moreover, since inclusive dijet production provides limited flavor discrimination, additional observables are needed to separate the OAM contributions from different quark flavors.

In this paper, we perform the first feasibility study of extracting the small-$x$ quark and gluon OAM distributions from elastic dijet and dihadron production in longitudinally polarized electron-hadron collisions at the EIC. 
We extend the elastic dijet formalism to dihadron production by introducing fragmentation functions, providing additional sensitivity to the flavor dependence of the quark OAM distributions.
Using the small-$x$ LCOT framework, we generate projected DSA pseudodata for proton, deuteron, and $^3{\rm He}$ targets at Bjorken $x_B < 0.01$.
These are incorporated into a global analysis, together with existing polarized deep-inelastic scattering (DIS)~\cite{Adamiak:2021ppq}, semi-inclusive DIS (SIDIS)~\cite{Adamiak:2023yhz}, and polarized proton-proton scattering data~\cite{JAMCollaborationSmall-xAnalysisGroup:2025tfa}, as well as projected EIC DIS and SIDIS measurements.
Our analysis explores the extent to which future elastic dijet and dihadron measurements can constrain the quark and gluon OAM distributions and reduce uncertainties on the helicity PDFs at small $x$.
We find that the projected measurements provide sensitivity sufficient to extract the OAM distributions, although sizable uncertainties remain, while also offering significant improvements in the determination of the small-$x$ helicity distributions.

The remainder of this paper is organized as follows.
In Sec.~\ref{sec:formalism} we summarize the small-$x$ formalism for helicity PDFs and OAM distributions, including their relations to polarized dipole amplitudes, moment amplitudes, and their small-$x$ evolution.
In Sec.~\ref{sec:dijets} we review the elastic dijet DSA calculation and present the azimuthal harmonics needed for the numerical analysis, including the unpolarized cross section.
The formalism is extended to elastic dihadron production in Sec.~\ref{sec:dihadrons}, where fragmentation functions provide additional sensitivity to the quark flavor dependence.
In Sec.~\ref{sec:numerical_imp} we describe the numerical implementation of the small-$x$ evolution equations, including the initial conditions and treatment of infrared sensitivity.
The feasibility study is presented in Sec.~\ref{sec:feas}, where we first update previous global analyses of polarized DIS, SIDIS, and $pp$ data using the complete large-$N_c$ \& $N_f$ helicity evolution and the additional inhomogeneous contributions.
We then assess the impact of projected EIC elastic dijet and dihadron DSA measurements on the extraction of quark and gluon OAM distributions, and the reduction of uncertainties in the small-$x$ helicity PDFs.
Finally, we summarize our conclusions and discuss future prospects in Sec.~\ref{sec:conclusions}. \\

\section{Polarized small-$x$ formalism}
\label{sec:formalism}

In this section we summarize the ingredients of the polarized small-$x$ formalism relevant to the present analysis of OAM distributions.
The details of the derivations and operator definitions can be found in Ref.~\cite{Becker:2026xcp} and references therein.

\subsection{Helicity}

Spin-dependent observables at small parton momentum fractions $x$ are constructed from polarized dipole amplitudes, which are traces of the product of an eikonal Wilson line and a polarized Wilson line~\cite{Kovchegov:2017lsr, Kovchegov:2018znm, Cougoulic:2022gbk}.
There are two classes of polarized Wilson lines: 
{\it type-1}, which lead to the polarized dipole amplitudes $Q_{q\, 10}(s)$ and $\widetilde{G}_{10}(s)$, and 
{\it type-2}, which are related to the polarized dipole amplitude $G^i_{10}(s)$, with $i=1,2$ the transverse vector index.
In addition, the quantity $\widetilde Q_{q\, 10}(s)$ is introduced in \cite{Borden:2024bxa}, which is not a dipole amplitude, but rather is related to the quark helicity TMD in the transverse position space.
Here $s$ is the center-of-mass energy squared for the dipole--proton scattering, and the indices ``1'' and ``0'' in the subscripts label the 2-vectors $\un x_1$ and $\un x_0$ describing the transverse locations of the Wilson lines in the dipole.

The hPDFs and small-$x$ evolution equations are expressed in terms of the impact parameter integrated dipole amplitudes $Q_q(x_{10}^2, s), \widetilde{G} (x_{10}^2, s), $ and $G_2(x_{10}^2, s)$, along with the impact parameter integrated quantity $\widetilde Q_q (x_{10}^2, s)$. 
These latter quantities depend only on the transverse size of the dipole, $x_{10}$, defined as $x_{ij} \equiv |\un{x}_i - \un{x}_j|$ for $i,j = 1, 0$ and with the two-vectors $\un{x}_i$ labeling the positions of the quark and antiquark in the transverse plane.  
The amplitudes $Q_q$ and $\widetilde Q_q$ also depend on the quark flavor $q=u,d,s$; in the present analysis, however, we restrict ourselves to light flavors only.
In addition to these amplitudes, there is a flavor nonsinglet polarized dipole amplitude, with a corresponding integrated amplitude $Q^{\mathrm{NS}}_q(x_{10}^2,s)$ \cite{Kovchegov:2016zex, Adamiak:2023okq, Kovchegov:2024wjs}.\footnote{The amplitude $Q^{\mathrm{NS}}_q$ was denoted by $G^{\mathrm{NS}}$ or $G^{\mathrm{NS}}_q$ in some earlier work~\cite{Kovchegov:2016zex, Adamiak:2023okq, JAMCollaborationSmall-xAnalysisGroup:2025tfa}.} 
In terms of the integrated dipole amplitudes, the $C$-even ($\Delta q^+ \equiv \Delta q + \Delta \bar q$) and $C$-odd ($\Delta q^- \equiv \Delta q - \Delta \bar q$) quark hPDFs and the gluon hPDF $\Delta g$ can be written at small $x$ as~\cite{Cougoulic:2022gbk, Borden:2024bxa}
\begin{subequations}\label{hPDFs}
\begin{align} 
    \Delta q^+ (x, Q^2) 
    &= \frac{1}{\as \pi^2} \, {\widetilde Q}_q \left( x_{10}^2 = \frac{1}{Q^2}, s = \frac{Q^2}{x} \right) 
\notag \\
    & = \frac{1}{\as \pi^2} \, {\widetilde Q}^{(0)}_q \left( x_{10}^2 = \frac{1}{Q^2}, s = \frac{Q^2}{x} \right)  
    - \frac{N_c}{2\pi^3} 
    \int \limits^{1/\Lambda^2}_{1/Q^2} \frac{\dd x_{10}^2}{x_{10}^2}
    \int \limits^{1/(Q^2 x_{10}^2)}_{x/(Q^2 x_{10}^2)} \frac{\dd z}{z}  
    \Big[ Q_q(x_{10}^2,zs) + 2\, G_2(x_{10}^2,zs)  \Big],
\label{Dq+} \\ 
    \Delta q^-(x,Q^2) 
    & = \frac{1}{\as \, \pi^2} \, {\widetilde Q}^{\mathrm{NS}}_q \left( x_{10}^2 = \frac{1}{Q^2}, s = \frac{Q^2}{x} \right) 
\notag \\
    & = \frac{1}{\as \pi^2} \, {\widetilde Q}^{\mathrm{NS} \, (0)}_q \left( x_{10}^2 = \frac{1}{Q^2}, s = \frac{Q^2}{x} \right)
    + \frac{N_c}{2\pi^3} 
    \int \limits^{1/\Lambda^2}_{1/Q^2} \frac{\dd x_{10}^2}{x_{10}^2}
    \int \limits^{1/(Q^2 x_{10}^2)}_{x/(Q^2 x_{10}^2) } 
    \frac{\dd z}{z}  Q^{\mathrm{NS}}_q (x_{10}^2,zs) ,
\label{Dq-} \\
    \Delta g(x,Q^2) 
    &= \frac{2N_c}{\alpha_s\pi^2}\,
    G_2 \!\left(x_{10}^2 = \frac{1}{Q^2}, s = \frac{Q^2}{x}\right),
    \label{Dg}
\end{align}
\end{subequations}
where $z$ is the longitudinal (minus) momentum fraction of the dipole momentum carried by the soft quark or antiquark, and $\Lambda$ is an IR cutoff, chosen here to be $\Lambda = 1$~GeV.
The coupling $\alpha_s = \alpha_s(Q^2)$ runs with $Q^2$, and is given by the one-loop expression,
\begin{align} \label{pspace_rc}
    \as(Q^2) = \frac{12\pi}{(11 N_c - 2 N_f)} 
    \frac{1}{\ln \!\big( Q^2/{\Lambda_{\mathrm{QCD}}^2} \big)},
\end{align}
with $\Lambda_{\mathrm{QCD}} = 241\, \mathrm{MeV}$.
The position space expression for the coupling we use is given below in \eq{xspace_rc}. 
The amplitude $\widetilde{Q}_q^\mathrm{NS}$ is defined analogously to $\widetilde{Q}_q$, except with a difference between quark and antiquark polarized interactions~\cite{Borden:2024bxa}. 
In the following we shall work in a frame where the proton moves in the light-cone ``$+$'' direction.

The main observables that couple to $Q_q, G_2,\widetilde{G},$ and $Q^{\mathrm{NS}}_q$, which have been studied in previous phenomenological analyses, are the polarized inclusive DIS $g_1$ structure functions of the proton and neutron, and the polarized SIDIS structure functions $g_1^h$ for the production of hadrons $h$~\cite{Adamiak:2021ppq, Adamiak:2023yhz}, along with asymmetries in single-inclusive jet production in polarized proton-proton collisions \cite{JAMCollaborationSmall-xAnalysisGroup:2025tfa}. 
The inclusive $g_1$ structure function is expressed in terms of the polarized dipole amplitudes as~\cite{Cougoulic:2022gbk, Borden:2024bxa,  Adamiak:2023yhz, JAMCollaborationSmall-xAnalysisGroup:2025tfa},
\begin{subequations} \label{g1_structs}
    \begin{align}
    g_1 (x,Q^2) &=  \sum_q \frac{e_q^2}{2 \pi^2 \as} \, {\widetilde Q}^{(0)}_q\!\!\left( x_{10}^2 = \frac{1}{Q^2}, s = \frac{Q^2}{x} \right) 
\notag \\ \label{dis_g1}
    & \hspace{1cm} -\sum_q \frac{N_c\,e_q^2}{4\pi^3}\int\limits_{\Lambda^2/s}^1\,\frac{\dd z}{z}\int\limits_{1/zs}^{\mathrm{min}\{1/zQ^2,\, 1/\Lambda^2\}}\,\frac{\dd x_{10}^2}{x_{10}^2}\,\Big[Q_q(x_{10}^2,zs) + 2G_2(x_{10}^2,zs)\Big],
    \end{align}
while the semi-inclusive $g_1^h$ structure function is given by
    \begin{align}
    g_1^h(x, Q^2, z_h) 
    &= \sum_q
       D_1^{h/q}(z_h,Q^2) \Bigg[ \frac{e_q^2}{4 \pi^2 \as} \,
       \left( {\widetilde Q}^{(0)}_q \left( x_{10}^2 = \frac{1}{Q^2}, s = \frac{Q^2}{x} \right) + {\widetilde Q}^{\mathrm{NS} \, (0)}_q \left( x_{10}^2 = \frac{1}{Q^2}, s = \frac{Q^2}{x} \right) \right) 
\notag\\
    & \hspace{1.5cm} - 
    \frac{N_c\,e_q^2}{8 \pi^3}
    \int\limits_{\Lambda^2/s}^1\,\frac{\dd z}{z}\int\limits_{1/zs}^{\mathrm{min}\{1/zQ^2,\, 1/\Lambda^2\}}\,\frac{\dd x_{10}^2}{x_{10}^2}\,
    \Big[Q_q(x_{10}^2,zs) 
    + 2\,G_2(x_{10}^2,zs) - Q^{\mathrm{NS}}_{q}(x_{10}^2, zs) 
    \Big] \Bigg]
\notag\\
    & + \sum_q
       D_1^{h/\bar{q}}(z_h,Q^2) \, \Bigg[ \frac{e_{\bar{q}}^2}{4 \pi^2 \as} \,\left( {\widetilde Q}^{(0)}_q \left( x_{10}^2 = \frac{1}{Q^2}, s = \frac{Q^2}{x} \right) - {\widetilde Q}^{\mathrm{NS} \, (0)}_q \left( x_{10}^2 = \frac{1}{Q^2}, s = \frac{Q^2}{x} \right) \right) 
\notag\\
   & \hspace{1.5cm} -  \frac{N_c\,e_{\bar{q}}^2}{8 \pi^3}
    \int\limits_{\Lambda^2/s}^1\,\frac{\dd z}{z}\int\limits_{1/zs}^{\mathrm{min}\{1/zQ^2,\, 1/\Lambda^2\}}\,\frac{\dd x_{10}^2}{x_{10}^2}\,
    \Big[ Q_q(x_{10}^2,zs) 
    + 2\,G_2(x_{10}^2,zs) + Q^{\mathrm{NS}}_{q}(x_{10}^2, zs) 
    \Big] \Bigg], 
\label{sidis_g1}
\end{align}
\end{subequations}
where the sums are over quark flavors $q = u,d,s$.
In \eq{sidis_g1}, $z_h$ is related to the fraction of the fragmenting parton's light-cone momentum carried by the hadron $h$, and $D^{h/q}_1(z_h,Q^2)$ is the collinear quark $q \to h$ fragmentation function.
The two sums in \eq{sidis_g1} represent the produced hadron originating from either a quark or an antiquark.
Unlike in previous studies~\cite{Kovchegov:2015pbl, Kovchegov:2018znm, Adamiak:2021ppq, Cougoulic:2022gbk, Adamiak:2023yhz, JAMCollaborationSmall-xAnalysisGroup:2025tfa}, here we also include the inhomogeneous terms ${\widetilde Q}^{(0)}_q$ and ${\widetilde Q}^{\mathrm{NS} \, (0)}_q$ into the expressions for $g_1$ and $g_1^h$ in Eqs.~(\ref{g1_structs}).
These were neglected in previous work as small unevolving corrections, but can in principle also affect the phenomenology. 
The derivation of these inhomogeneous terms is given in Appendix~\ref{sec:g1} (see also Chirilli~\cite{Chirilli:2026vij, Chirilli:2026pkv}).

The single-inclusive jet production cross section for polarized $p p \to \mathrm{jet} + X$ scattering in the center-of-mass frame is given, in the gluons-only sector, by \cite{Kovchegov:2024aus, JAMCollaborationSmall-xAnalysisGroup:2025tfa} 
\begin{align}\label{num_xperp}
    \frac{\dd\Delta\sigma^{pp\to \mathrm{jet} X}}{\dd^2p_T \dd y} 
    &=\frac{C_F}{\pi^4}\frac{1}{s\,p_T^2}
    \int \dd^2x_\perp\,
    e^{-i\bm{p}_\perp\cdot\, \bm{x}_\perp}\,
    \frac{1}{\alpha_s(1/x_{\perp}^2)}
    \notag \\
    &\times \big( 4\, \widetilde{G}_{P}\;\,2\, G_{2,P}\big)\,
    \Big( x_{\perp}^2,\sqrt{s}\,p_T\,e^{-y} \Big)
    \begin{pmatrix}
        \frac{1}{4}\lvec{\nabla}_{\perp}\cdot\rvec{\nabla}_\perp & \lvec{\nabla}_{\perp}^2+\lvec{\nabla}_{\perp}\cdot\rvec{{\nabla}}_{\perp} \\
    \rvec{{\nabla}}_{\perp}^2+\lvec{{\nabla}}_{\perp}\cdot\rvec{{\nabla}}_{\perp} & 0
    \end{pmatrix}
    \begin{pmatrix}
        4\, \widetilde{G}_{T} \\
        2\, G_{2,T}
    \end{pmatrix}
    \Big( x_{\perp}^2, \sqrt{s}\,p_T\,e^y \Big)\,,
\end{align}
where $C_F$ is the fundamental Casimir operator of SU($N_c$), $\bm{p}_\perp$ is the transverse momentum of the jet, with $p_T \equiv |\bm{p}_\perp|$, and $y = \thalf \ln (p^-/p^+)$ is the rapidity, where the light-cone momenta are defined by $p^\pm = \frac{1}{\sqrt{2}} (p^0 \pm p^3)$.
With this definition, the target moves in the ``+'' direction, while the projectile moves in the ``$-$'' direction. 
The cross section (\eq{num_xperp}) is written in a projectile-target symmetric form, where the subscripts ``$T$'' or ``$P$'' on the dipole amplitudes $G$ and $\widetilde{G}$ indicate whether the scattering is on the target or projectile proton, respectively.

In writing \eq{num_xperp}, we have followed Ref.~\cite{JAMCollaborationSmall-xAnalysisGroup:2025tfa} in assuming that in the  double logarithmic approximation (DLA) in $\ln (1/x)$ and in the narrow jet approximation \cite{Kaufmann:2015hma} the jet function can be approximated at leading order by $\mathcal{J} \simeq \delta(1 - z_j)$, where $z_j$ is the transverse momentum fraction of the parton momentum carried by the jet.
Higher-order corrections to $\mathcal{J}$ generate logarithmic contributions, but these can be understood as DGLAP evolution with the $\as \, \ln R$ resummation parameter (with $R$ the jet radius), which are outside of the DLA.
To maintain consistency with the previous analysis \cite{JAMCollaborationSmall-xAnalysisGroup:2025tfa}, we leave these resummation effects for future work.
The DLA evolution equations for $Q_q, \widetilde{G}, G_2$, $\widetilde{Q}_q$, $Q^{\mathrm{NS}}_{q}$, and $\widetilde{Q}^{\mathrm{NS}}_q$ at large $N_c \& N_f$ and with running coupling are listed in Eqs.~\eqref{eq_LargeNcNf} of Appendix~\ref{sec:AppendixB}.

\subsection{Orbital angular momentum}

The definitions and small-$x$ evolution of the OAM distributions in the large-$N_c \& N_f$ limit have been derived in Ref.~\cite{Becker:2026xcp}, building on earlier work~\cite{Kovchegov:2023yzd, Manley:2024pcl, Kovchegov:2024wjs}; here we present only the main conclusions of Ref.~\cite{Becker:2026xcp} relevant for the current feasibility study.

The quark and gluon OAM distributions are given in terms of the moment amplitudes via~\cite{Kovchegov:2023yzd, Kovchegov:2024wjs, Becker:2026xcp}
\begin{subequations}\label{OAMs}
\begin{align}
\label{quark_OAM}
    L_{q^+}(x,Q^2) 
    &= \frac{2}{\as \pi^2}\, \widetilde{I}_q \left(x_{10}^2
    = \frac{1}{Q^2}, s = \frac{Q^2}{x} \right), 
\\
\label{gluon_OAM}
L_g(x,Q^2) &= - \frac{2 \, N_c}{\as \pi^2} 
    \Big[ 2\, I_4 + 3\, I_5\Big]\!\!\left(x_{10}^2 = \frac{1}{Q^2}, \, s = \frac{Q^2}{x} \right).
\end{align}
\end{subequations}

The moment amplitudes $I_4$, $I_5$, and $\widetilde{I}_q$ only appear in the OAM distributions and not in the hPDFs. 
Furthermore, they do not appear in either the $g_1$ and $g_1^h$ structure functions in \eqs{g1_structs} or in the jet production cross section in \eq{num_xperp}. 
Instead, it was shown~\cite{Kovchegov:2024wjs, Becker:2026xcp} that the moment amplitudes $I_{3q}$, $I_4$, $I_5$, and $I_E$ explicitly appear in the DSA in elastic dijet production in $ep$ collisions. 
Our main goal in this work is to assess the feasibility of extracting these moment amplitudes from the elastic dijet observable and, via \eqs{OAMs}, the OAM distributions.

The moment amplitudes also satisfy their own system of integro-differential evolution equations in the DLA, which closes in the large-$N_c$ or large-$N_c\&N_f$ limits.
The large-$N_c$ equations for the moment amplitudes $I_{3q}$, $I_4$, and $I_5$ can be found in Eq.~(A1) of Ref.~\cite{Kovchegov:2024wjs} (see also \cite{Kovchegov:2023yzd}); their corrected version that includes $I_E$ has been constructed in Ref.~\cite{Becker:2026xcp} (see Eqs.~(84) therein). 
At large $N_c\&N_f$, the moment amplitudes mix with $Q_q$, $\widetilde{G}$, $G_2$, and $\widetilde{Q}_q$ under evolution, as discussed in Ref.~\cite{Becker:2026xcp}.
Including the running-coupling corrections while employing the prescription from Ref.~\cite{Kovchegov:2021lvz}, the evolution equations for the moment amplitudes can be written at large $N_c\&N_f$ as \\ \\ \\ \\
\begin{align}    \notag 
    \begin{pmatrix}
        I_{3q} \\
        \tI_3\\ 
        I_4 \\
        I_5 \\
        \widetilde{I}_q \\
        I_E
    \end{pmatrix} (x_{10}^2, & \, zs)
    = \begin{pmatrix}
        I_{3q}^{(0)} \\
        \tI_3^{(0)}  \\ 
        I_4^{(0)} \\
        I_5^{(0)} \\
        \widetilde{I}_q^{(0)} \\
        0
    \end{pmatrix} (x_{10}^2, zs) 
    \\ \notag & \hspace{-1.5cm}
    + \frac{N_c}{2\pi} \iUV \, \as\left(\frac{1}{x_{21}^2}\right)
    \begin{pmatrix}
        -1 & 2 & -2 & 1 & -1  \\
        - \frac{1}{4N_c} \sum_q   & 1 & \frac{N_f}{2N_c} -2  & 1- \frac{N_f}{4N_c} & \frac{N_f}{4N_c} - 1  \\
        0 & 0 & 0 & 0  &  0  \\ 
        0 & 0 & 0 & 0  &  0  \\ 
        0 & 0 & 0 & 0  &  0  \\
        0 & 0 & 0 & 0  &  0 
    \end{pmatrix}
    \begin{pmatrix}
        \Gamma_{3q} \\ 
        \tGm_3 \\
        \Gamma_4 \\
        \Gamma_5 \\
        \Gamma_2 
    \end{pmatrix}(x_{10}^2, x_{21}^2 ,z^\prime s)
    \\ \notag & \hspace{-1.5cm}
    + \frac{N_c}{4\pi} \iIR \, \as\left(\frac{1}{x_{21}^2}\right)
    \begin{pmatrix}
        0 & 4 & -4 & 2 & 0 & -2  \\
        -\frac{1}{2N_c} \sum_q  & 4 & \frac{N_f}{N_c} - 4 & 2 - \frac{N_f}{2N_c}   & \frac{1}{2N_c} \sum_q  & -2 \\
        0 & 0 & 4 & 2  & 0 & -1 \\
        0 & -2 & 2 & -1 & 0 & 1  \\
        -1 & 0 & 2 & -1 & 1 & 0 \\
        0 & -4 & 4 & -2 & 0 & 2
    \end{pmatrix}
    \begin{pmatrix}
        I_{3q} \\ 
        \tI_3 \\
        I_4 \\
        I_5 \\
        \widetilde{I}_q \\
        I_E
    \end{pmatrix}(x_{21}^2 ,z^\prime s)
    \notag
    \\ & \hspace{-1.5cm}
        + \frac{N_c}{4\pi} \iIR \, \as\left(\frac{1}{x_{21}^2}\right)
    \begin{pmatrix}
         0 & -4  & -6 & 0   \\
         0 & -4  & \frac{N_f}{2N_c} -6 & \frac{1}{4 N_c} \sum_q \\
         0 &  -2 & -3 & 0 \\
         0 & 4 & 7 & 0 \\
         1 & 0 & 3 & \thalf \\
         0 & 4 & 6 & 0
    \end{pmatrix}
    \begin{pmatrix}
        Q_q \\
        \tG \\
        G_2 \\
        \widetilde{Q}_q
    \end{pmatrix}(x_{21}^2 ,z^\prime s).
        \label{mom_eqn_rc}
\end{align}
The $\Gamma_i$ functions appearing in \eq{mom_eqn_rc} are auxiliary amplitudes known as ``neighbor'' amplitudes, and are required for lifetime ordering in the evolution (see Ref.~\cite{Kovchegov:2015pbl} for details).
The inhomogeneous terms or initial condition for $I_E$ in \eqs{mom_eqn_rc} are identically zero in the quasi-classical sub-eikonal approximation to all orders in $\as$~\cite{Becker:2026xcp}.

Similarly to the helicity evolution, the moment amplitudes (with the exception of $\widetilde{I}_q$) have corresponding neighbor amplitudes (see Ref.~\cite{Becker:2026xcp} for explicit definitions). The moment neighbor amplitudes obey the following large-$N_c\&N_f$ DLA evolution equations~\cite{Becker:2026xcp}
\begin{align}  \notag 
    \begin{pmatrix}
        \Gamma_{3q} \\
        \tGm_3\\ 
        \Gamma_4 \\
        \Gamma_5 
    \end{pmatrix} &(x_{10}^2, x_{21}^2, z's)
    = \begin{pmatrix}
        I_{3q}^{(0)} \\
        \tI_3^{(0)}  \\ 
        I_4^{(0)} \\
        I_5^{(0)} 
    \end{pmatrix} (x_{10}^2, z's) 
    \\ \notag & \hspace{-1cm}
    + \frac{N_c}{2\pi} \inUV \, \as \left(\frac{1}{x_{32}^2}\right)
    \begin{pmatrix}
        -1 & 2 & -2 & 1 & -1 \\
        - \frac{1}{4N_c} \sum_q   & 1 & \frac{N_f}{2N_c} -2  & 1- \frac{N_f}{4N_c} & \frac{N_f}{4N_c} - 1  \\
        0 & 0 & 0 & 0  &  0  \\ 
        0 & 0 & 0 & 0  &  0  
    \end{pmatrix}
    \begin{pmatrix}
        \Gamma_{3q} \\ 
        \tGm_3 \\
        \Gamma_4 \\
        \Gamma_5 \\
        \Gamma_2 
    \end{pmatrix}(x_{10}^2, x_{32}^2 ,z^{\prime \prime} s) 
    \notag  \\ & \hspace{-1cm}
    + \frac{N_c}{4\pi}
    \inIR \, \as \left(\frac{1}{x_{32}^2}\right)
     \begin{pmatrix}
        0 & 4 & -4 & 2 & 0 & -2 \\
        -\frac{1}{2N_c} \sum_q  & 4 & \frac{N_f}{N_c} - 4 & 2 - \frac{N_f}{2N_c}   & \frac{1}{2N_c} \sum_q  & -2\\
        0 & 0 & 4 & 2  & 0 & -1  \\
        0 & -2 & 2 & -1 & 0 & 1 
    \end{pmatrix}
     \begin{pmatrix}
        I_{3q} \\ 
        \tI_3 \\
        I_4 \\
        I_5 \\
        \widetilde{I}_q \\
        I_E
    \end{pmatrix} (x_{32}^2 ,z^{\prime\prime} s)
    \notag 
    \\ & \hspace{-1cm}
    + \frac{N_c}{4\pi}
    \inIR \, \as \left(\frac{1}{x_{32}^2}\right)
    \begin{pmatrix}
         0 & -4  & -6 & 0   \\
        0 & -4  & \frac{N_f}{2N_c} -6 & \frac{1}{4 N_c} \sum_q \\
         0 &  -2 & -3 & 0 \\
         0 & 4 & 7 & 0 
    \end{pmatrix}
    \begin{pmatrix}
        Q_q \\
        \tG \\
        G_2 \\
        \widetilde{Q}_q
    \end{pmatrix} (x_{32}^2 ,z^{\prime\prime} s).
    \label{neigh_mom_eqn_rc}
\end{align}

Note that, while the expressions for the $g_1$ and $g_1^h$ structure functions in \eqref{g1_structs} are valid at large $N_c\&N_f$, the production cross section \eqref{num_xperp} for the polarized $pp$ collisions was derived in a pure-gluon limit \cite{Kovchegov:2024aus}. 
The phenomenology based on this formula, carried out in a preliminary study \cite{JAMCollaborationSmall-xAnalysisGroup:2025tfa}, was performed only in the large-$N_c^{+q}$ limit: this is the large-$N_c \& N_f$ limit, but with internal quark loops neglected.
This is the closest one can be to the true large-$N_c$ limit, while still distinguishing the contributions of different quark flavors.
The drawback of such an approach in our case is that the evolution equation for $\widetilde{I}_q$ would decouple from the rest of the equations in the large-$N_c^{+q}$ limit (see \eqs{mom_eqn_rc}).
Since $\widetilde{I}_q$ does not enter any of the observables, having its evolution equation decoupled from the rest would prevent constraining its initial condition $\widetilde{I}_q^{(0)}$, and hence $\widetilde{I}_q$ itself, leading to an unconstrained quark OAM distribution.
This precludes us from employing the large-$N_c^{+q}$ limit in the present study.

To proceed beyond this complication, we adopt two different approaches to our data analysis.
One is to restrict the analysis to polarized DIS and SIDIS only, taking advantage of the full large-$N_c \& N_f$ evolution equations \eqref{mom_eqn_rc} and \eqref{neigh_mom_eqn_rc} (along with \eqs{eq_LargeNcNf}). 
In Eqs.~\eqref{mom_eqn_rc} and \eqref{neigh_mom_eqn_rc}, $\widetilde{I}_q$ mixes with other quantities, and may be partially constrained by the EIC pseudodata that will be generated. 
The alternative approach is to state that, in a given scheme and for a proton target, the initial conditions for the impact parameter integrated dipole amplitudes and the moment amplitudes are fundamental properties of QCD.
To maximally constrain the extraction of these initial conditions from the future EIC data, we should use \eqs{g1_structs} for the polarized DIS+SIDIS analysis with all the amplitudes found using the large-$N_c \& N_f$ evolution equations \eqref{mom_eqn_rc}, \eqref{neigh_mom_eqn_rc}, and \eqref{eq_LargeNcNf}, along with the production cross section \eqref{num_xperp} for polarized $pp$ collisions.
For the latter, the dipole and moment amplitudes are calculated using the large-$N_c^{+q}$ version of the same evolution equations, but, importantly, with the same initial conditions as for polarized DIS and SIDIS.
While somewhat less orthodox, this approach should provide a stronger constraint on the initial conditions for the dipole and moment amplitudes, and hence on the amplitudes themselves.
The hPDFs and OAM distributions in the second approach will still be calculated using the large-$N_c \& N_f$-evolved amplitudes, while employing the initial conditions determined by the above procedure. 

In the large-$N_c^{+q}$ limit with running coupling, the evolution equations for the moment amplitudes are obtained from \eqs{mom_eqn_rc} and \eqref{neigh_mom_eqn_rc} by putting $N_f=0$ and discarding flavor sums $\sum_q$, thus removing internal quark loops, while retaining the external quark flavors (in the amplitudes $Q_q, Q^{\mathrm{NS}}_q$, $\widetilde{Q}_q$, $\widetilde{I}_q$, and $I_{3q}$). 
Note, however, that Eqs.~\eqref{mom_eqn_rc} and \eqref{neigh_mom_eqn_rc} are not closed, and must be solved simultaneously with (or after solving) the evolution equations \eqref{eq_LargeNcNf} for $Q_q, \widetilde{G},$ $G_2$, and $\widetilde{Q}_q$.
The large-$N_c^{+q}$ limit version of equations \eqref{eq_LargeNcNf}
can be found in Eqs.~(8) of \cite{JAMCollaborationSmall-xAnalysisGroup:2025tfa}. It can also be straightforwardly obtained from Eqs.~\eqref{eq_LargeNcNf} by setting $N_f=0$ and dropping the sums over flavors (replacing those by zeros). The numerical solution of these large-$N_c \& N_f$ and large-$N_c^{+q}$ equations is described in Sec.~\ref{sec:numerical_imp} (see also Refs.~\cite{Adamiak:2021ppq, Adamiak:2023yhz, Adamiak:2023okq, JAMCollaborationSmall-xAnalysisGroup:2025tfa, Kovchegov:2023yzd, Cougoulic:2022gbk}).

\section{Polarized elastic dijet production at small $x$}
\label{sec:dijets}

Having established the connection between the quark and gluon OAM distributions and the polarized dipole moment amplitudes in Sec.~\ref{sec:formalism}, we now turn to the elastic dijet production process in longitudinally polarized $ep$ collisions at small $x$.
We first review the DSA and its azimuthal dependence, which provides sensitivity to the moment amplitudes relevant for extracting the OAM distributions.
We then summarize the corresponding unpolarized cross section, which enters the denominator of the DSA, and identify the independent azimuthal harmonics that will be used in our numerical analysis.

\subsection{Angular integrated form of DSA numerator}

\begin{figure}[ht!]
    \centering
    \includegraphics[width=0.45\linewidth]{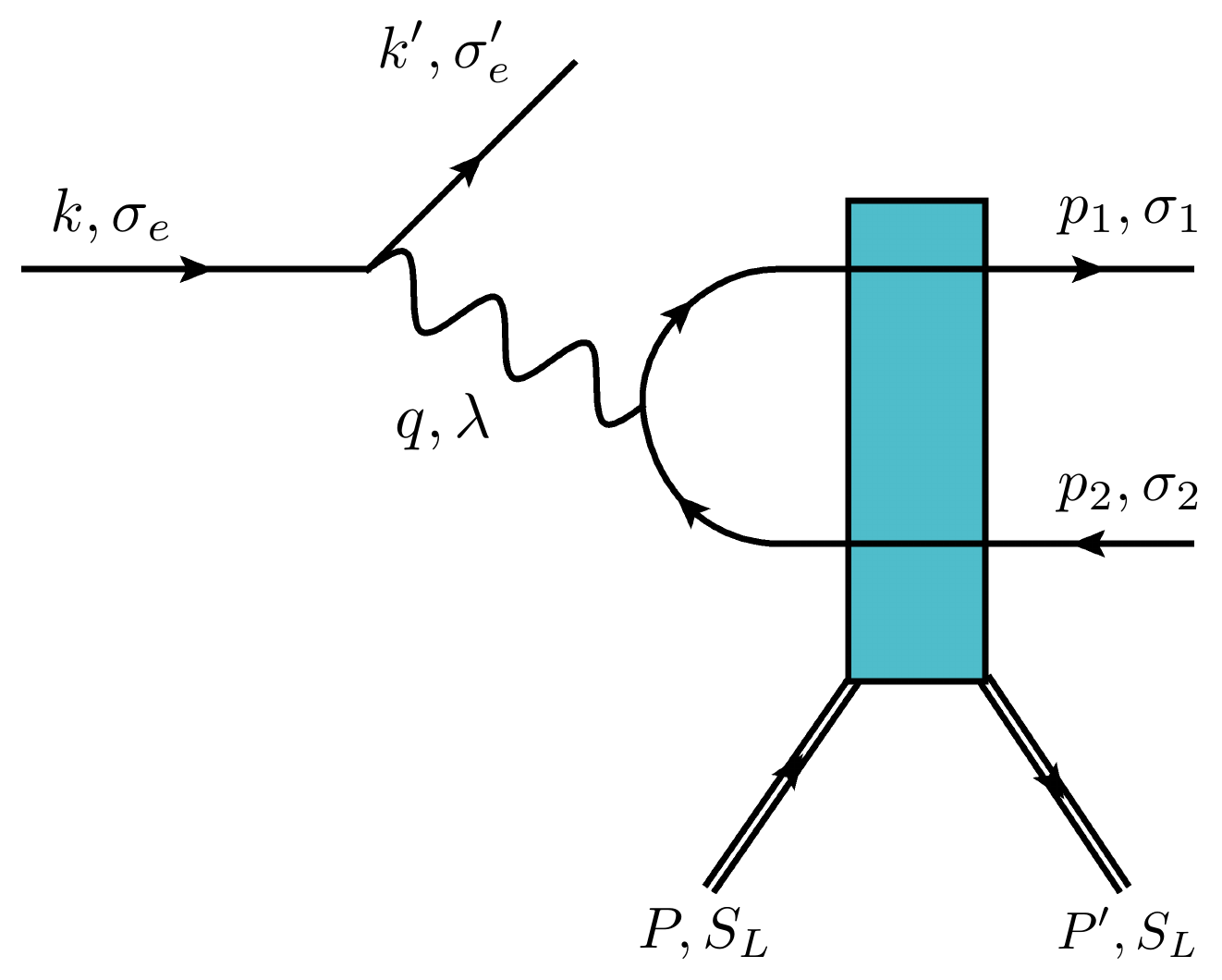}
    \caption{One of the diagrams for elastic dijet electroproduction at small $x$ and to leading order in $\alpha$. The shaded rectangle indicates the proton shockwave. The proton helicity is the same before and after the interaction since the spin-flip process is suppressed in the eikonal and subeikonal approximations.}
    \label{fig:dijet}
\end{figure}

The DSA in $ep$ scattering in the dipole/Trento frame~\cite{Bacchetta:2004jz} is given by~\cite{Diehl:2005pc}
\begin{align}\label{DSA_def}
 A_{LL}^{\gamma^*} &= \frac{ \dd\sigma (++) - \dd\sigma(+-) }{\dd\sigma (++) + \dd\sigma(+-)} \equiv \frac{ \dd\sigma (++) - \dd\sigma(+-) }{2 \, \dd\sigma^\textrm{unpol}} \equiv   \frac{\dd \sigma^{\rm DSA} }{\dd\sigma^\textrm{unpol}},
\end{align}
where the (differential) cross section $\dd \sigma (\sigma_e, S_L)$ depends on the helicities of the electron ($\sigma_e = \pm 1$) and the proton ($S_L = \pm 1$). 
The elastic dijet process at small-$x$ is illustrated in \fig{fig:dijet}, where, to contribute to the DSA, either the quark or the antiquark line should interact sub-eikonally with the shock wave, either in the amplitude or in the complex-conjugate amplitude.

We take the proton target to be moving along the $x^+$ direction and the virtual photon moving along the $x^-$ direction. 
In the dipole/Trento frame \cite{Bacchetta:2004jz, Braun:2005rg, Mantysaari:2020lhf, Bhattacharya:2022vvo,Bhattacharya:2024sck,Hatta:2016aoc, Kovchegov:2024wjs, Kovchegov:2024aus}, the virtual photon has no transverse momentum, $\un q=  0$.
Additionally, the proton moves primarily along the light-cone ``+'' direction such that $P^\mu \approx (P^+, 0^-, \un 0)$. 
The transverse momentum transfer to the proton is $\un P' - \un P = -\un \Delta$, where $\Delta$ is the sum of the jet momenta. 
In the following we neglect the ``+'' and ``$-$'' momentum transfer to the proton. 

In addition, we define the following invariants,
\begin{align}
        Q^2 = - q^2, \hspace{0.5cm}
        y = \frac{P \cdot q}{P \cdot k}, 
        \hspace{0.5cm}
        x = \frac{Q^2}{2 P \cdot q}, 
        \hspace{0.5cm}
        t = \Delta^2 \approx -\Delta_\perp^2,
\end{align}
along with the center-of-mass energy squared of the $ep$ collision, $s=(P+k)^2$, and center-of-mass-energy squared of the photon-proton collision, $W^2 = (P+q)^2$.
Note that in Sec.~\ref{sec:formalism} we denoted the dipole-proton center-of-mass energy squared (the ``energy scale" of the dipole and moment amplitudes) by either $s$ or $z s$. 
Below, for the dijet production process at hand, the energy scale in the dipole and moment amplitudes will be $W^2$. 
(We will consider ${\cal O}(1)$ values of $z$.) 
The variable energy scale $s$ should not be confused with the \textit{electron}-proton center-of-mass energy squared $s=(P+k)^2\approx W^2/y$. 
Below, we will mainly use $s$ to denote the electron--proton center-of-mass energy squared, with the exception of Sec.~\ref{sec:num_tech}, where $s$ will again be the variable energy scale in the evolution equations, whose numerical solution is described in that Section. 
Hence, $s$ in the arguments of the dipole or moment amplitudes, along with the variable redefinitions like those in \eq{log_vars}, will continue to be the energy scale. 
Everywhere else, $s$ will be the $ep$ center-of-mass energy squared.

We define the light-cone momentum fraction of the quark jet by $z=p^-_1/q^-= 1- p_2^- /q^-$, where $p_1 (p_2)$ is the momentum of the quark (antiquark) jet, and define $p = (1-z)\, p_1 - z \,p_2$ to be the relative momentum of the jets.
In this frame the transverse momentum transfer is related to the jet momenta by $\un \Delta \equiv \un p_1 + \un p_2$. 
For brevity, we define $\overline{Q} \equiv Q \sqrt{z(1-z)}$. 

The elastic dijet production cross section in the longitudinally polarized $ep$ high energy collisions was derived in Ref.~\cite{Kovchegov:2024wjs}. 
In the companion paper \cite{Becker:2026xcp}, the expression was corrected to include the contribution of $I_E$ and the internal azimuthal integrations have been carried out as well.
The result is given by 
\begin{align} 
\frac{\dd \sigma^{\mathrm{DSA}}}{\dd y \, \dd Q^2  \, \dd z \, \dd p_\perp \, \dd t\, \, \dd\phi_{kp} \, \dd\phi_{\Delta p} } &= \sum_{q} 
\Bigg[ 
            (2-y)   \, C_{\mathrm{TT}\, q} + 
            \sqrt{2(1-y)} \,
            \cos \phi_{kp}\, \,C_{\mathrm{LT}\,q}^{\cos\phi_{kp}} 
+ \frac{\sqrt{-t}}{p_\perp} (2-y) \cos \phi_{\Delta p}\, 
            C_{\mathrm{TT}\,q}^{\cos \phi_{\Delta p} }
\notag
\\ 
& \hspace{-1cm}
+ \frac{\sqrt{-t}}{p_\perp}  \sqrt{2(1-y)}  
  \bigg(
              \cos\phi_{kp} \cos\phi_{\Delta p}\,  C_{\mathrm{LT}\,q}^{\cos\phi_{kp} \cos\phi_{\Delta p}}
            +  \sin\phi_{kp} \sin\phi_{\Delta p}\,  C_{\mathrm{LT}\,q}^{\sin\phi_{kp} \sin\phi_{\Delta p}}
  \bigg)
\Bigg],
\label{angles2}
\end{align}
where $\phi_{kp} = \phi_k - \phi_p$ and $\phi_{\Delta p} = \phi_{\Delta} - \phi_p$ denote the differences between the azimuthal angles of $\un k $ and $\un p$, and $\un \Delta$ and $\un p$, respectively. 
Here we assume light flavor jets are experimentally indistinguishable and therefore sum over flavors~$q$. 
Note, however, that there are proposals to separate light flavor jets at the EIC \cite{Lee:2022kdn, Kang:2020fka}. 
Additionally, in arriving at \eq{angles2} it was assumed that the quark and antiquark jets are also experimentally indistinguishable \cite{Kovchegov:2024wjs}.
In the following we also assume that the (light) quarks are massless.

There are two contributions to the DSA in \eq{angles2}, as found in Ref.~\cite{Kovchegov:2024wjs}.
The first is the contribution from the case where both photons (one in the amplitude, one in the complex conjugate amplitude) are transversely polarized.
We will refer to this term as the ``TT" contribution. 
The second term arises from the interference between a longitudinally polarized photon and a transversely polarized photon, which we denote by the ``LT" contribution.
The corresponding coefficients are labeled $C_{TT}$ and $C_{LT}$ in \eq{angles2}.

The TT coefficients are given by~\cite{Becker:2026xcp}
\begin{subequations} \label{tt_coeffs}
    \begin{align}
        C_{\mathrm{TT}\,q} &= -  \frac{\alpha^2 e_q^2 N_c^2\, p_\perp}{16  \pi^5 W^2} N_{(11)}^{(00)}
        \Bigg\{ 
        (1- 2 z)^2 {Q_q}_{(11)}^{(00)} + 2\Big[z^2 + (1-z)^2 \Big]  {G_2}_{(11)}^{(00)}
        \Bigg\}, \label{tt_coeffs_a}
        \\ 
         \notag 
        C^{\cos\phi_{\Delta p}}_{\mathrm{TT}\,q} &=  
       \frac{\alpha^2 e_q^2 N_c^2\, p_\perp}{16  \pi^5 W^2} (1-2z) 
\\ \notag &\times 
         \Bigg\{  \frac{(1- 2z)^2}{2} N^{(10)}_{(01)} \, {Q_q}_{(11)}^{(00)}
         + \big[ z^2 + (1-z)^2 \big] N^{(10)}_{(01)}\, {G_2}_{(11)}^{(00)}
         + N_{(11)}^{(00)}  \left[ \frac{1}{2} {Q_q}_{(11)}^{(00)} + {I_{3q}}_{(11)}^{(00)} - {I_{3q}}^{(10)}_{(01)}\right]
\\
         & \hspace{0.4cm}
         -\Big[z^2 + (1-z)^2 \Big] N_{(11)}^{(00)} \Bigg[ {Q_q}_{(21)}^{(10)} + 2\, {G_2}_{(21)}^{(10)}
         + {I_4}^{(10)}_{(21)} 
         + {I_4}^{(01)}_{(10)} 
         - {I_5}_{(11)}^{(00)}
         + {I_5}^{(01)}_{(10)}
         + {I_5}^{(10)}_{(01)}
         \Bigg]
\notag \\ 
        & \hspace{0.4cm}
        + \, N_{(11)}^{(00)} \, \Bigg[ 2z(1-z) \, {I_E}_{(11)}^{(00)}
                 + \Big[ z^2 + (1-z)^2 \Big] \, {I_E}^{(01)}_{(10)} \Bigg]
        \Bigg\},
\label{tt_coeffs_b}
    \end{align}
\end{subequations}
where $\alpha$ is the electromagnetic coupling, and
we have defined the Fourier-Bessel transform of a dipole amplitude as 
\begin{align} \label{fb_transform}
    F^{(cd)}_{(ab)}(p_\perp, \overline{Q}, W^2) \equiv \int \displaylimits^\infty_0 \dd x_{12} \, x_{12}\, \left( p_\perp x_{12} \right)^c \left( \overline{Q} x_{12} \right)^d J_a(p_\perp x_{12} ) K_b \left( \overline{Q} \, x_{12} \right)\, F(x_{12}^2, W^2),
\end{align}
with $F(x_{12}^2, W^2)$ any dipole amplitude, polarized or unpolarized, integrated over impact parameters. 
For brevity, we will suppress the arguments of the Fourier-Bessel transforms $F^{(cd)}_{(ab)}$ in the remainder of this paper.
Here $x_{12}$ denotes the transverse size of a dipole. 
Furthermore, $N^{(cd)}_{(ab)}$ appearing in \eqs{tt_coeffs} is the Fourier-Bessel transform of the unpolarized dipole amplitude $N(x_{12}^2, W^2)$ \cite{Kovchegov:1999yj}, integrated over all impact parameters (see the discussion below \eqs{dcoeffs} for details).

The LT coefficients are \cite{Becker:2026xcp} 
\begin{subequations} \label{lt_coeffs}
    \begin{align}
         C_{\mathrm{LT}\,q}^{\cos\phi_{kp}} 
         &= -  \frac{\alpha^2  e_q^2 N_c^2 \sqrt{z(1-z)} \, p_\perp }{ 8 \sqrt{2} \,\pi^5  W^2} (1-2z) 
        \Bigg\{ 
        N_{(00)}^{(00)} \Big[ 2 \, {G_2}_{(11)}^{(00)} -  {Q_q}_{(11)}^{(00)} \Big]
        - N_{(11)}^{(00)}  {Q_q}_{(00)}^{(00)} 
        \Bigg\}, \label{lt_coeffs_a}
\\ \notag
        C_{\mathrm{LT} \,q}^{\cos\phi_{kp} \cos\phi_{\Delta p}} 
        &=  -  \frac{\alpha^2  e_q^2 N_c^2 \sqrt{z(1-z)} \, p_\perp }{ 8 \sqrt{2} \,\pi^5  W^2}
         \Bigg\{  N_{(00)}^{(00)} \Big[
         {I_{3q}}_{(11)}^{(00)}  - {I_{3q}}_{(01)}^{(10)}
         \Big]
         - \frac{1}{2}(1-2z)^2 \Big[N_{(11)}^{(00)} - N_{(01)}^{(10)}  \Big] {Q_q}_{(00)}^{(00)}
\\ \notag & \hspace{-2cm}
         + N_{(11)}^{(00)} {I_{3q}}^{(10)}_{(10)} + \Big[ z^2 + (1-z)^2 \Big] N_{(00)}^{(00)} \Big[ {Q_q}_{(01)}^{(10)} - {Q_q}_{(11)}^{(00)} \Big]
         - \Big[ z^2 + (1-z)^2 \Big] N_{(11)}^{(00)}  {Q_q}_{(10)}^{(10)}
\\ & \hspace{-2cm}
         - \frac{1}{2} (1-2z)^2 N_{(10)}^{(10)} {Q_q}_{(11)}^{(00)} 
         + (1-2z)^2 N_{(00)}^{(00)} \Bigg[3 \, {G_2}_{(11)}^{(00)}  - 2 \,{G_2}^{(10)}_{(01)}  + {I_4}^{(10)}_{(21)}  + {I_4}^{(01)}_{(10)} 
         - {I_5}_{(11)}^{(00)} + {I_5}^{(10)}_{(01)}  + {I_5}^{(01)}_{(10)}
         \Bigg] 
\notag \\ & \hspace{-2cm} 
         + (1-2z)^2 N_{(10)}^{(10)} {G_2}_{(11)}^{(00)} 
         + \, N_{(00)}^{(00)} 
           \Bigg[ 4 z(1-z) \, {I_E}_{(11)}^{(00)}
                + (1-2z)^2 \, {I_E}^{(01)}_{(10)} 
           \Bigg]
         \Bigg\}, 
\\ \notag 
        C_{\mathrm{LT} \,q}^{\sin\phi_{kp} \sin\phi_{\Delta p}} &=  - \frac{\alpha^2  e_q^2 N_c^2 \sqrt{z(1-z)} \, p_\perp }{ 8 \sqrt{2} \,\pi^5  W^2}
         \Bigg\{  \Big[ z^2 + (1-z)^2 \Big] N_{(00)}^{(00)} {Q_q}_{(11)}^{(00)} - N_{(00)}^{(00)} {I_{3q}}_{(11)}^{(00)}
\\ \notag & \hspace{-2cm}
        + \frac{1}{2}(1-2z)^2 N_{(11)}^{(00)} {Q_q}_{(00)}^{(00)}
         + (1-2z)^2 N_{(11)}^{(00)} \Bigg[ 
        {G_2}^{(10)}_{(10)} + {I_4}^{(11)}_{(11)}
        + {I_4}^{(20)}_{(00)} + {I_5}^{(11)}_{(11)}
        - {I_5}^{(10)}_{(10)} + {I_5}^{(20)}_{(00)}
        \Bigg]
\\ \notag & \hspace{-2cm}
        + (1-2z)^2 N_{(00)}^{(00)}\Bigg[ {G_2}^{(10)}_{(01)} - 3\, {G_2}_{(11)}^{(00)} 
         - 2 \, {I_4}_{(11)}^{(00)} + {I_4}^{(10)}_{(01)} - {I_4}^{(20)}_{(11)} - {I_4}^{(01)}_{(10)}
        + {I_4}^{(11)}_{(00)}
        + {I_5}_{(11)}^{(00)} 
        - {I_5}^{(20)}_{(11)} - {I_5}^{(01)}_{(10)} + {I_5}^{(11)}_{(00)} 
        \Bigg]
\\ \notag 
        &
        \hspace{-2cm}
        + \, N_{(00)}^{(00)} \Bigg( 4 z(1-z) \Big[ {I_E}_{(11)}^{(00)} - {I_E}^{(10)}_{(21)} \Big]
        + (1-2z)^2 \Big[ {I_E}^{(01)}_{(10)} - {I_E}^{(11)}_{(20)} \Big] \Bigg)
\\ \label{c_sinsin}
        & \hspace{-2cm}
        - (1-2z)^2 \, N_{(11)}^{(00)} \Bigg[ 2 \, {I_E}^{(10)}_{(10)} - {I_E}^{(11)}_{(11)} \Bigg]
       \Bigg\}.
    \end{align}
\end{subequations}
Equations~(\ref{tt_coeffs}) and (\ref{lt_coeffs}), along with \eq{angles2}, are the form of the DSA numerator we will use in the analysis below.
To analyze the asymmetry itself, we also need expressions for the unpolarized cross section in the denominator of \eq{DSA_def}, which we turn to next.

\subsection{Unpolarized elastic dijet production} 

Unpolarized elastic and inelastic dijet production in $ep$ collisions has been extensively studied at small~$x$~\cite{Bartels:1996tc, Dominguez:2010xd, Dominguez:2011br, Dominguez:2011wm, Metz:2011wb, Dumitru:2015gaa, Altinoluk:2015dpi, Hatta:2016dxp, Dumitru:2016jku, Boussarie:2016ogo, Dumitru:2018kuw, Mantysaari:2019csc, Salazar:2019ncp, Hatta:2019ixj, Mantysaari:2019hkq, Fujii:2020bkl, Hatta:2020bgy, Boer:2021upt,Zhao:2021kae, Boussarie:2021ybe, vanHameren:2021sqc, Caucal:2021ent, Iancu:2021rup, Taels:2022tza, Iancu:2022lcw, Caucal:2022ulg, Altinoluk:2022jkk, Rodriguez-Aguilar:2023ihz, Altinoluk:2023qfr, Caucal:2023fsf, Caucal:2023nci, Kar:2023jkn, Shao:2024nor, Caucal:2024bae, Agostini:2024xqs, Altinoluk:2024zom, Caucal:2025xxh, Kar:2026vzk}. 
The leading order picture of dijet production in the leading logarithmic approximation (LLA), including saturation effects, has been thoroughly analyzed \cite{Bartels:1996tc, Dominguez:2010xd, Dominguez:2011br, Dominguez:2011wm, Metz:2011wb, Dumitru:2015gaa, Altinoluk:2015dpi, Hatta:2016dxp, Dumitru:2018kuw, Mantysaari:2019csc, Salazar:2019ncp, Fujii:2020bkl, Mantysaari:2019hkq, Boussarie:2021ybe, vanHameren:2021sqc, Iancu:2021rup, Iancu:2022lcw, Kar:2023jkn}.
Furthermore, there has been a considerable effort in recent years to quantify various higher-order corrections to this process~\cite{Dumitru:2016jku, Boussarie:2016ogo, Hatta:2019ixj,Hatta:2020bgy, Zhao:2021kae, Caucal:2021ent,Taels:2022tza, Caucal:2022ulg,Altinoluk:2022jkk, Rodriguez-Aguilar:2023ihz, Altinoluk:2023qfr, Caucal:2023fsf, Caucal:2023nci, Shao:2024nor, Caucal:2024bae, Agostini:2024xqs, Altinoluk:2024zom, Caucal:2025xxh}.
In this work, we focus exclusively on the LLA result, which we reproduce below and use for the unpolarized denominator in the DSA in \eq{DSA_def}.
We leave the inclusion of higher-order effects for future work. 
One of the contributing diagrams for elastic dijet electroproduction at small $x$ is shown in \fig{fig:dijet}, where now both the quark and antiquark lines interact eikonally with the shock wave.

Analogously to the polarized case considered in Ref.~\cite{Kovchegov:2024wjs}, the unpolarized elastic dijet production cross section can be expressed in terms of the virtual photon-proton cross sections via \cite{Bartels:1996tc, Kovchegov:2024wjs}
\begin{align} 
\notag 
E_{k'} z(1-z)\, 
\frac{\dd \sigma^{\mathrm{unpol}}}{\dd^3 k'\, \dd^2 p\, \dd^2 \Delta \, \dd z}
&= \frac{\alpha \, z(1-z)}{4 \pi^2  \, Q^2 \, y} \, 
\Bigg\{ 
\big[ 1 + (1-y)^2 \big]  \,
\sum_{\lambda = \pm 1} 
\frac{\dd\sigma^{\gamma^* p \to q \bar{q} p'}_{\lambda \lambda}}{\dd^2 p\, \dd^2 \Delta\, \dd z}  
+
4 (1-y)\, \frac{\dd \sigma^{\gamma^* p \to q\bar{q} p'}_{00}}{\dd^2p\, \dd^2 \Delta \, \dd z} 
\\ 
\label{unpol_dijets}
&  \hspace{-1.5cm}
- 2 (1-y) \, \sum_{\lambda = \pm 1} 
e^{-2 i \lambda \, \phi} \, 
\frac{\dd \sigma^{\gamma^* p \to q\bar{q} p'}_{\lambda, - \lambda}}
     {\dd^2 p\, \dd^2 \Delta \, \dd z} 
+ (2-y) \, \sqrt{2 (1-y)} \,  
\sum_{\lambda = \pm 1} \lambda \, 
\left[ 
  e^{i \lambda \, \phi}\, 
  \frac{\dd\sigma^{\gamma^* p \to q\bar{q} p'}_{0\lambda}}{\dd^2 p \, \dd^2 \Delta \, \dd z} 
  + \cc 
\right] 
\Bigg\} .
\end{align}
For the denominator of the DSA, we do not need the angular dependence of the unpolarized cross section. 
Integrating Eq.~(\ref{unpol_dijets}) over azimuthal angles we then obtain
\begin{align} \label{unpol_dijets_aa}
\frac{\dd \sigma^{\mathrm{unpol}}}{\dd y\, \dd Q^2\, \dd z\, \dd p_\perp \dd t} &= \frac{\alpha}{4} \frac{p_\perp}{Q^2 \, y}  
 \int \displaylimits^{2\pi}_0 \dd\phi_{\Delta p }\Bigg\{ \left[ 1 + (1-y)^2 \right] \sum_{\lambda = \pm 1} \frac{\dd \sigma^{\gamma^* p \to q \bar{q} p'}_{\lambda \lambda}}{\dd^2 p\, \dd^2 \Delta \, \dd z} 
 + 4 (1-y) \,  \, \frac{\dd\sigma^{\gamma^* p \to q\bar{q} p'}_{00}}{\dd^2p \, \dd^2 \Delta \, \dd z} \Bigg\}.
\end{align}
The first term in the braces in Eq.~(\ref{unpol_dijets_aa}) comes from both photons being transversely polarized, and therefore the TT term in Eq.~(\ref{unpol_dijets}).
The second term is from both photons longitudinally polarized, and hence the LL term.
Starting with the TT term, to eikonal order we have,
\begin{align}
\notag 
    z (1-z)\sum_\lambda \frac{\dd\sigma_{\lambda \lambda}^{\gamma^* p \to \bar{q}q p'}}{\dd^2 p_1 \dd^2 p_2 \dd z} 
    &= \sum_q \frac{1}{(2\pi)^5} \int \dd^2 x_1 \, \dd^2 x_2 \, \dd^2 x_{1'} \, \dd^2 x_{2'} \, e^{-i \un p_1 \cdot \un x_{11'} - i \un p_2 \cdot \un x_{22'}} 
    \\[-0.3cm] 
     \label{TT_unpol}
    & \hspace{2.3cm}  \times 
    \Phi^{\prime [2]q}_{\mathrm{TT}}(\un x_{12}, \un x_{1'2'}, z)\, N_{12}(W^2)\, N_{1'2'}(W^2),
\end{align}
where we have also included, in addition to the process in \fig{fig:dijet}, the contribution where the photon fluctuates into the quark-antiquark pair after the shockwave interaction.
Furthermore, we have employed the wave function overlap from Eq.~(65a) of \cite{Kovchegov:2024wjs}, defined as
\begin{align}
    \Phi^{\prime [2]q}_{\mathrm{TT}}(\un x_{12}, \un x_{1'2'}, z) = 
         \frac{2 \, \alpha e_q^2 N_c^2 Q^2}{\pi} z^2 (1-z)^2 \left[ z^2 + (1-z)^2 \right] \frac{\un x_{12} \cdot \un x_{1'2'} }{x_{12} x_{1'2'}} K_1 \left(x_{12} \overline{Q}  \right) K_1 \left(x_{1'2'} \overline{Q}  \right).
\end{align}
In arriving at \eq{TT_unpol}, we have also assumed that the unpolarized dipole amplitude $N$ is real in the high-energy kinematics considered here. 
Similar to the numerator of the DSA \cite{Kovchegov:2024wjs}, one can show that the corrections to this approximation are at least ${\cal O}(\as^2)$ suppressed and therefore are outside of the precision of our calculation. 
Additionally, we have again summed over light quark flavors and assumed that the quark and antiquark jets are interchangeable.

The LL term is computed similarly, 
\begin{align} 
 \notag 
    z (1-z) \frac{\dd\sigma_{00}^{\gamma^* p \to \bar{q}q p'}}{\dd^2 p_1 \dd^2 p_2 \dd z} 
    &= \sum_q \frac{1}{2(2\pi)^5} \int \dd^2 x_1 \, \dd^2 x_2 \, \dd^2 x_{1'} \, \dd^2 x_{2'} \, e^{-i \un p_1 \cdot \un x_{11'} - i \un p_2 \cdot \un x_{22'}} 
    \\[-0.3cm]
    \label{LL_unpol}
    & \hspace{2.5cm} \times 
    \Phi^{[2] q}_{\mathrm{LL}}(\un x_{12}, \un x_{1'2'}, z)\, N_{12}(W^2)\, N_{1'2'}(W^2),
\end{align}
with the wave function overlap given by \cite{Kovchegov:2024wjs}
\begin{align}
    \Phi^{[2]q}_{\mathrm{LL}}(\un x_{12}, \un x_{1'2'} , z) &= 
        \frac{8 \, \alpha e_q^2 N_c^2 Q^2}{\pi} z^3 (1-z)^3 K_0 \left(x_{12} \overline{Q}  \right) K_0 \left(x_{1'2'} \overline{Q}  \right).
\end{align}
Since we are interested in the region of small $\Delta_\perp$, we follow Ref.~\cite{Kovchegov:2024wjs} and expand the Fourier phase in \eqs{TT_unpol} and (\ref{LL_unpol}) to the lowest order in $\Delta_\perp$ using 
\begin{align} \label{phase_expand}
    e^{-i \un p_{1} \cdot \un x_{1 1^\prime} - i \un p_2 \cdot \un x_{2 2^\prime}} = e^{-i \un p \cdot (\un x_{12} - \un x_{1^\prime 2^\prime})}
    e^{-i \un \Delta \cdot (\un b - \un b')} =
    e^{-i \un p \cdot (\un x_{12} - \un x_{1^\prime 2^\prime})} \Big[ 
    1 + \mathcal{O}(\Delta_\perp) \Big],
\end{align}
and assuming that  $p_\perp, Q \gg \Lambda_{\mathrm{QCD}}   \gg \Delta_\perp$. We then integrate over $\un b = z\, \un x_1 + (1-z)\, \un x_2$ and $\un b' = z\, \un x_{1'} + (1-z)\, \un x_{2'}$, and finally integrate over the azimuthal angle $\phi_{\Delta p}$.
Analogously to \eq{angles2}, we decompose the unpolarized cross section as (see Refs.~\cite{Altinoluk:2015dpi, Boer:2021upt} for similar decompositions)
\begin{align} \label{unpol_decomp}
     \frac{\dd \sigma^{\mathrm{unpol}}}{\dd y \, \dd Q^2  \, \dd z \, \dd p_\perp \, \dd t} 
     &= \sum_q \left[ \frac{1+ (1-y)^2}{y}
        D_{\mathrm{TT}\,q} 
      + \frac{4\,(1-y) }{y}
        D_{\mathrm{LL}\,q}
        \right],
\end{align}
where the coefficients obtained after the $\Delta_\perp$ expansion and integration over azimuthal angles are
\begin{subequations} \label{dcoeffs}
    \begin{align}
    D_{\mathrm{TT}\,q} &= 
        \frac{\alpha^2 e_q^2 N_c^2 z (1-z) \left[ z^2 + (1-z)^2\right] \, p_\perp}{8 \pi^3} 
    \Big[ N_{(11)}^{(00)} \Big]^2,
    \\ 
    D_{\mathrm{LL}\,q} &=
    \frac{\alpha^2  e_q^2 N_c^2 z^2 (1-z)^2 }{4 \pi^3} 
    \Big[ N_{(00)}^{(00)} \Big]^2.
    \end{align}
\end{subequations}
Similarly to \eqs{tt_coeffs} and \eqref{lt_coeffs}, we have used the Fourier-Bessel transform in \eq{fb_transform}. 
Equation~\eqref{unpol_decomp} along with the coefficients in \eqs{dcoeffs} give the angular-integrated form of the denominator of the DSA that we will use below.

To evaluate either the numerator or the denominator of the DSA in \eqs{angles2} and (\ref{unpol_decomp}), respectively, one needs the unpolarized dipole amplitude $N$.
The LLA evolution of $N$ is given by the Balitsky-Kovchegov (BK) evolution equation in the large-$N_c$ limit \cite{Balitsky:1995ub, Balitsky:1998ya, Kovchegov:1999yj, Kovchegov:1999ua} and by the Jalilian-Marian-Iancu-McLerran-Weigert-Leonidov-Kovner (JIMWLK) evolution equation for finite $N_c$ \cite{Jalilian-Marian:1997ubg, Jalilian-Marian:1997jhx, Weigert:2000gi, Iancu:2001ad, Iancu:2000hn, Ferreiro:2001qy}. 
Here we will employ the large-$N_c$ limit and therefore use the BK equation to evolve $N(x_{12}^2, W^2)$.
Further, we employ the Balitsky prescription for the running-coupling version of the BK equation (rcBK) \cite{Balitsky:2006wa} ({\it cf.} Refs.~\cite{Gardi:2006rp, Kovchegov:2006vj}), which is given by 
\begin{align} 
\notag 
    \frac{\partial {\cal N} (x_{10}^2, W^2)}{\partial \ln (W^2/\Lambda^2)}  &= \int \dd^2 x_2 \, K^{\mathrm{Bal.}}_{\mathrm{rc}}(\un x_{10}, \un x_{20}, \un x_{21}) 
    \\ 
    \label{rcBK}
    & \times 
    \Big[ {\cal N} (x_{12}^2, W^2) + {\cal N}(x_{20}^2, W^2) - {\cal N}(x_{10}^2, W^2) - {\cal N}(x_{12}^2, W^2) \, {\cal N}(x_{20}^2, W^2)\Big],
\end{align}
where we have averaged over impact parameters such that one can neglect impact-parameter dependence (while assuming that the averaging can be done independently for each dipole amplitude in the nonlinear term).
The impact-parameter averaged dipole amplitude ${\cal N}$ is defined by 
\begin{align}
    {\cal N} (x_{10}^2, W^2) \equiv \frac{N (x_{10}^2, W^2)}{\sigma_0 /2},
\end{align}
where the parameter $\sigma_0 /2$ represents the transverse area of the proton. 
The kernel in \eq{rcBK} is given by \cite{Balitsky:2006wa}
\begin{align}
    K_{\mathrm{rc}}^{\mathrm{Bal.}}(\un x_{10}, \un x_{20}, \un x_{21}) = \frac{\as(1/x_{10}^2) N_c}{2\pi^2} 
    \left[
        \frac{x_{10}^2}{x_{20}^2 x_{21}^2} + \frac{1}{x_{21}^2} 
        \bigg( \frac{\as(1/x_{21}^2)}{\as(1/x_{20}^2)} -1 
        \bigg)
        +  \frac{1}{x_{20}^2} 
        \bigg( \frac{\as(1/x_{20}^2)}{\as(1/x_{21}^2)} -1 
        \bigg)
    \right], 
\end{align}
where we take the running coupling in position space to be 
\begin{align} \label{xspace_rc}
    \as(1/x_{10}^2) = \frac{12\pi}{11 N_c - 2 N_f} 
    \ln^{-1}\bigg( \frac{4 \,C^2}{x_{10}^2 \Lambda^2_{\mathrm{QCD}}} \bigg),
\end{align}
with $C^2$ a parameter that reflects the uncertainty from the Fourier-transform.

For the initial condition of \eq{rcBK}, ${\cal N}(x_{10}^2, W^2_0)$, we employ a generalized McLerran-Venugopalan model \cite{McLerran:1993ka, McLerran:1994vd, McLerran:1993ni}, 
\begin{align} \label{gMV}
    N(x_{10}^2, W^2_0) = \frac{\sigma_0}{2} \, {\cal N}(x_{10}^2, W^2_0)  = \frac{\sigma_0}{2} 
    \left\{ 1 - \exp\bigg[ - \frac{\left( x_{10}^2  Q_{s_0}^2 \right)^\gamma }{4} \ln \left( \frac{1}{x_{10} \Lambda_{\rm QCD}} + e \right) \bigg] 
    \right\},
\end{align}
where $\sigma_0, Q_{s_0}^2, \gamma$, along with $C^2$ in \eq{xspace_rc}, are parameters to be extracted from data. 
In this study we use the best fit values from Albacete {\it et al.}~\cite{Albacete:2009fh}, given in Table \ref{tab:mv}. 
There is a newer result \cite{Casuga:2023dcf} that uses a Bayesian setup to obtain the posterior distributions of the fit parameters, however, we leave the inclusion of this for future work, and in the analysis below we fix the initial conditions of the unpolarized dipole amplitude to the parameters in Table~\ref{tab:mv}. 
To solve \eq{rcBK} numerically and extract the unpolarized dipole amplitude, we utilize the code \texttt{rcbk}\footnote{The \texttt{rcbk} code is available here: \href{https://github.com/hejajama/rcbk}{https://github.com/hejajama/rcbk.} We thank Heikki Mäntysaari for 
allowing us to use this code.}.

\begin{table}[h]
    \centering
    \caption{Best fit parameters from Ref.~\cite{Albacete:2009fh} for the generalized McLerran-Venugopalan model in \eq{gMV}.}
    \begin{tabular}{c|c|c|c}
         \hline 
         ~$\sigma_0$ [mb] \,\, & \,\,  $Q_{s0}^2\,\, \left[\mathrm{GeV}^2 \right]$ \,\, & \,\, $\gamma$ \,\,& \,\,$C^2$ \\
         \hline 
         32.77 & 0.15 & ~1.13~ & ~6.5~ \\
         \hline 
    \end{tabular}
    \label{tab:mv}
\end{table}

\subsection{Summary of the DSA azimuthal harmonics}

In the notation of Ref.~\cite{Bacchetta:2004jz}, we define the azimuthal harmonics of the DSA as
\begin{align} \label{az_mom}
    \langle f(\phi_{kp}, \phi_{\Delta p}) \rangle  = \frac{\int \dd\phi_{kp} \, \dd\phi_{\Delta p}
    f(\phi_{kp}, \phi_{\Delta p}) \, d \sigma^{\mathrm{DSA}}
    }{\int \dd\phi_{kp} \, \dd\phi_{\Delta p}\, d\sigma^{\mathrm{unpol}}}.
\end{align}
Using \eq{angles2} in the numerator of \eq{az_mom} and \eq{unpol_decomp} in the denominator, we identify five distinct azimuthal harmonics in the elastic dijet DSA,
\begin{subequations} \label{correlations}
\begin{align} \label{dsa_1}
     \left\langle 1 \right\rangle  &= \frac{2\pi}{D}
    \int \dd X \, 
    (2-y )\,  \sum_q C_{\mathrm{TT}\,q}, 
\\ \label{cos_lt}
      \left\langle \cos \phi_{kp}  \right\rangle  &= \frac{\pi}{D}
        \int \dd X \, 
        \sqrt{2(1-y)}\, 
        \sum_q C_{\mathrm{LT}\,q}^{\cos\phi_{kp}},
\\ \label{cos}
    \left\langle \cos \phi_{\Delta p} \right\rangle  &= 
    \frac{\pi }{D}
    \int \dd X \,
    (2-y) \frac{\sqrt{-t}}{p_\perp}
        \sum_q \,C_{\mathrm{TT}\,q}^{\cos \phi_{\Delta p} },
\\ \label{coscos}
    \left\langle  \cos \phi_{kp}  \cos \phi_{\Delta p} \right\rangle  &= 
    \frac{\pi}{2\, D}
        \int \dd X \,
        \sqrt{2(1-y)} \frac{\sqrt{-t}}{p_\perp}
        \sum_q \,
         C_{\mathrm{LT}\,q}^{\cos\phi_{kp} \cos\phi_{\Delta p}},
\\ \label{sinsin}
    \left\langle \sin \phi_{kp}   \sin \phi_{\Delta p} \right\rangle  &= 
    \frac{\pi}{2\, D}
    \int \dd X \,
    \sqrt{2(1-y)}  \frac{\sqrt{-t}}{p_\perp}
         \sum_q\, C_{\mathrm{LT}\,q}^{\sin\phi_{kp} \sin\phi_{\Delta p}},
\end{align}
\end{subequations}
where the denominator is given by 
\begin{align} \label{denom_def}
    D = 
    \int \dd X \, 
    \sum_q \,
        \left\{ \frac{1+ (1-y)^2}{y}
        D_{\mathrm{TT}\,q} 
    + \frac{4(1-y) }{y}
        D_{\mathrm{LL}\,q}
        \right\}. 
\end{align}
The coefficients in \eqs{correlations} and \eqref{denom_def} are defined in \eqs{tt_coeffs}, (\ref{lt_coeffs}) and \eqref{dcoeffs}. 
The measure $\dd X$ denotes integration over a subset of the variables $Q^2, y, t, z$, and $p_\perp$. 
As will be seen below, for the purpose of extracting the OAM distributions via \eqs{correlations}, it is advantageous to integrate over $Q^2$ only, such that $\dd X = \dd Q^2$.

As can be seen from \eqs{tt_coeffs} and (\ref{lt_coeffs}), only azimuthal harmonics 
$\left\langle \cos \phi_{\Delta p} \right\rangle$,
$\left\langle  \cos \phi_{kp}  \cos \phi_{\Delta p} \right\rangle$ and
$\left\langle \sin \phi_{kp}   \sin \phi_{\Delta p} \right\rangle$
explicitly depend on the moment amplitudes and therefore may allow one to extract the OAM distributions. 
By contrast, the harmonics
$\left\langle 1 \right\rangle$ and 
$\left\langle \cos \phi_{kp} \right\rangle$ 
depend on $Q_q$ and $G_2$ alone, and can therefore only constrain the hPDFs.
Consequently, below we will assess the feasibility of extracting the hPDFs as well as the OAM distributions from the complete set of azimuthal harmonics in \eqs{correlations}.

\section{Elastic dihadron production at small $x$}
\label{sec:dihadrons}

Before assessing the impact of elastic dijet measurements on the hPDFs and OAM distributions, one should consider whether there are enough observables to constrain the entire set of moment amplitudes.
To extract the quark and gluon OAM distributions---including the individual quark flavor contributions---requires nine independent moment amplitudes: $I_{3u}$, $I_{3d}$, $I_{3s}$, $\widetilde{I}_3$, $I_4$, $I_5$, $\widetilde{I}_u$, $\widetilde{I}_d$, and $\widetilde{I}_s$. 
Clearly, the three independent harmonics, which depend on the moment amplitudes in \eqs{correlations}, are insufficient alone to constrain these nine amplitudes. 
To provide additional constraints, we can use additional targets beyond the proton, such as the deuteron or $^3\mathrm{He}$.
Using isospin symmetry, this can provide an additional three harmonics which depend on the moment amplitudes distinctly from \eqs{correlations}.
However, since \eqs{correlations} contain $C_{\mathrm{TT}\,q}$ and $C_{\mathrm{LT}\,q}$ with the ($e_q^2$-weighted) sums over light flavors, we effectively only have two linear combinations of the three flavor-dependent moment amplitudes $I_{3u}$, $I_{3d}$, and $I_{3s}$. 
To constrain the strange amplitude $I_{3s}$ therefore requires yet another observable.
In this study, in addition to the elastic dijet process, we also include elastic dihadron production in polarized $ep$ collisions.
In particular, we will consider elastic $\pi^+ \pi^-$ and $K^+ K^-$ production channels, which will again come with three harmonics for each, thereby tripling the number of observables and generating different linear combinations for different flavor contributions.

Following Refs.~\cite{Collins:2011zzd, Altarelli:1979kv, Fucilla:2022wcg, Caucal:2024nsb, Bergabo:2022tcu, Zheng:2014vka}, we use collinear factorization to write the elastic dihadron cross section as 
\begin{align} \label{dh0}
    \frac{\dd \sigma^{e p \to e p h_1 h_2}}{\dd z_{h_1} \dd z_{h_2} \dd^2 p_{h_1} \dd^2 p_{h_2}} &= \sum_q\int \limits^1_{z_{h_1}} \frac{\dd z_1}{z_1} 
    \int \limits^1_{z_{h_2}} \frac{\dd z_2}{z_2}\,  
    D^{h_1/q}_1\!\!\left(\frac{z_{h_1}}{z_1}  \right)
    D^{h_2/\bar{q}}_1\!\!\left(\frac{z_{h_2}}{z_2} \right)
    \frac{\dd\sigma^{e p \to e p q \bar{q}}}{\dd z_1 \dd z_2\, \dd^2 p_1 \, \dd^2 p_2} 
    \Bigg|_{\substack{\un p_1 = (z_1/z_{h_1})\, \un p_{h_1} \\ 
                      \un p_2 = (z_2/z_{h_2})\, \un p_{h_2}}}
    + (h_1 \leftrightarrow h_2),
\end{align}
where $\un p_{h_i}$ are the hadron momenta, and $z_{h_i}$ are the hadron light-cone momentum fractions of the virtual photon ($z_{h_i} = p_{h_i}^- / q^-$).
The $D_1$ functions are the  collinear fragmentation functions, and $z_{1} (z_2)$ is the ``minus'' light-cone momentum fraction of the quark (antiquark) with respect to the photon. 
In \eq{dh0}, the term $(h_1 \leftrightarrow h_2)$ denotes the contribution where the antiquark fragments into hadron $h_1$ and the quark fragments into hadron $h_2$.
We must sum the two contributions (quark fragmenting into $h_1$, antiquark fragmenting into $h_2$, and vice versa) since we cannot distinguish them in the detector.
Furthermore, since the hadrons should be approximately collinear to the partons, we have used \cite{Collins:2011zzd, Altarelli:1979kv, Fucilla:2022wcg, Caucal:2024nsb, Bergabo:2022tcu, Zheng:2014vka}
\begin{align}
    \un p_{i} = \frac{z_i}{z_{h_i}}\, \un p_{{h_i}},
\end{align}
for $i = 1,2$.

To utilize the partonic dijet calculation from Ref.~\cite{Kovchegov:2024wjs},  we must, strictly speaking, restrict ourselves to the case where a hadron is identified in each jet of the dijet configuration. 
We assume the invariant mass of the dihadron is large enough to neglect contributions where this is produced from the fragmentation of a single quark (or antiquark), which would involve dihadron fragmentation functions~\cite{Collins:1993kq, Jaffe:1997hf, Bianconi:1999cd, Radici:2001na, Courtoy:2012ry, Pitonyak:2023gjx, Cocuzza:2023vqs, Rogers:2024nhb, Pitonyak:2025lin, Mahaut:2025hie}.
However, even in this case, \eq{dh0} is not fully correct:~instead of just the collinear fragmentation functions, the partonic cross section need to be convoluted with the fragmenting jet functions (FJFs), which describe the fragmentation of a parton into a hadron within a jet \cite{Procura:2009vm, Jain:2011xz, Procura:2011aq, Kang:2016ehg, Kaufmann:2015hma, Ritzmann:2014mka}. 
For consistency with our treatment of jet functions in both the elastic dijet process in Sec.~\ref{sec:dijets} and the single-inclusive jet production cross section in $pp$ collisions \cite{JAMCollaborationSmall-xAnalysisGroup:2025tfa}, we approximate the jet functions by their leading order expressions $\mathcal{J}(z_j) \sim \delta(z_j -1)$, justifying neglecting higher-order corrections as non-DLA. 
At this order, the FJFs reduce to the usual collinear fragmentation functions and \eq{dh0} is valid \cite{Procura:2009vm, Jain:2011xz, Procura:2011aq, Kang:2016ehg, Kaufmann:2015hma, Ritzmann:2014mka}. 
We again leave inclusion of higher-order terms in the jet functions for future work.

Using the fact that the partonic cross section at small $x$ is proportional to $\delta(z_1 + z_2 - 1)$, the dihadron differential cross section in \eq{dh0} simplifies to 
\begin{align} \label{dh1}
    \frac{\dd \sigma^{e p \to e p h_1 h_2}}{\dd z_{h_1} \dd z_{h_2} \dd^2 p_{h_1} \dd^2 p_{h_2}} = \sum_q\int \limits^{1-z_{h_2}}_{z_{h_1}} \frac{\dd z}{z (1-z)} 
    D^{h_1/q}_1\!\left(\frac{z_{h_1}}{z}  \right)
    D^{h_2/\bar{q}}_1\!\left(\frac{z_{h_2}}{1-z} \right)
    \frac{\dd\sigma^{e p \to e p q \bar{q}}}{\dd z\, \dd^2 p_1\, \dd^2 p_2}  \Bigg|_{\substack{\un p_1 = (z/z_{h_1})\, \un p_{h_1}~~~~~ \\ \un p_2 = ((1-z)/z_{h_2})\, \un p_{h_2}}}
    \!\!+\, (h_1 \leftrightarrow h_2),
\end{align}
where we have relabeled $z_1 \to z$.
To make contact with the symmetric partonic dijet cross section in Ref.~\cite{Kovchegov:2024wjs}, we symmetrize in $z_{h_1}, \,\un p_{h_1} \longleftrightarrow z_{h_2}, \, \un p_{h_2}$, and define
\begin{align} \label{sym_combo}
    \frac{\dd \sigma^{ep \to e p h_1 h_2}_{\mathrm{sym}}}{\dd z_{h_1} \dd z_{h_2} \dd^2 p_{h_1} \dd^2 p_{h_2}}
    \equiv \frac{1}{2} 
    \left( \frac{\dd \sigma^{e p \to e p h_1 h_2}}{\dd z_{h_1} \dd z_{h_2} \dd^2 p_{h_1} \dd^2 p_{h_2}}  + \frac{\dd \sigma^{e p \to e p h_1 h_2}}{\dd z_{h_2} \dd z_{h_1} \dd^2 p_{h_2} \dd^2 p_{h_1}} 
    \right),
\end{align}
and \cite{Kovchegov:2024wjs}
\begin{align}
    \frac{\dd\sigma^{ep \to e p q \bar{q}}_{\mathrm{sym}}}{\dd z\, \dd^2 p_1\, \dd^2 p_2}
    \equiv \frac{1}{2} 
    \left( \frac{\dd\sigma^{e p \to e p q \bar{q}}}{\dd z\, \dd^2 p_1\, \dd^2 p_2} + \frac{\dd\sigma^{e p \to e p q \bar{q}}}{\dd (1-z)\, \dd^2 p_2\, \dd^2 p_1}
    \right).
\end{align}
Restricting ourselves to the case where $h_1$ and $h_2$ are oppositely charged particle pairs, such as $\pi^+$ and $\pi^-$, allows us to take $D^{h_1/q}_1(z) = D^{h_2/\bar{q}}_1(z)$, in which case \eq{dh1} simplifies to 
\begin{align} 
    \notag 
    \frac{\dd \sigma^{e p \to e p h_1 h_2}_{\mathrm{sym}}}{\dd z_{h_1} \dd z_{h_2} \dd^2p_h\, \dd^2 \Delta} 
    &= \sum_q\int \limits^{1-z_{h_2}}_{z_{h_1}} \!\!\frac{\dd z}{z (1-z)} 
    \left[ D^{h_1/q}_1\!\left(\frac{z_{h_1}}{z}  \right)
    D^{h_2/\bar{q}}_1\!\left(\frac{z_{h_2}}{1-z} \right) + D^{h_1/\bar{q}}_1\!\left(\frac{z_{h_1}}{z}  \right)
    D^{h_2/q}_1\!\left(\frac{z_{h_2}}{1-z} \right)  \right] \\
    \label{dh333}
    & \hspace{3.1cm}
    \times \, 
    \frac{\dd\sigma^{e p \to e p q \bar{q}}_{\mathrm{sym}}}{\dd z\, \dd^2p \, \dd^2 \Delta}  \Bigg|_{\un p = z (1-z)/(z_{h_1} z_{h_2} )\, {\un p}_h} ,
\end{align}
where $\un p_h = z_{h_2} \un p_{h_1} - z_{h_1} \un p_{h_2}$ is the relative transverse momentum of the hadrons, and $\un \Delta = \un p_1 + \un p_2$ is the transverse momentum transferred to the target, which can be measured from the recoil of the outgoing proton.
With this change of variables, one has $\dd^2 p_{h_1} \dd^2 p_{h_2} = \dd^2 p_h\, \dd^2 \Delta$.

Using \eq{dh333} and the expressions for the elastic dijet azimuthal harmonics in \eqs{correlations}, we can write the corresponding azimuthal harmonics of elastic dihadron production as 
\begin{subequations} \label{dh_correlations}
\begin{align} \label{dh_dsa_1}
     \left\langle 1 \right\rangle  
     &= \frac{2\pi}{D}
     \int \dd X \, 
     (2-y)\, \sum_q C_{\mathrm{TT}\,q} \otimes D^{h_1/q}_1 D^{h_2/\bar{q}}_1, 
\\ \label{dh_cos_lt}
     \left\langle \cos \phi_{kp_h}  \right\rangle  
     &= \frac{\pi}{D}
        \int \dd X \, 
        \sqrt{2(1-y)}\, 
        \sum_q
        C_{\mathrm{LT}\,q}^{\cos\phi_{kp}} \otimes D^{h_1/q}_1 D^{h_2/\bar{q}}_1,
\\ \label{dh_cos}
    \left\langle \cos \phi_{\Delta p_{h}} \right\rangle  
    &= \frac{\pi }{D}
    \int \dd X \,
    (2-y) \frac{\sqrt{-t}}{p_\perp}
    \sum_q
        C_{\mathrm{TT}\,q}^{\cos \phi_{\Delta p} }\otimes D^{h_1/q}_1 D^{h_2/\bar{q}}_1,
\\ \label{dh_coscos}
    \left\langle  \cos \phi_{kp_h}  \cos \phi_{\Delta p_h} \right\rangle  
    &= \frac{\pi}{2\, D}
        \int \dd X \,
        \sqrt{2(1-y)} \frac{\sqrt{-t}}{p_\perp}
        \sum_q
        C_{\mathrm{LT}\, q}^{\cos\phi_{kp} \cos\phi_{\Delta p}}\otimes D^{h_1/q}_1 D^{h_2/\bar{q}}_1,
\\ \label{dh_sinsin}
    \left\langle \sin \phi_{kp_h}   \sin \phi_{\Delta p_h} \right\rangle  &= 
    \frac{\pi}{2\, D}
    \int \dd X \,
    \sqrt{2(1-y)}  \frac{\sqrt{-t}}{p_\perp}
    \sum_q C_{\mathrm{LT}\,q}^{\sin\phi_{kp} \sin\phi_{\Delta p}} \otimes D^{h_1/q}_1 D^{h_2/\bar{q}}_1,
\end{align}
\end{subequations}
with the denominator given by 
\begin{align} \label{dh_denom_def}
    D = \int \dd X \, 
        \sum_q 
        \bigg\{ \Big[ 1+ (1-y)^2 \Big] 
        D_{\mathrm{TT}\,q} + 4(1-y) D_{\mathrm{LL}\,q}
        \bigg\} \otimes D^{h_1/q}_1 D^{h_2/\bar{q}}_1. 
\end{align}
The convolutions in \eqs{dh_correlations} and \eqref{dh_denom_def} are defined by
\begin{align}
    f(z, \un p) \otimes D^{h_1/q}_1 D^{h_2/\bar{q}}_1 
    \equiv \!\!\int \limits^{1-z_{h_2}}_{z_{h_1}} \!\!\! \frac{\dd z}{z (1-z)} &\! \left[ 
    D^{h_1/q}_1\!\Big(\frac{z_{h_1}}{z} \Big)
    D^{h_2/\bar{q}}_1\!\Big(\frac{z_{h_2}}{1-z} \Big) + D^{h_1/\bar{q}}_1\!\Big(\frac{z_{h_1}}{z}  \Big)
    D^{h_2/q}_1\!\Big(\frac{z_{h_2}}{1-z} \Big)  
    \right] \!
    f(z, \un p)  \Bigg|_{\un p = \frac{z (1-z)}{z_{h_1} z_{h_2}} \, {\un p}_h},
\end{align}
as given in \eq{dh333}. 
Note that the azimuthal angles in \eqs{dh_correlations} will be measured with respect to the final-state momentum $\un p_h$ instead of the partonic-level momentum $\un p$ used for the dijet harmonics in \eqs{correlations}.
However, since in the collinear approximation for fragmentation the two momenta are parallel, all angles remain the same.

\section{Numerical implementation} \label{sec:numerical_imp}

Having established the formalism for elastic dijet and dihadron production in Secs.~\ref{sec:dijets} and \ref{sec:dihadrons}, we now describe the numerical implementation needed to evaluate the corresponding DSAs and extract the quark and gluon OAM distributions. 
We first outline the numerical solution of the small-$x$ evolution equations for the polarized dipole and moment amplitudes, including their discretization, parametrization of the initial conditions, and constraints among the associated parameters.
We then discuss the treatment of infrared sensitivity in the Fourier-Bessel transforms entering the azimuthal harmonics, and the regularization procedure to suppress unphysical oscillations in the calculated observables. 
These ingredients provide the basis for the feasibility study that will be presented in Sec.~\ref{sec:feas}.

\subsection{Overall discretization and numerical technique}
\label{sec:num_tech}

To solve the evolution equations in \eqs{mom_eqn_rc} and \eqref{neigh_mom_eqn_rc} (as well as in \eqs{eq_LargeNcNf}) numerically, we follow the standard approach \cite{Kovchegov:2016weo, Kovchegov:2020hgb, Cougoulic:2022gbk, Adamiak:2023okq, JAMCollaborationSmall-xAnalysisGroup:2025tfa, Adamiak:2023yhz, Adamiak:2021ppq} to write the equations using logarithmic variables $\eta^{(n)}, s_{ij}$ (see below), discretize on an evenly spaced grid in $\eta^{(n)}$ and $s_{ij}$, and compute the integrals using left-handed Riemann sums. 
Following Refs.~\cite{Kovchegov:2016weo, Kovchegov:2020hgb, Cougoulic:2022gbk, Adamiak:2023okq, JAMCollaborationSmall-xAnalysisGroup:2025tfa, Adamiak:2023yhz, Adamiak:2021ppq}, we employ the logarithmic variables 
\begin{align} \label{log_vars}
    \eta^{(n)} = \sqrt{\frac{N_c}{2\pi}} \, \ln \frac{z^{(n)} s}{\Lambda^2} \hspace{0.5cm} {\rm and}\ \ \ \ 
    s_{ij} = \sqrt{\frac{N_c}{2\pi}} \ln \frac{1}{x_{ij}^2 \Lambda^2},
\end{align}
where $\eta^{(n)} = \{ \eta, \eta^\prime, \eta^{\dprime} \}$ corresponds to $z^{(n)} = \{ z, z^\prime, z^{\dprime} \}$ in \eqs{mom_eqn_rc}, \eqref{neigh_mom_eqn_rc}, and \eqref{eq_LargeNcNf}. 
Similarly, $s_{ij} = \{ s_{10}, s_{21}, s_{32} \}$ corresponds to $x_{ij} = \{ x_{10}, x_{21}, x_{32} \}$.
The resulting dipole and neighbor amplitudes are denoted by $G[i,j] \equiv G(s_{10} = i \delta, \eta = j\delta)$ and $\Gamma[i, k, j] \equiv \Gamma(s_{10} = i \delta, s_{21} = k \delta, \eta' = j \delta)$ for a given grid spacing $\delta$, where $i, j,k \in \mathbb{Z}$.

Since the evolution equations are only valid at small $x$, we must introduce a constraint to enforce this region of applicability.
Following Refs.~\cite{Adamiak:2021ppq, Adamiak:2023yhz, JAMCollaborationSmall-xAnalysisGroup:2025tfa}, we introduce a cutoff $x_0$ such that for $x>x_0$ the dipole amplitudes are given by their initial conditions, while for $x < x_0$ they are determined by the DLA evolution equations.
In terms of the logarithmic variables, we impose $\eta - s_{10} > \sqrt{\frac{N_c}{2\pi}} \ln (1/x_0) \equiv y_0$.
We also extend this constraint to daughter dipoles, such that $\eta^\prime - s_{21} > y_0$ and $\eta^{\dprime} - s_{32} > y_0$.
Furthermore, the perturbative calculation used in deriving the evolution equations breaks down for dipole sizes $x_{10}^2 \gtrsim 1/\Lambda^2$, where $\Lambda = 1$~GeV is introduced as an infrared scale characterizing the proton. 
Here, we enforce the validity of the perturbative calculation by requiring $x_{ij}^2 < 1/\Lambda^2$ for all dipoles. 
In terms of the logarithmic variables, this condition becomes $s_{ij} > 0$. 

The discretized large-$N_c \& N_f$ evolution equations \eqref{mom_eqn_rc} and \eqref{neigh_mom_eqn_rc} for the moment amplitudes with running coupling are listed explicitly in Appendix~\ref{sec:AppendixC}, along with the corresponding expressions for the hPDFs and OAM distributions. 
As~discussed in Sec.~\ref{sec:formalism} above, \eqs{mom_eqn_rc} and \eqref{neigh_mom_eqn_rc} are not closed, and need to be solved simultaneously with the evolution equations \eqref{eq_LargeNcNf} for the amplitudes $Q_q, \widetilde{G}$, $G_2$, and $\widetilde{Q}_q$. 
The discretized versions of these equations are given in Eqs.~(17) of~Ref.~\cite{Adamiak:2023yhz}.\footnote{Equation~(17) from Ref.~\cite{Adamiak:2023yhz} do not include the contributions from or evolution equation for $\widetilde{Q}_q$. 
However, these contributions can be readily inferred from \eqs{eq_LargeNcNf} of the current paper. 
See also the companion paper \cite{Becker:2026xcp}.}

For the initial conditions of both the polarized dipole and moment amplitudes, we follow the previous JAM small-$x$ analyses~\cite{Adamiak:2021ppq, Adamiak:2023yhz, JAMCollaborationSmall-xAnalysisGroup:2025tfa} and use forms that are linear expansions in the logarithmic variables,
\begin{align} \label{ic_ansatz}
    F^{(0)}(s_{10}, \eta) = a \, \eta + b \, s_{10} + c,
\end{align}
where $F$ denotes any one of the 23 amplitudes $\{ Q_q, \widetilde{Q}_q, \widetilde{G}, G_2, \widetilde{Q}^{\mathrm{NS}}_q, Q^{\mathrm{NS}}_q, I_{3q}, \widetilde{I}_q, \widetilde{I}_3, I_4, I_5 \} $, where $q = u, d, s$.
The resulting 69 $a,b,c$ parameters are to be inferred from experimental data. 
These parameters will be labeled by the subscripts, superscripts, and tildes of the amplitudes that they correspond to.

\begin{figure}[ht!]
    \centering
    \includegraphics[scale=0.57]{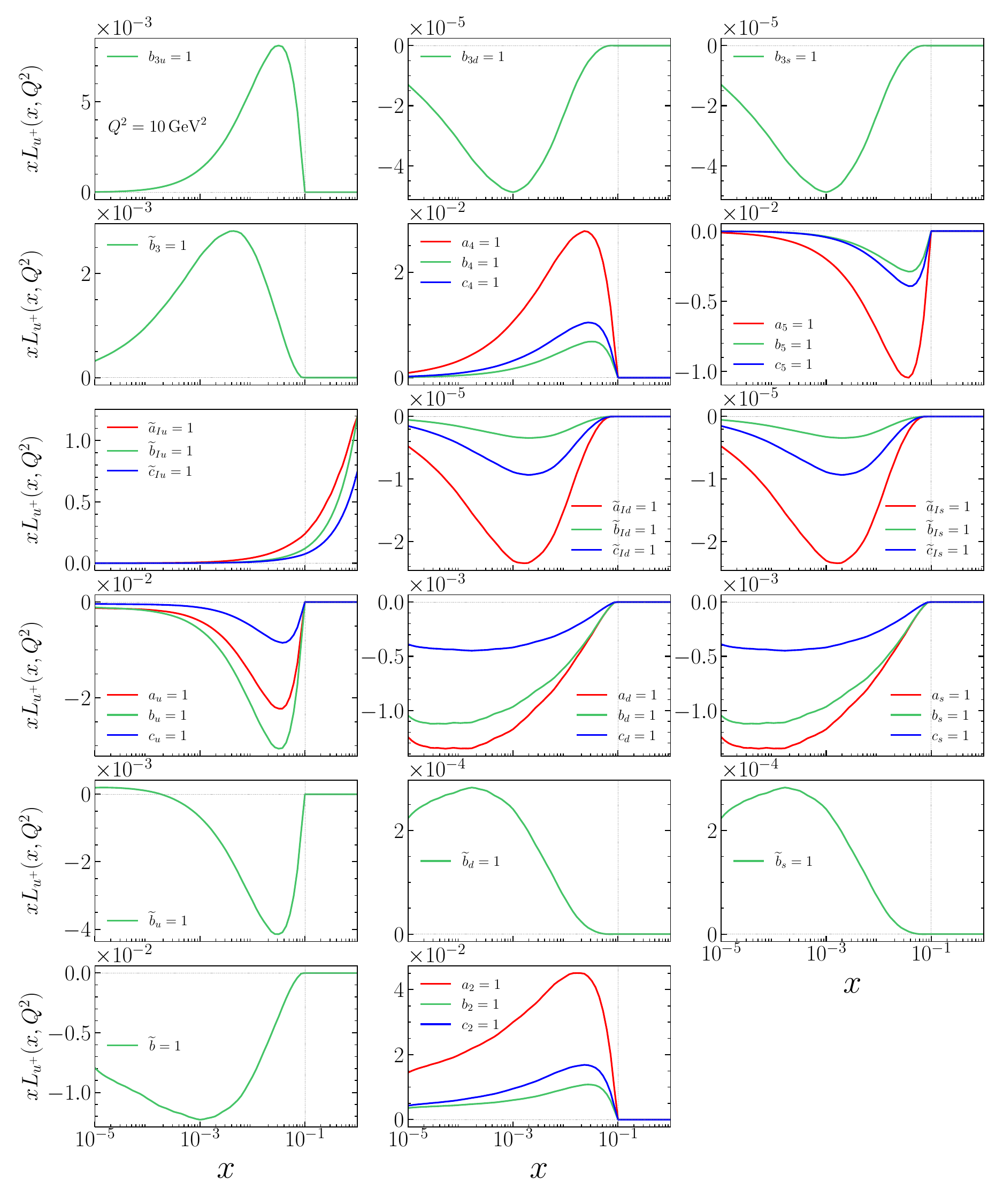}
    \caption{Basis up-quark OAM distributions $x L_{u^+}$ computed for $x \in [10^{-5}, 1]$ and $Q^2 = 10$~GeV$^2$. The legend indicates which curve corresponds to which parameter being nonzero (and equal to 1) in the initial conditions of Eqs.~\eqref{mom_eqn_rc}, \eqref{neigh_mom_eqn_rc}, and \eqref{eq_LargeNcNf}. The $a$ and $c$ coefficients for the amplitudes $I_{3u}, I_{3d}, I_{3s}, \widetilde{I}_3, \widetilde{Q}_u, \widetilde{Q}_d, \widetilde{Q}_s$, and $\widetilde{G}$ are constrained via \eqs{ccf}. Since they are not independent of the other basis distributions, we do not explicitly show them. The dashed vertical lines at $x= x_0 = 0.1$ indicate where the DLA evolution begins.}
    \label{fig:basis}
\end{figure}

In the DLA, the OAM distributions depend linearly on the polarized dipole and moment amplitudes, and can therefore be expressed as a linear combination of ``basis" OAM distributions \cite{Adamiak:2023yhz}.
Each basis OAM distribution is obtained by setting only one parameter in \eq{ic_ansatz} equal to $1$, and the rest to zero, solving the corresponding evolution equations, and using them to compute the OAM distributions in \eqs{logv_dists}.
(An exception is the parameters for the nonsinglet amplitudes, which do not mix with the other amplitudes in \eq{ic_ansatz} under evolution and do not enter the expressions for the OAM distributions.)
The basis OAM distributions are illustrated for $L_{u^+}$ in \fig{fig:basis}.

For $x \gtrsim 0.1$, $L_{u^+}(x,Q^2)$ is dominated by the $\widetilde{I}_u$ parameters, as is clear from Eq.~\eqref{quark_OAM}. 
For $x < 0.1$, the gluonic amplitudes $G_2$, $I_4$, and $I_5$ along with the quark amplitudes $I_{3q}$ and $\widetilde{I}_u$ contribute the most, with the rest of the amplitudes contributing relatively little at all values of $x$.

Note that not all the parameters in \eq{ic_ansatz} are independent.
In Ref.~\cite{Borden:2024bxa}, the following relations were discovered: 
\begin{subequations} \label{cc2}
\begin{align}
    \widetilde{G} \!\left( s_{10} = \eta, \eta \right) &= - \frac{1}{2N_c} \sum_q Q_q \!\left( s_{10} = \eta, \eta \right),
    \\
     \widetilde{Q}_q \!\left( s_{10} = \eta, \eta \right) &= - 2 \, Q_q \!\left( s_{10} = \eta, \eta \right),
     \\
    \widetilde{Q}_q^{\mathrm{NS}} \!\left( s_{10} = \eta, \eta \right) &=  2 \, Q_q^{\mathrm{NS}} \!\left( s_{10} = \eta, \eta\right),
\end{align}
\end{subequations}
which reflect the coincident operator structures between the amplitudes on the left- and right-hand sides in the collinear ($x_{10}^2 \to 1/(zs)$) limit. 
Similarly, the corresponding moment amplitudes have the same coincident operator structures in the collinear limit and therefore obey the relations,
\begin{subequations}
\begin{align} \label{cc3}
    \widetilde{I}_3 \!\left( s_{10} = \eta, \eta \right) &= - \frac{1}{2 N_c}\sum_q I_{3q}\! \left( s_{10} = \eta, \eta\right),
    \\  \label{cc3_b}
    \widetilde{Q}_q \!\left( s_{10} = \eta, \eta \right) &= - 4 \,I_{3q} \!\left( s_{10} = \eta, \eta \right).
\end{align}
\end{subequations}
Note that in \eq{cc3_b}, $\widetilde{Q}_q$ appears because there is no independent moment amplitude corresponding to $\widetilde{Q}_q$~\cite{Becker:2026xcp}. 
Since evolution begins only when $\eta - s_{10} > y_0$, all of the amplitudes in \eqs{cc2} and \eqref{cc3} can be replaced by their respective initial conditions. 
Using the ansatz (\ref{ic_ansatz}) in \eqs{cc2} and \eqref{cc3}, and equating $\eta$-dependent and $\eta$-independent terms, yields the following constraints amongst the initial condition parameters, \\
\begin{subequations} \label{ccf}
    \begin{align}
        \widetilde{a} + \widetilde{b} = -\frac{1}{2N_c} \sum_q \left( a_q + b_q \right), 
        &\qquad 
        \widetilde{c} = -\frac{1}{2N_c} \sum_q c_q, 
       \\
        \widetilde{a}_3 + \widetilde{b}_3 = -\frac{1}{2N_c} \sum_q \left( a_{3q} + b_{3q} \right), 
        &\qquad 
        \widetilde{c}_3 = -\frac{1}{2N_c} \sum_q c_{3q}, 
        \\ 
        \widetilde{a}_q + \widetilde{b}_q = - 4\left( a_{3q} + b_{3q} \right),
    &\qquad 
        \widetilde{c}_q = - 4\, c_{3q},
        \\ 
        \widetilde{a}_q + \widetilde{b}_q = -2\, \left( a_q + b_q \right),
        &\qquad 
         \widetilde{c}_q = -2\, c_q,
         \\
        \widetilde{a}_q^{\mathrm{NS}} + \widetilde{b}_q^{\mathrm{NS}} = 2\, \left( a_q^{\mathrm{NS}} + b_q^{\mathrm{NS}} \right),
        &\qquad 
         \widetilde{c}_q^{\mathrm{NS}} = 2\, c_q^{\mathrm{NS}}.
    \end{align}
\end{subequations}
With these constraints, we find that only 47 of the 69 parameters in \eq{ic_ansatz} are independent, which are to be extracted from existing experimental data and pseudodata.

\subsection{IR safety}

The basic ingredients in the azimuthal harmonics in \eqs{correlations} are the Fourier-Bessel transforms of the dipole amplitudes defined in \eq{fb_transform}. Since the polarized dipole amplitudes are only defined for dipole sizes $x_{12} < 1/\Lambda$, \eq{fb_transform} should more accurately be replaced by
\begin{align} \label{fb_transformIR}
    F^{(cd)}_{(ab)}(p_\perp, \overline{Q}, W^2)
    = \int \displaylimits^{1/\Lambda}_0 \dd x_{12} \, x_{12}\, \left( p_\perp x_{12} \right)^c \left( \overline{Q} x_{12} \right)^d J_a(p_\perp x_{12} ) K_b \left( \overline{Q} \, x_{12} \right)\, F(x_{12}^2, W^2).
\end{align}
For large values of $p_\perp$, however, the oscillations in the integrand induced by the Bessel function $J_a(p_\perp x_{12})$ will not decay sufficiently with increasing $x_{12}$ before $x_{12}$ reaches the upper cutoff of $1/\Lambda$.
This generates unphysical oscillations in the $p_\perp$ spectrum, producing a ``ringing" phenomenon that is typical for Fourier transforms of sharp boundaries.

\begin{figure}[b]
    \centering
     \includegraphics[scale=0.7]{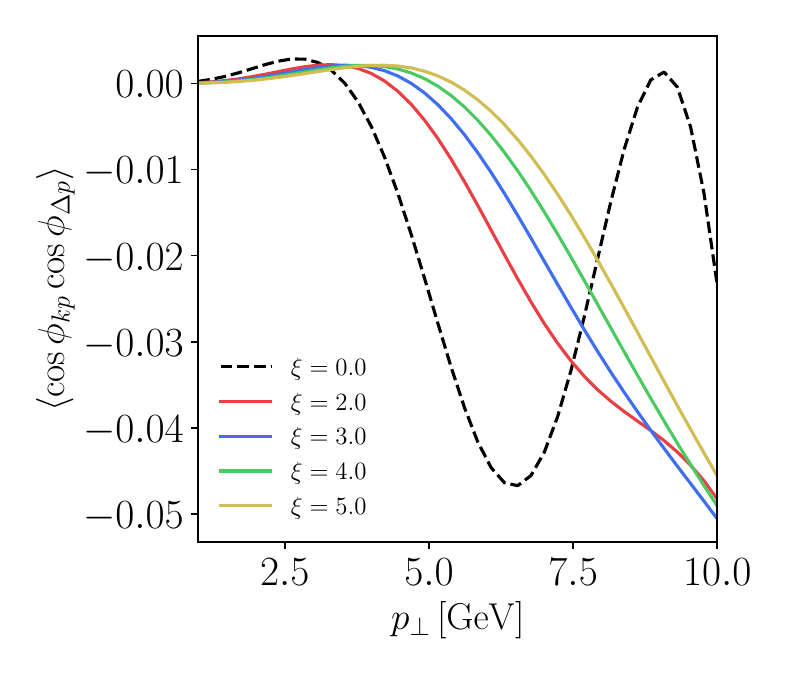}
    \caption{Transverse momentum $p_\perp$ spectrum of the $\langle \cos\phi_{kp} \cos \phi_{\Delta p}\rangle$ harmonic in \eq{coscos}, for $\sqrt{s}=40$~GeV, $z=0.2$, $y=0.2$, $t=-0.1$~GeV$^2$, and integrated over $Q^2\in [1, 100]$~GeV$^2$, for an IR cutoff $\Lambda=1$~GeV. The black dashed line is the unregulated spectrum (for which $\xi = 0$), and the colored lines represent various nonzero values of the regulator $\xi$ in \eq{fb_reg}.}
    \label{fig:vary_xi}
\end{figure}

A possible method of removing these oscillations is to impose a minimum value of $\overline{Q}$ above which we analyze the dijet cross section. 
For increasing $\overline{Q}$, the modified Bessel function $K_b(\overline{Q} x_{12})$ strongly suppresses the oscillations before $x_{12}$ reaches the upper cutoff, thereby removing the sensitivity to the IR region of the integral in \eq{fb_transformIR} and eliminating the oscillations.
This method is theoretically appealing, as a kinematical cut does not introduce any additional model dependence.
However, a relatively large value of $\overline{Q} \sim 4$~GeV is required to remove the oscillations in the kinematic region of interest for this study, which restricts the analysis to $Q^2 > 64$~GeV$^2$.
Since the DSA $A_{LL}^{\gamma^*}$ is a decreasing function of $Q^2$, in the $Q^2 > 64$~GeV$^2$ region the signal-to-noise ratio is much lower than in the moderate-$Q^2$ region.
The low statistics of the high $Q^2$ region would then severely restrict our ability to constrain the OAM distributions.

An alternative method of removing the oscillations is to introduce a Gaussian regulator in the integrand of \eq{fb_transformIR}, such that the Fourier-Bessel transform for polarized dipole amplitudes becomes, 
\begin{align} \label{fb_reg}
       F^{(cd)}_{(ab)}(p_\perp, \overline{Q}, W^2)
       = \int \displaylimits^{1/\Lambda}_0 \dd x_{12}\, x_{12}\, 
       \left( p_\perp x_{12} \right)^c 
       \left( \overline{Q} x_{12} \right)^d J_a(p_\perp x_{12} ) K_b 
       \left( \overline{Q} \, x_{12} \right)\, 
       F(x_{12}^2, W^2)\, e^{-\xi \,\Lambda^2 x_{12}^2 }\ ,
\end{align}
where we have introduced a new dimensionless parameter $\xi$.
In \fig{fig:vary_xi}, we illustrate the impact of the Gaussian regulator on the $p_\perp$ spectrum of the $\langle \cos\phi_{kp} \cos \phi_{\Delta p}\rangle$ harmonic in \eq{coscos}.
The variation in the $p_\perp$ spectrum is found to be small for a range of sufficiently large $\xi$ values.
This suggests that introducing a Gaussian regulator will not impact the results of our feasibility study, and throughout the rest of this paper we will use \eq{fb_reg} with $\xi = 3.0$.

\section{Feasibility study of OAM extraction at the EIC}
\label{sec:feas}

Having presented the theoretical framework and numerical implementation in the preceding sections, we now investigate the feasibility of extracting the quark and gluon OAM distributions at small $x$ from projected EIC measurements of elastic dijet and dihadron production. 
We first update previous global analyses of polarized DIS, SIDIS, and $pp$ scattering data to incorporate the complete small-$x$ helicity evolution and additional contributions to the polarized structure functions, establishing a consistent baseline for the present study.
We then describe the generation and fitting of EIC pseudodata for the dijet and dihadron DSAs, together with projected polarized DIS and SIDIS measurements.
Finally, we quantify the impact of these measurements on the extraction of the small-$x$ quark and gluon OAM distributions and helicity PDFs, including their contributions to the proton spin sum rule.

\subsection{Updated global analysis of helicity PDFs}
\label{sec:baseline}

Before generating and fitting EIC dijet and dihadron pseudodata, we first revisit the small-$x$ hPDF extractions from the JAMsmallx-23~\cite{Adamiak:2023yhz} and JAMsmallx-25~\cite{JAMCollaborationSmall-xAnalysisGroup:2025tfa} analyses. 
This step is necessary because the present analysis modifies both of the previous analyses~\cite{Adamiak:2023yhz,JAMCollaborationSmall-xAnalysisGroup:2025tfa} in two related ways. 
First, whereas earlier work neglected the polarized amplitudes $\widetilde{Q}_q^{(0)}$ and $\widetilde{Q}^{\mathrm{NS} \, (0)}_q$ in \eqs{Dq+} and \eqref{Dq-}, here we retain these terms throughout, both in the large-$N_c\&N_f$ helicity evolution in \eq{eq_LargeNcNf} and in the quark hPDFs in \eqs{Dq+} and \eqref{Dq-}.
Second, we include the in-shock-wave contributions to the $g_1$ and $g_1^h$ structure functions derived in Appendix~\ref{sec:g1}, which correspond to the inhomogeneous terms proportional to $\widetilde{Q}^{(0)}_q$ and $\widetilde{Q}^{\mathrm{NS}\,(0)}_q$ in \eqs{dis_g1} and \eqref{sidis_g1}. In addition to these updated formulas, we include the parameters of $\widetilde{Q}_q^{(0)}$ and $\widetilde{Q}^{\mathrm{NS}\,(0)}_q$ from \eq{ic_ansatz} in the fit, along with the constraints from \eqs{ccf}. 
Of these 18 new parameters (for three light flavors), 12 are constrained via \eqs{ccf}, which further constrains two parameters for the $\widetilde{G}^{(0)}$ initial condition, giving an additional 4 parameters compared to previous work~\cite{Adamiak:2023yhz, JAMCollaborationSmall-xAnalysisGroup:2025tfa}.

In our present approach, the polarized (SI)DIS fit is performed using the large-$N_c \& N_f$ evolution equations \eqref{eq_LargeNcNf}, while the polarized $pp$ data are analyzed using the large-$N_c^{+q}$ limit of those equations.
This is done since the existing theoretical calculation of the small-$x$ particle production cross section in polarized $pp$ collisions only includes the gluon contribution~\cite{Kovchegov:2024aus}. 
When analyzing the polarized DIS, SIDIS and $pp$ data simultaneously, we use these two different versions of small-$x$ helicity evolution, but constrain the same set of initial condition parameters from \eq{ic_ansatz}. 

We use the same DIS, SIDIS and $pp$ world data as in the JAMsmallx-25 analysis~\cite{JAMCollaborationSmall-xAnalysisGroup:2025tfa}, and perform fits with and without the inclusion of the $pp$ dataset to compare also with the earlier JAMsmallx-23 analysis~\cite{Adamiak:2023yhz}.

We apply the same cuts on the DIS and SIDIS data as in Refs.~\cite{Adamiak:2023yhz, JAMCollaborationSmall-xAnalysisGroup:2025tfa}, namely, $x < 0.1$ and $Q^2 > m_c^2 = 1.69\,\mathrm{GeV}^2$, 
and restrict the jet data to $p_T \geq 8\,\mathrm{GeV}$ and momentum fraction of the parton from the target or projectile $x_{T,P} < 0.1$.
After these cuts, 122 DIS, 104 SIDIS, and 14 $pp$ data points survive, for a total of $N_{\mathrm{pts}} = 226$ without jet data and $N_{\mathrm{pts}} = 240$ including the jet data.

The analysis is performed within the JAM Bayesian Monte Carlo framework~\cite{Sato:2016tuz, Sato:2019yez, Moffat:2021dji}, with an ensemble of 500 replicas constructed by resampling every data point within its quoted uncertainty and minimizing $\chi^2$ against the fluctuated set. 
The breakdown of $\chi^2/N_{\mathrm{pts}}$ per reaction in Table~\ref{tab:chi2} shows that the DIS and SIDIS-only fit yields an overall $\chi^2/N_{\mathrm{pts}} = 0.85$, with 0.94 for the DIS data and 0.74 for the SIDIS data. 
Adding $pp$ data slightly lowers the overall $\chi^2/N_{\mathrm{pts}}$ to 0.80, with 0.83 across the combined DIS and SIDIS sets and 0.35 for the $pp$ data alone.

\begin{table}[h]
\caption{Values of $\chi^2/N_{\rm pts}$ by reaction for fits with and without $pp$ data (``This work'') compared with the previous JAMsmallx-23  \cite{Adamiak:2023yhz} and JAMsmallx-25 \cite{JAMCollaborationSmall-xAnalysisGroup:2025tfa} results.}
\centering
\setlength{\tabcolsep}{10pt}
\begin{tabular}{l | c | cc | cc}
\hline
\multicolumn{1}{c}{} & \multicolumn{1}{c}{} & \multicolumn{2}{c}{DIS\,+\,SIDIS only} & \multicolumn{2}{c}{DIS\,+\,SIDIS\,+\,$pp$} 
\\ \hline
&  & JAMsmallx-23 & \textbf{This work} & JAMsmallx-25 & \textbf{This work} 
\\
Reaction & $N_{\mathrm{pts}}$ & $\chi^2/N_{\mathrm{pts}}$ & $\chi^2/N_{\mathrm{pts}}$ & $\chi^2/N_{\mathrm{pts}}$  & $\chi^2/N_{\mathrm{pts}}$ \\ \hline
~~DIS   & 122 & 1.09 & \textbf{0.94} & 1.00 & \textbf{0.97} 
\\
~~SIDIS & 104 & 0.84 & \textbf{0.74} & 1.04 & \textbf{0.66} 
\\
\textbf{Total} (DIS+SIDIS) & 226 & 0.98 & \textbf{0.85} & --- & \textbf{0.83}   
\\ \hline
~~$pp$ & 14  & ---   & \textbf{---}   & 0.36 & \textbf{0.35} 
\\ 
\textbf{Total} (all data)  & 240 & ---   & --- & 0.98 & \textbf{0.80} 
\\ \hline
\end{tabular}
\label{tab:chi2}
\end{table}

Both fits compare favorably with the corresponding previous JAMsmallx-23 \cite{Adamiak:2023yhz} and JAMsmallx-25~\cite{JAMCollaborationSmall-xAnalysisGroup:2025tfa} analyses, with the former reporting a slightly higher overall $\chi^2/N_{\mathrm{pts}} = 0.98$, with 1.09 for DIS and 0.84 for SIDIS. 
Comparing with the JAMsmallx-25 global analysis~\cite{JAMCollaborationSmall-xAnalysisGroup:2025tfa}, our combined DIS and SIDIS $\chi^2/N_{\mathrm{pts}}$ value of 0.83 improves on the 1.02 reported there, while the $pp$ description is  essentially unchanged (0.35 versus 0.36). 
This confirms that retaining $\widetilde{Q}_q$, $\widetilde{Q}^{\mathrm{NS}}_q$, and the in-shockwave $g_1$ terms preserves the quality of the description of the world data obtained in the previous JAMsmallx fits~\cite{Adamiak:2023yhz, JAMCollaborationSmall-xAnalysisGroup:2025tfa}. 

\begin{figure}[b]
    \centering
     \includegraphics[scale=0.58]{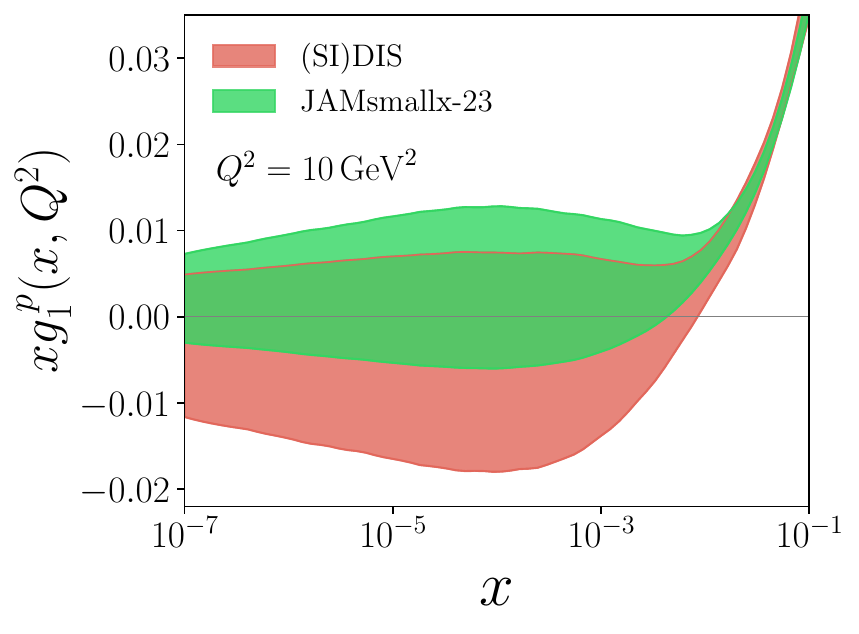}
    \includegraphics[scale=0.58]{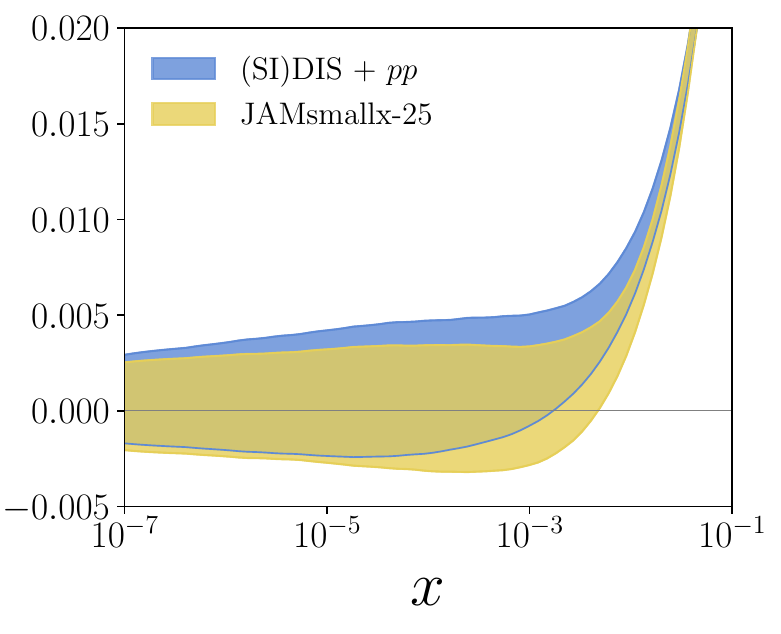}
    \caption{Comparison of the extracted $xg_1^p(x,Q^2)$ structure function at $Q^2=10$~GeV$^2$ with the previous JAMsmallx-23 DIS+SIDIS \cite{Adamiak:2023yhz} (left) and JAMsmallx-25 DIS+SIDIS+$pp$ \cite{JAMCollaborationSmall-xAnalysisGroup:2025tfa} (right) analyses. The bands are 68\% credible intervals.}
    \label{fig:g1_compare}
\end{figure}

In \fig{fig:g1_compare}, we compare the extracted $xg_1^p(x,Q^2)$ structure function from the present analysis with the previous JAMsmallx analyses~\cite{Adamiak:2023yhz, JAMCollaborationSmall-xAnalysisGroup:2025tfa}. 
In the DIS and SIDIS-only case the two extractions agree for $x > 10^{-2}$. 
For $x < 10^{-2}$, our updated baseline develops a markedly wider uncertainty band and contains more negative than positive replicas of $xg_1^p$ (whereas the JAMsmallx-23 extraction~\cite{Adamiak:2023yhz} contains more positive replicas).
For the fit that also includes $pp$ data, our extraction and the JAMsmallx-25 extraction~\cite{JAMCollaborationSmall-xAnalysisGroup:2025tfa} are much more consistent, with the updated baseline having slightly more negative replicas for $10^{-4} < x< 10^{-2}$. 
The minimal widening of the uncertainty band compared to the previous JAMsmallx-25 extraction reflects the increased constraining power of including the $pp$ data.

\begin{figure}[t!]
    \centering
    \includegraphics[scale=0.56]{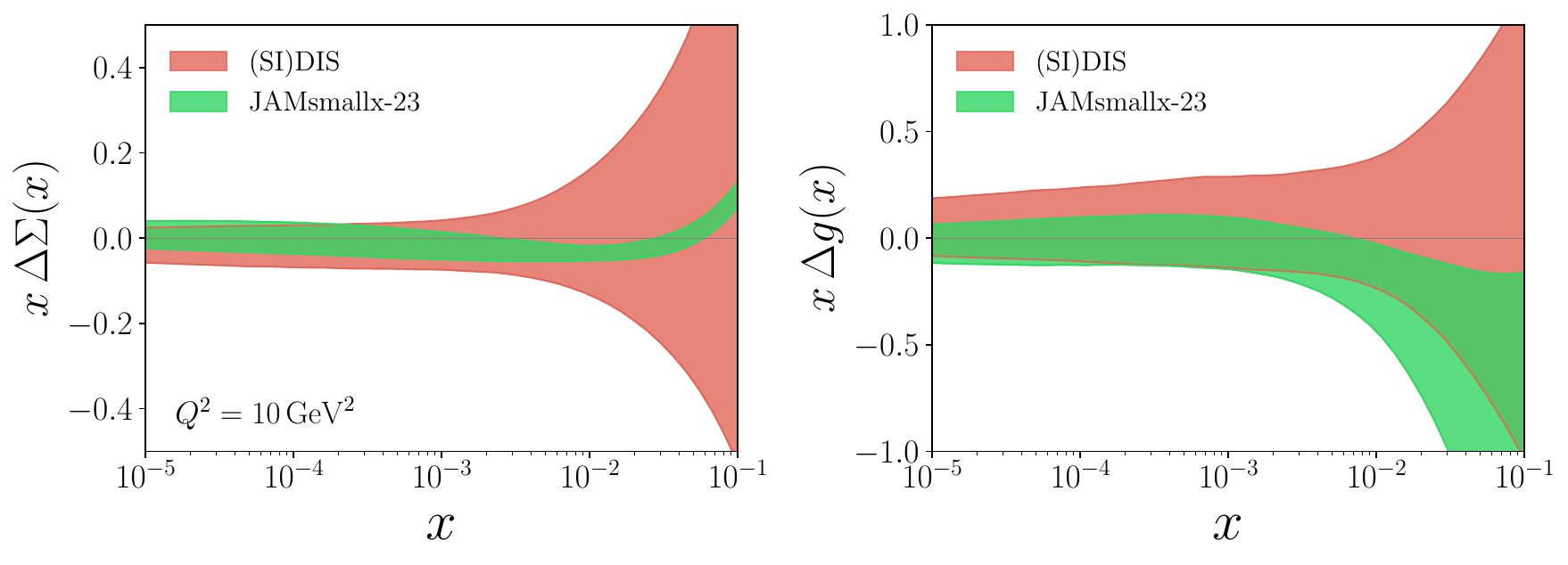}
    \includegraphics[scale=0.56]{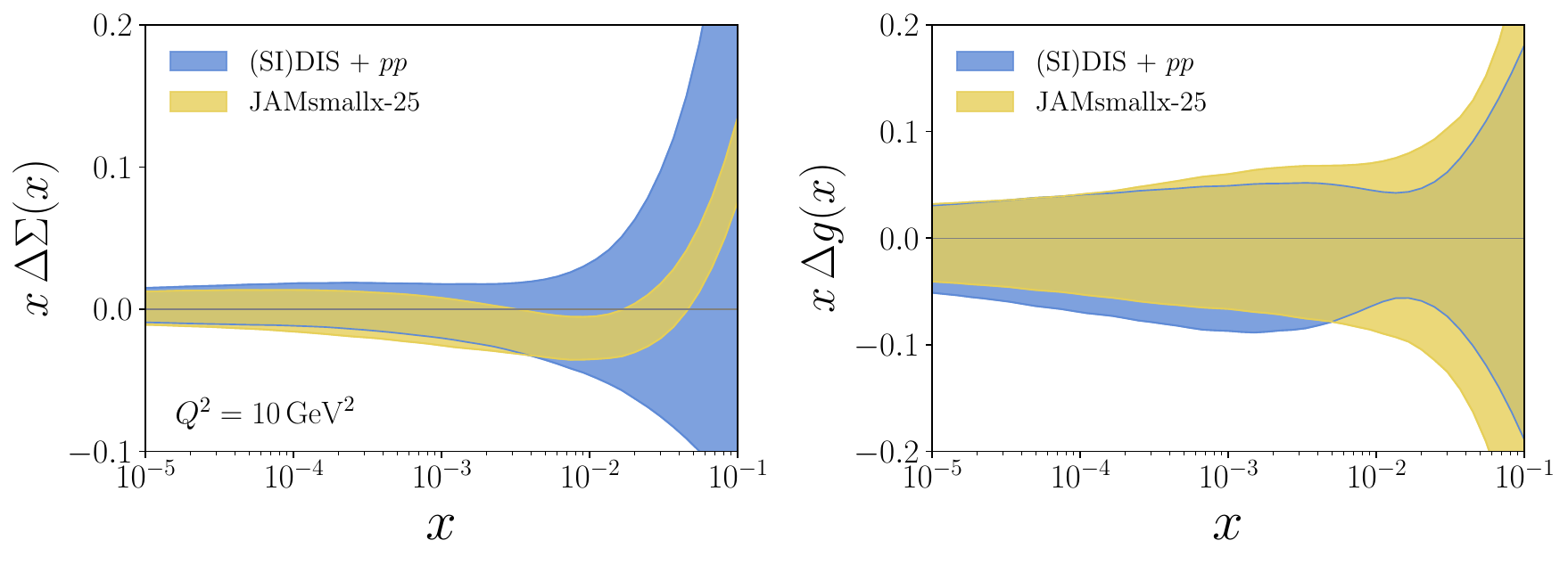}
    \vspace*{-0.2cm}
    \caption{Updated baseline quark singlet $x \Delta\Sigma$ (left column) and gluon $x \Delta g$ (right column) hPDFs versus $x$ at $Q^2=10$~GeV$^2$, compared with the previous JAMsmallx-23~\cite{Adamiak:2023yhz} DIS+SIDIS only (top row) and JAMsmallx-25~\cite{JAMCollaborationSmall-xAnalysisGroup:2025tfa} DIS+SIDIS+$pp$ (bottom row) fit results. All bands are 68\% credible intervals.\\}
    \label{fig:baseline_hpdfs}
\end{figure}

\begin{figure}[h!]
    \centering
    \includegraphics[scale=0.55]{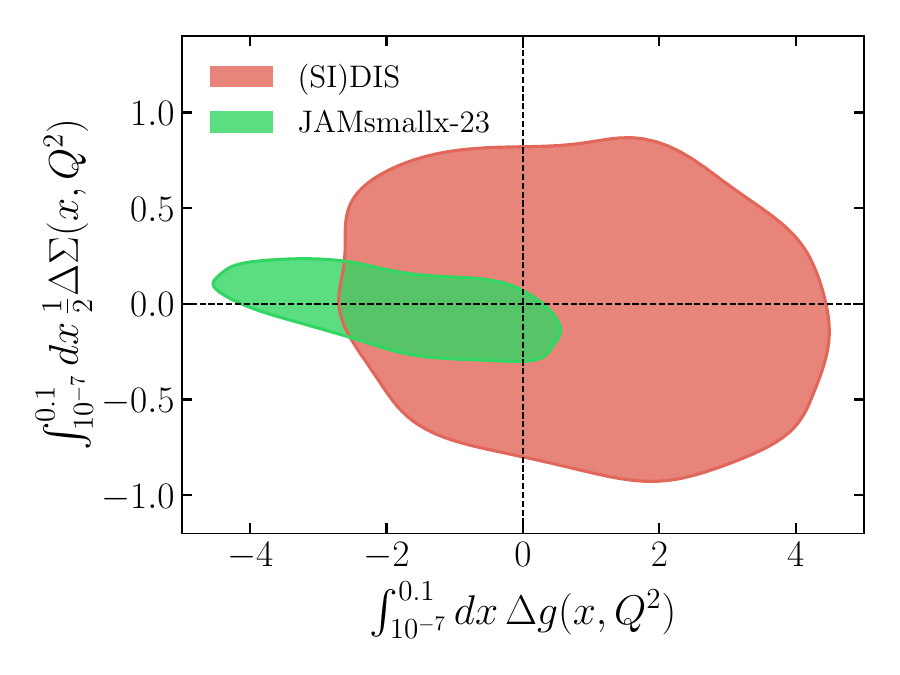}
    \includegraphics[scale=0.55]{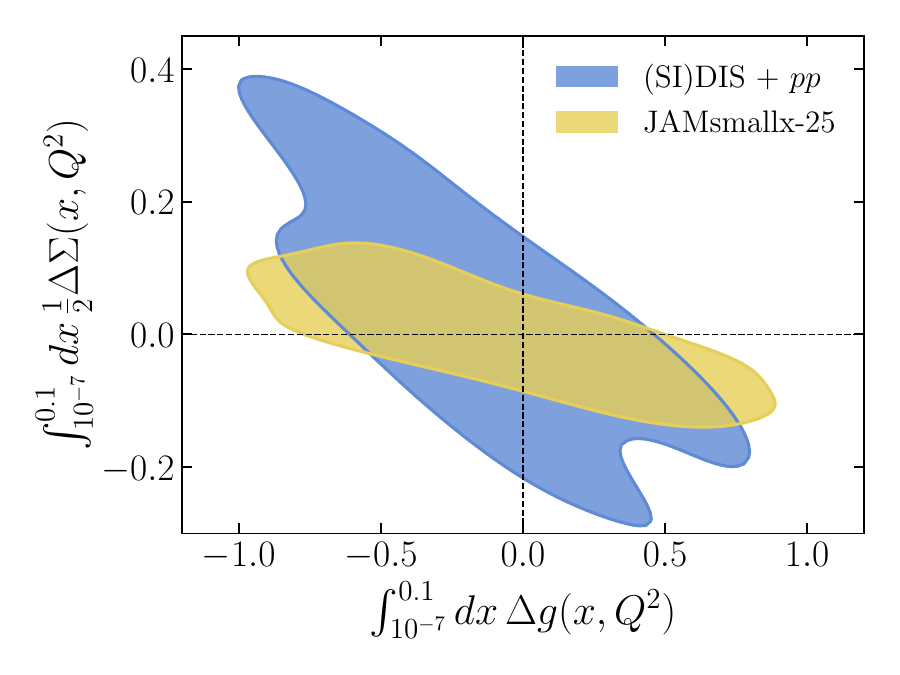}
    \vspace*{-0.2cm}
    \caption{Truncated integrals of $\thalf \Delta \Sigma$ and $\Delta g$ for the range $10^{-7} < x < 0.1$ at $Q^2 = 10$~GeV$^2$ for the DIS+SIDIS only analysis (left panel, red) compared with the JAMsmallx-23~\cite{Adamiak:2023yhz} fit (green), and for the combined DIS+SIDIS+$pp$ analysis (right panel, blue) compared with the JAMsmallx-25~\cite{JAMCollaborationSmall-xAnalysisGroup:2025tfa} result (yellow). All contours represent the 68\% credible regions.}
    \label{fig:baseline_contour}
\end{figure}

The updated flavor-singlet quark and gluon hPDFs are shown in \fig{fig:baseline_hpdfs}, compared to the previous analyses~\cite{Adamiak:2023yhz, JAMCollaborationSmall-xAnalysisGroup:2025tfa}. 
The quark sector exhibits the most profound effect, since $\Delta q^+$ now includes the initial condition from $\widetilde{Q}_q$ rather than the evolution term alone, and the $g_1$ and $g_1^h$ inhomogeneous terms reweight the pull of the DIS and SIDIS data on all the initial conditions.
In particular, the singlet quark hPDF $\Delta\Sigma$ has a much larger uncertainty band for $x > 10^{-3}$ compared to the previous JAMsmallx-23 extraction~\cite{Adamiak:2023yhz}, reflecting the enhanced sensitivity to the initial conditions from \eq{Dq+}. 
This effect is more pronounced without the inclusion of $pp$ data.
The gluon hPDF is affected only indirectly through evolution and accordingly its uncertainty changes the least.
In the DIS+SIDIS fit, the uncertainty on the gluon hPDF grows slightly compared to the previous extraction~\cite{Adamiak:2023yhz}, with more positive replicas over the $10^{-7} < x < 0.1$ range considered here. 
For the fit including $pp$ data, the slight narrowing of the uncertainty band in the present fit compared with the JAMsmallx-25~\cite{JAMCollaborationSmall-xAnalysisGroup:2025tfa} results can be attributed to the difference between the large-$N_c\&N_f$ evolution employed here and the large-$N_c^{+q}$ evolution employed in Ref.~\cite{JAMCollaborationSmall-xAnalysisGroup:2025tfa}.
A similar effect was observed including only DIS and SIDIS data \cite{JAMCollaborationSmall-xAnalysisGroup:2025tfa}.

To quantify the small-$x$ helicity content of the proton in the updated baseline, \fig{fig:baseline_contour} shows the truncated quark and gluon helicity moments, integrated over $10^{-7} < x < 0.1$ at $Q^2 = 10$~GeV$^2$, again compared with the JAMsmallx-23~\cite{Adamiak:2023yhz} and JAMsmallx-25~\cite{JAMCollaborationSmall-xAnalysisGroup:2025tfa} results.\footnote{The previous extractions of helicity distributions in Refs.~\cite{Adamiak:2023yhz, JAMCollaborationSmall-xAnalysisGroup:2025tfa} used the base-10 logarithmic integration variable, $\int \dd x\, \Delta f(x,Q^2) = \ln 10 \, \int \dd\big(\log_{10}x\big)\, x \,\Delta f(x, Q^2)$, but omitted the factor of $\ln 10$, which is included here.}
For the DIS+SIDIS only fit, the uncertainty bands in both the gluon and quark sectors grow considerably due to the uncertainty from the additional parameters.
We find that more replicas are consistent with a positive total integrated gluon helicity contribution. 
Including $pp$ data, the range of integrated gluon helicity contributions carries an uncertainty of similar size, while the range of integrated quark helicity contributions approximately doubles in size.
The increased uncertainty in the quark sector reflects the greater sensitivity of the quark hPDFs to the initial conditions via \eq{Dq+}. 

In the following sections, all comparisons that quantify the impact of the EIC dijet and dihadron measurements, as well as of the EIC DIS and SIDIS pseudodata, will be made relative to the baseline fit described here.
Any observed reduction in uncertainty will therefore reflect genuinely new information rather than arising from a change in the underlying formalism.

\subsection{Generation and analysis of EIC pseudodata}

Our central goal is to assess the feasibility of extracting the small-$x$ quark and gluon OAM distributions from EIC dijet and dihadron measurements.
This requires constraining the full set of polarized dipole and moment amplitudes that enter both the expressions for the OAM distributions in \eqs{OAMs} and their corresponding small-$x$ evolution equations.
Here we describe the generation and subsequent fitting procedure for EIC pseudodata based on the dijet and dihadron DSA azimuthal harmonics in Secs.~\ref{sec:dijets} and \ref{sec:dihadrons}.

To begin we first compute the relevant uncertainty projections for the elastic dijet and dihadron production processes. 
Although the systematic uncertainty associated with these processes is difficult to ascertain without a dedicated detector simulation, the statistical uncertainty is straightforward.
One can show that,
\begin{align} \label{all_err}
    \delta \big(A_{LL}\big)_{\rm stat}\ 
    \equiv\ \delta 
    \left( 
    \frac{N^{\uparrow\uparrow} - N^{\uparrow\downarrow}}
         {N^{\uparrow\uparrow} + N^{\uparrow\downarrow}} 
    \right)_{\rm stat} 
    =\ \sqrt{\frac{1 - A_{LL}^2}{N^{\uparrow\uparrow} + N^{\uparrow\downarrow}}}\ \approx\ \sqrt{\frac{1}{\dd\sigma^{\mathrm{unpol}}\, \mathcal{L}_\mathrm{int}}},
\end{align}
where $\mathcal{L}_{\mathrm{int}}$ is the projected integrated luminosity (including all beam spin configurations), and we have assumed that $A_{LL} \ll 1$ to obtain the last approximate equality. 
For the angular modulations in \eqs{cos_lt}--(\ref{sinsin}), the statistical uncertainty in \eq{all_err} must be multiplied by $\sqrt{2}$~\cite{Ma:2009xyz}. 
In the calculation of the statistical uncertainty below, we will employ the unpolarized cross section in \eq{unpol_decomp}.

The parameters for the moment amplitudes in \eq{ic_ansatz} are randomly sampled in the range $[-20, 20]$, consistent with the ranges of the parameter values obtained in the baseline fit described in Sec.~\ref{sec:baseline} and a Born level calculation of the moment amplitude parameters.
To ensure the selected parameters are physically reasonable, we impose an additional constraint on the small-$x$ truncated integrals of the total OAM distributions, requiring that the moment amplitude parameters used to produce the pseudodata satisfy 
\begin{equation}
\label{oam_const}
\left| \ \int \displaylimits_{10^{-7}}^{0.01} 
\dd x \sum_q L_{q^+}(x, Q^2)
\right|\ <\ 1, 
\qquad\qquad
\left| \ \int \displaylimits_{10^{-7}}^{0.01} 
\dd x\, L_g(x,Q^2) 
\right|\ <\ 1
\end{equation}
at $Q^2=10$~GeV$^2$.
Enforcing the constraints (\ref{oam_const}) ensures that the sampled initial conditions yield realistic total OAM contributions, while remaining agnostic about the detailed shape of the individual distributions. 
Since the dijet and dihadron calculations described in Secs.~\ref{sec:dijets} and \ref{sec:dihadrons} are only valid for $x \leq 0.01$ due to the applicability range of the unpolarized amplitude $N$, we choose $x=0.01$ as the upper limit of the integrals in Eq.~(\ref{oam_const}).

Within the range of initial condition parameters allowed by \eq{oam_const}, the integrated quark and gluon OAM distributions can take qualitatively different small-$x$ behaviors. To assess whether our feasibility conclusions depend on the (presently unknown) magnitude of the small-$x$ OAM, we generate pseudodata for three representative scenarios:
\begin{itemize}
\item 
{\bf Scenario A}
{\bf (large positive OAMs):}
the quark and gluon OAM integrals approximately saturate the upper bounds of Eq.~(\ref{oam_const}), 
\\ \hspace*{3cm}
$\int \dd x \sum_q L_{q^+} \approx \int \dd x\, L_g \approx 1$;
\item 
{\bf Scenario B}
{\bf (large negative OAMs)}: 
the integrals approximately saturate the lower bounds of Eq.~(\ref{oam_const}), 
\\ \hspace*{3cm}
$\int \dd x \sum_q L_{q^+} \approx \int \dd x\, L_g \approx -1$; 
\item 
{\bf Scenario C}
{\bf (small OAMs)}: 
both integrals are close to zero.
\end{itemize}
These three scenarios represent three distinct choices within the range of physically reasonable small-$x$ OAM behaviors consistent with current constraints. 
Note that while we enforce one of the scenarios above on the parameters to generate pseudodata, we do not enforce these constraints on the integrated OAM distributions when fitting the pseudodata.
In parallel, we generate supplementary DIS and SIDIS pseudodata  under the three $g_1^p$ scenarios of the JAMsmallx-25 analysis~\cite{JAMCollaborationSmall-xAnalysisGroup:2025tfa} --- ``positive $g_1^p$'', ``negative $g_1^p$'', and ``all $g_1^p$'' --- which correspond, respectively, to the means of the asymptotically positive, asymptotically negative, and all replicas. 
These three scenarios span qualitatively distinct small-$x$ asymptotics that are consistent with currently available data (see \fig{fig:g1_compare}).
The $g_1^p$ scenarios span the helicity sector uncertainty and the OAM scenarios probe the OAM-sector uncertainty, as the two sectors are independent in our formalism. 
We therefore generate pseudodata for each of the $3 \times 3$ possible combinations, yielding 9 joint fits that we compare in the following section to assess the robustness of our conclusions. 
As mentioned above, we generate pseudodata and perform the fits with and without the existing polarized $pp$ data, since the theory formalism for the latter currently does not include quark contributions.

We generate EIC dijet and dihadron pseudodata at center-of-mass energies $\sqrt{s}=30$ and 40~GeV for proton, deuteron, and $^3\mathrm{He}$ beams with an integrated luminosity of 100~fb$^{-1}$.
Higher-energy configurations are not included because the dijet and dihadron DSAs decrease with increasing $\sqrt{s}$, so measurements at the upper end of the planned EIC energy range would yield very small azimuthal harmonics with correspondingly limited discriminating power.
We have verified this by extending the pseudodata generation to include energies up to $\sqrt{s}=140$~GeV and found no noticeable change in our conclusions.

We fix the momentum transfer squared to $t = -0.1$~GeV$^2$ since this falls within the acceptance of the EIC far-forward detectors and should maximize the efficiency of forward-proton tagging in the Roman pots \cite{AbdulKhalek:2021gbh}.\footnote{We thank Brian Page for communicating to us the results of his Monte Carlo study, which allowed us to reach this conclusion.} 
This choice is also statistically favorable, as fixing $t$ rather than integrating over it produces a relative error for the azimuthal harmonics of the DSA scaling as $\sim 1/\sqrt{|t|}$ rather than $\sim 1/|t|$. 
Indeed, since $\delta A_{LL} / A_{LL} = \sqrt{\dd \sigma^\textrm{unpol}}/ (\sqrt{\mathcal{L}}\, \dd\sigma^{\rm DSA})$, for a differential in $t$ measurement at small $|t|$ one has $\dd\sigma^\textrm{unpol} \sim t^0$, while for the angular harmonics $\dd\sigma^{\rm DSA} \sim \sqrt{|t|}$, such that $\delta A_{LL} / A_{LL} \sim 1/\sqrt{|t|}$. 
For an integrated in $t$ measurement, in a small-$|t|$ range, we have $\dd\sigma^\textrm{unpol} \sim |t|$, $\dd\sigma^{\rm DSA} \sim |t|^{3/2}$, leading to $\delta A_{LL} / A_{LL} \sim 1/|t|$. 
Since we are interested in the kinematics where $|t|$ is small, we have $\Lambda/\sqrt{|t|} \ll \Lambda^2/|t|$, and therefore a $t$-binned measurement will yield a larger signal-to-noise ratio.

Similarly, to maximize discriminating power, we bin the dijet pseudodata in the momentum fraction $z \in \{0.2, 0.3, 0.4, 0.5\}$ and in the inelasticity $y \in \{0.2, 0.3, 0.43, 0.61, 0.86, 0.95\}$. 
The dihadron pseudodata are likewise binned in the hadron momentum fractions $(z_{h_1}, z_{h_2}) \in \{(0.2, 0.4), (0.3, 0.3), (0.3, 0.4), (0.4, 0.4)\}$ and in the same $y$ bins. 
The value of this multi-dimensional binning is best appreciated by contrasting it with $z$- or $y$-integrated measurements of the same observables.
Varying $z$ (or, analogously, $(z_{h_1}, z_{h_2})$) accesses different linear combinations of the polarized dipole and moment amplitudes, since the kinematic prefactors multiplying each amplitude depend nontrivially on these momentum fraction(s) (as can be seen in the coefficients in \eqs{tt_coeffs} and \eqref{lt_coeffs}).
A $z$- or $(z_{h_1}, z_{h_2})$-integrated measurement collapses these distinct projections onto a single, averaged sum.

Binning in $y$, on the other hand, lifts a parametric degeneracy between the $a$- and $c$-type initial condition parameters in \eq{ic_ansatz}.
The Fourier-Bessel transform in \eq{fb_transform} is dominated by the kinematic region in which the amplitudes are well-approximated by their initial conditions, given in \eq{ic_ansatz}. 
Since neither the $a$-type initial condition terms nor the $c$-type depend on the dipole size, they can be taken outside of the Fourier-Bessel transform in \eq{fb_transform}, so that at fixed $W^2 = sy$, only the combination $a \, \eta(W^2) + c$ enters the observables (here $\eta(W^2)$ is the $\eta$-variable, which depends on $W^2$). 
The $a$ and $c$ parameters for each polarized dipole amplitude are therefore inseparable in $y$-integrated measurements, where only the average value of the linear combination $a\,\eta(W^2)  + c$ is probed. 
By contrast, binning in $y$ changes the relative weighting of the two terms, thereby breaking the degeneracy. 
Changing $\sqrt{s}$ also breaks this degeneracy for fixed $y$, but, as noted above, the range of $\sqrt{s}$ which has discriminating power over the polarized dipole amplitudes is limited.

In both the dijet and dihadron pseudodata, we integrate over a range of $Q^2 \in [1, 100]$~GeV$^2$, and to ensure the applicability of the small-$x$ formalism (both polarized and unpolarized) require $x < 0.01$ across all kinematic bins.\footnote{While the polarized small-$x$ formalism describes data up to $x \sim 0.1$ \cite{Adamiak:2021ppq, Adamiak:2023yhz, JAMCollaborationSmall-xAnalysisGroup:2025tfa}, the elastic dijet and dihadron observables also require the unpolarized small-$x$ formalism, whose phenomenological fits are limited to $x < 0.01$~\cite{Albacete:2009fh}. This difference can be traced to their resummation parameters: polarized evolution resums double logarithms, $\alpha_s \ln^2(1/x)$, while the unpolarized evolution resums single logarithms, and for $\alpha_s \sim 0.2$ and $x \sim 0.1$ one has $\alpha_s \ln^2(1/x) \sim 1$.} 
For the dijet pseudodata, we restrict the analysis to \mbox{2~GeV $< p_\perp <$ 7~GeV}.
Below $p_\perp = 2$~GeV, we expect jet corrections to become important, and above $p_\perp = 7$~GeV the relative error of the dijet DSA grows substantially, reducing the potential impact of pseudodata above this threshold. 
Similarly, the dihadron pseudodata are bounded by \mbox{0.5~GeV $< p_h <$ 3.0~GeV}.
Below $p_h = 0.5$~GeV, detector efficiency becomes an important factor which we cannot reliably model here; above $p_h = 3.0$~GeV, the relative error of the dihadron pseudodata again becomes too large to be useful. 
Furthermore, we find that the kinematic region with the most constraining power satisfies $p_h \sim Q$, which justifies neglecting contributions from dihadron fragmentation functions.
Here we assume a binning of $\Delta p_\perp  = 1.0$~GeV for dijets and $\Delta p_h = 0.5$~GeV for dihadrons.

As mentioned above, to further constrain the polarized dipole amplitudes, we supplement the dijet and dihadron pseudodata with EIC pseudodata for inclusive polarized DIS and SIDIS. 
The DIS pseudodata span the kinematic region $10^{-4} < x < 10^{-1}$ and 1.69~GeV$^2 < Q^2 < 50$~GeV$^2$, at center-of-mass energies $\sqrt{s} \in \{29, 45, 63, 141\}$~GeV for proton beams (with 100~fb$^{-1}$ integrated luminosity) and $\sqrt{s} \in \{29, 66, 89\}$~GeV for deuteron and $^3\mathrm{He}$ beams (with 10~fb$^{-1}$).
The SIDIS pseudodata are generated for the proton at $\sqrt{s} = 141$~GeV.
Both sets assume a 2\% point-to-point uncorrelated systematic uncertainty, consistent with the {\it EIC Yellow Report} detector design~\cite{AbdulKhalek:2021gbh}. 
The upper $Q^2$ cut reflects the limit of applicability of our small-$x$ formalism, which does not include the DGLAP resummation needed to describe large $Q^2$ (beyond its small-$x$ part included in the DLA).

\begin{figure}[t]
    \centering
    \includegraphics[scale=0.65]{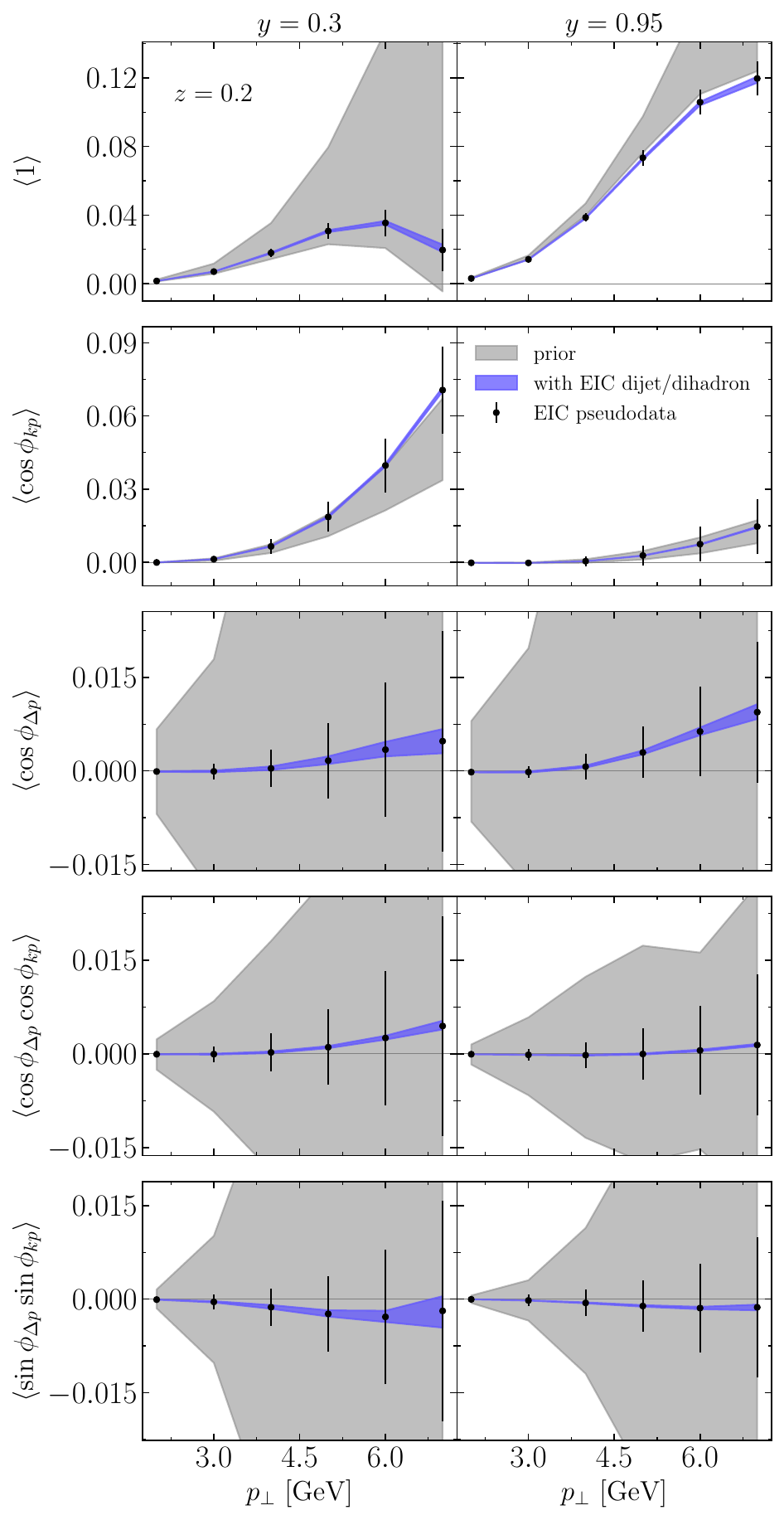}
    \caption{Representative fit to the elastic dijet pseudodata, for a proton target at $\sqrt{s}=30\,\mathrm{GeV}$ in the all-$g_1^p$, OAM scenario A. The five rows are the dijet DSA azimuthal harmonics of \eqs{correlations} plotted against $p_\perp$. The two columns are the inelasticity bins $y\in\{0.3,\,0.95\}$  at fixed momentum fraction $z=0.2$. The 68\% credible intervals of the fitted replica ensemble (blue bands) are compared with the prior distribution of replicas (gray bands) from the polarized dipole amplitudes constrained by the world polarized DIS, SIDIS, and $pp$ data together with the EIC DIS/SIDIS pseudodata, before the dijet/dihadron pseudodata are included, and with the generated pseudodata (black points) with the projected statistical uncertainties obtained from \eq{all_err}.}
    \label{fig:dijet_fit}
\end{figure}

\begin{figure}[t!]
    \centering
    \includegraphics[scale=0.55]{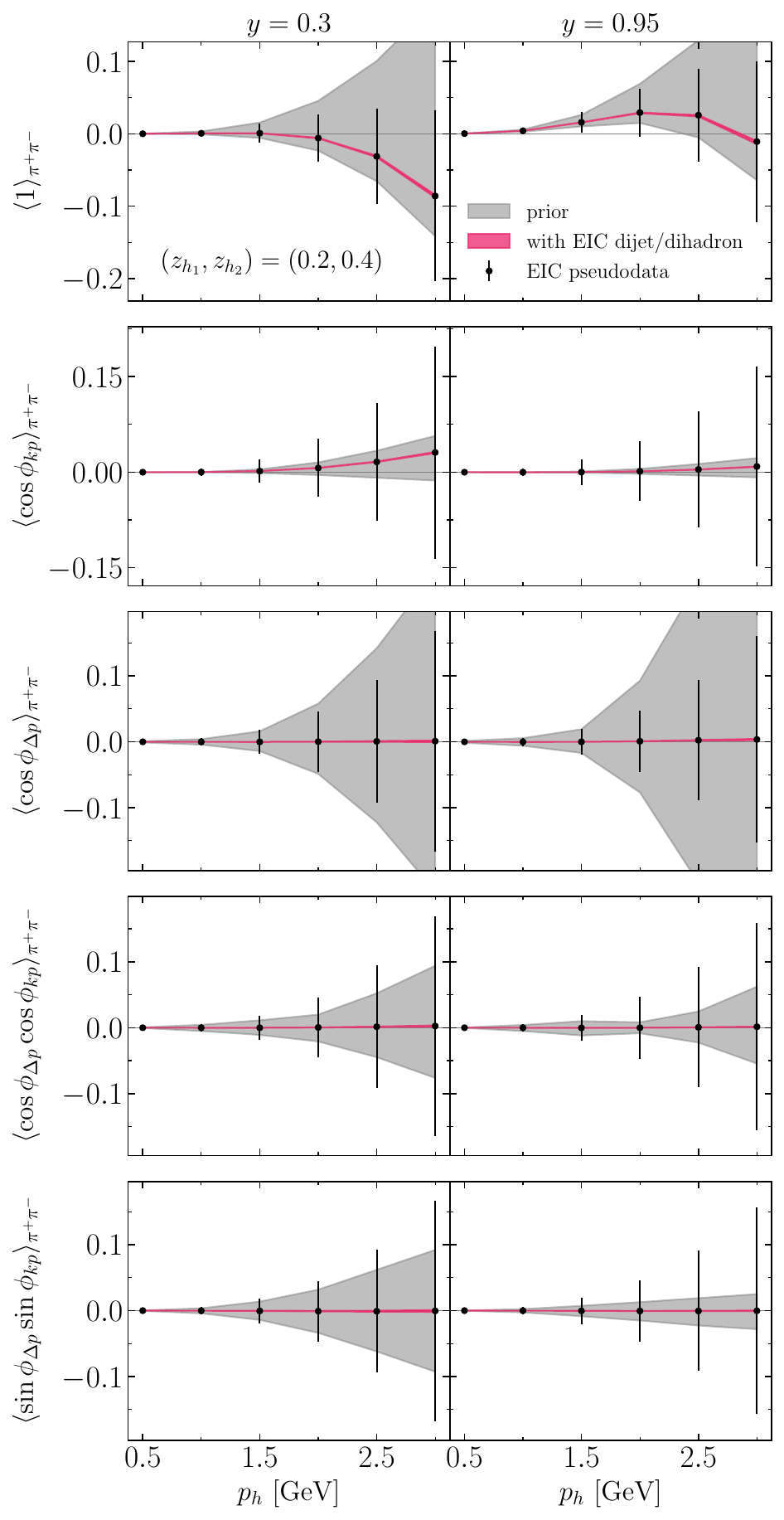}
    \includegraphics[scale=0.55]{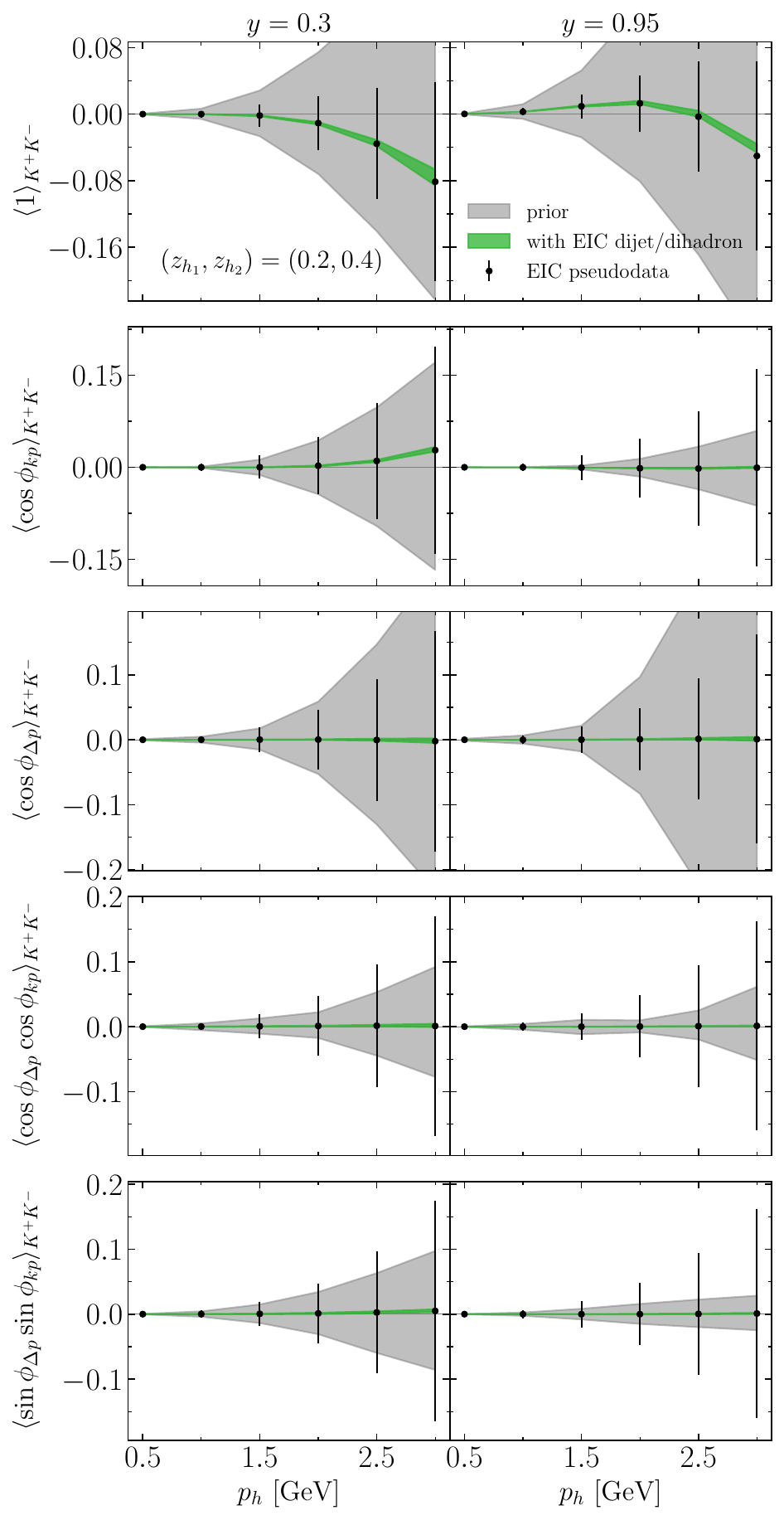}
    \caption{As in \fig{fig:dijet_fit}, but for dihadron production with $\pi^+\pi^-$ production (left) and $K^+K^-$ production (right). The rows are the dihadron DSA harmonics of \eqs{dh_correlations} versus the dihadron momentum $p_h$, and the fixed momentum-fraction bin is $(z_{h_1},z_{h_2})=(0.2,0.4)$.}
    \label{fig:dihadron_pipi_fit}
\end{figure}

We perform the fits within the JAM Bayesian Monte Carlo framework~\cite{Sato:2016tuz, Sato:2019yez, Moffat:2021dji}, sampling the 47 initial condition parameters in \eq{ic_ansatz} of the polarized dipole amplitudes. 
Rather than re-evolving the amplitudes for each parameter choice, we exploit the linearity of the helicity and moment small-$x$ DLA evolution equations. 
The evolved dipole amplitudes corresponding to any choice of initial conditions can be reconstructed as linear combinations of a fixed set of pre-evolved basis amplitudes~\cite{Adamiak:2023yhz}, each weighted by the corresponding initial condition parameter. 
These reconstructed amplitudes are then used to compute the observables entering the fit --- the polarized DIS and SIDIS $g_1$ and $g_1^h$ structure functions in \eqs{g1_structs}, the polarized $pp\to (\mathrm{jet})X$ cross section in \eq{num_xperp}, and the dijet and dihadron DSA numerators in \eqs{correlations} and \eqref{dh_correlations}.

For each fit, the initial condition parameters are optimized by a gradient descent $\chi^2$ minimization algorithm against the combined DIS, SIDIS, dijet, and dihadron pseudodata. 
Experimental uncertainties are propagated by 
resampling each pseudodata point within its quoted uncertainty, minimizing $\chi^2$ against the resulting fluctuated dataset, and recording the resulting optimized parameters as a single replica. 
Repeating this procedure 200 times yields a statistically stable ensemble of replicas from which we reconstruct the polarized dipole and moment amplitudes, hPDFs, and OAM distributions, along with their uncertainty bands. 
In addition to the pseudodata, we also include the world polarized small-$x$ data from DIS, SIDIS, and polarized $pp$ jet production, as in the JAMsmallx-25 analysis~\cite{JAMCollaborationSmall-xAnalysisGroup:2025tfa}.

The quality of the fit to the elastic dijet and dihadron pseudodata is illustrated in Figs.~\ref{fig:dijet_fit} and \ref{fig:dihadron_pipi_fit} for the dijet sample and for $\pi^+\pi^-$ and $K^+K^-$ dihadron production, respectively.
Each figure displays a representative configuration for a proton target, center-of-mass energy $\sqrt{s}=30$~GeV, and all-$g_1^p$ OAM scenario A; the remaining targets, energies, and scenarios are described equally well.
In each figure, we compare the results of our fit to the prior prediction using polarized dipole amplitudes constrained from the DIS, SIDIS, and polarized $pp$ data and the EIC DIS and SIDIS pseudodata.

Across every harmonic and most kinematic bins, the fitted band is also markedly narrower than the prior over most of the range, which shows that these harmonics carry genuine information about the polarized dipole and moment amplitudes beyond what the polarized DIS, SIDIS, and $pp$ data already provide.
The variation across the different $y$ columns makes the degeneracy-breaking of the preceding discussion visible directly, since the relative weighting of the $a$- and $c$-type initial conditions shifts with $y$ at fixed $\sqrt{s}$.

\subsection{Impact of EIC pseudodata on OAM and helicity distributions}

In this section we present the fits to the projected EIC pseudodata for DIS, SIDIS, and elastic dijet and dihadron production.
All distributions below are shown at $Q^2 = 10\,\mathrm{GeV}^2$, and all uncertainty bands correspond to 68\% credible intervals constructed from the ensemble of 200 fitted replicas.
We discuss the OAM distributions first, followed by the hPDFs, and finally the joint constraint they imply on the small-$x$ piece of the Jaffe-Manohar sum rule.

\begin{figure}[t]
    \centering
    \includegraphics[scale=0.45]{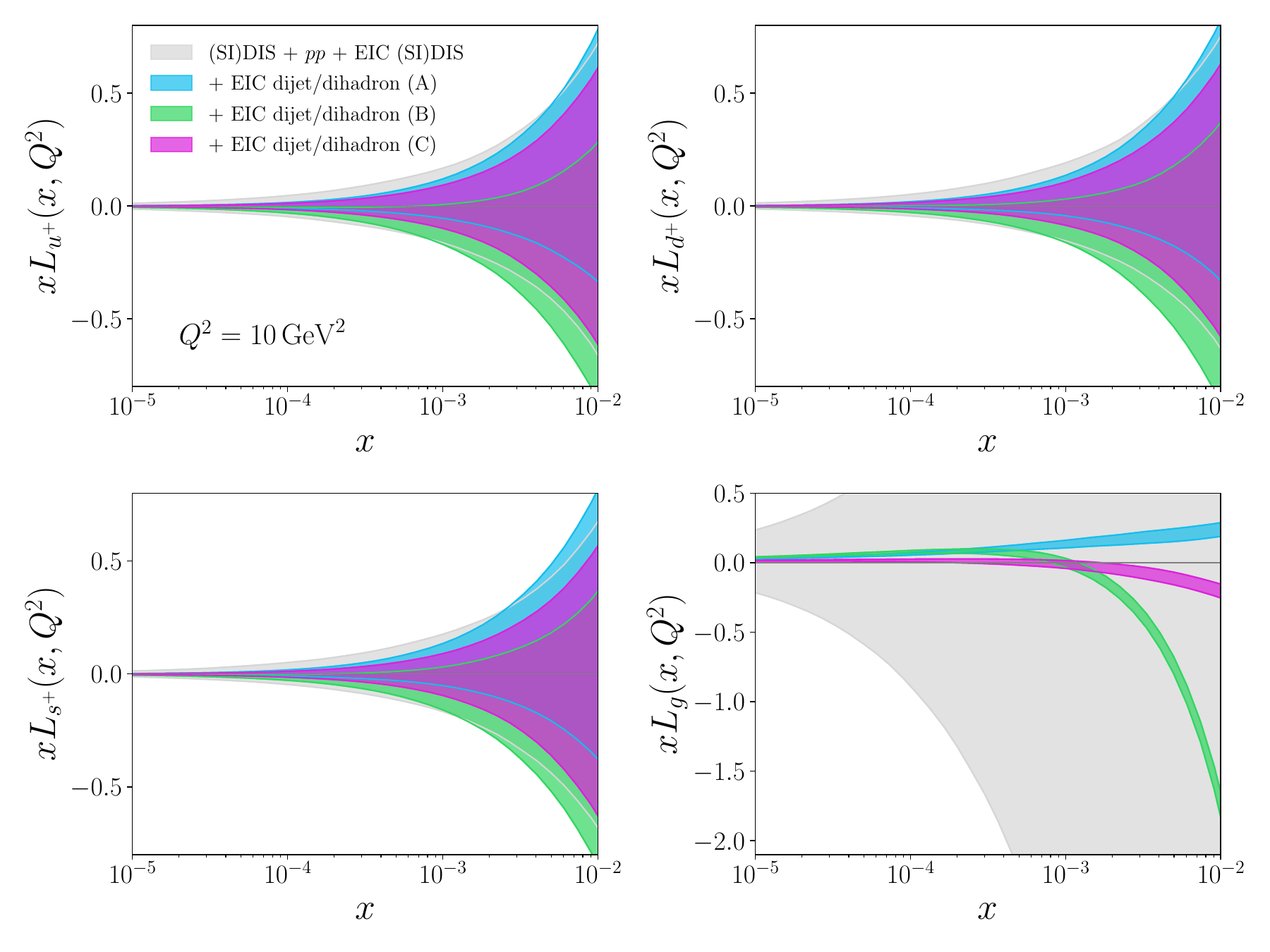}
    \vspace*{-0.5cm}
    \caption{OAM distributions for the $u^+$, $d^+$, $s^+$ and $g$ flavors, weighted by $x$, at $Q^2 = 10$~GeV$^2$ from our fit to the polarized DIS, SIDIS, and $pp$ data, as well as EIC DIS, SIDIS, dijet, and dihadron pseudodata. The fits without dijet and dihadron pseudodata (gray bands), with the OAM distributions constrained only by the $[-20, 20]$ range we chose for the parameters in \eq{ic_ansatz}, are compared with fits including dijet and dihadron pseudodata for the large-positive OAM (scenario A, blue bands), large-negative OAM (scenario B, green bands), and small OAM (scenario C, purple bands), with all bands given by the 68\% credible intervals.}
    \label{fig:xoam}
\end{figure}

The extracted OAM distributions, weighted by $x$, are shown in \fig{fig:xoam} for the $u^+$, $d^+$, $s^+$ and $g$ flavors, both with and without the EIC dijet and dihadron pseudodata included in the fit.
Without these data, no observable in the fit couples to the moment amplitudes; consequently, the moment amplitudes --- and through \eqs{OAMs} the OAM distributions --- remain almost completely unconstrained, as reflected by the broad gray bands in \fig{fig:xoam}. 
The narrowing of these bands as $x \to 0$ is not a constraint from data but a feature of the formalism:~the small-$x$ asymptotic behavior of the OAM distributions is such that $x\,L(x) \to 0$ for any choice of initial conditions \cite{Kovchegov:2023yzd, Manley:2024pcl}. 
Including the dijet and dihadron pseudodata produces a nontrivial extraction of all four OAM distributions shown in \fig{fig:xoam}. 
Crucially, the band widths in all three OAM scenarios are comparable across the full $x$ range, indicating that the size of the projected uncertainties is essentially independent of the underlying true value within the range of physically reasonable small-$x$ OAM behaviors. We also find the projected OAM uncertainties to be independent of the $g_1^p$ scenario used.

\begin{figure}[b]
    \centering
    \includegraphics[scale=0.45]{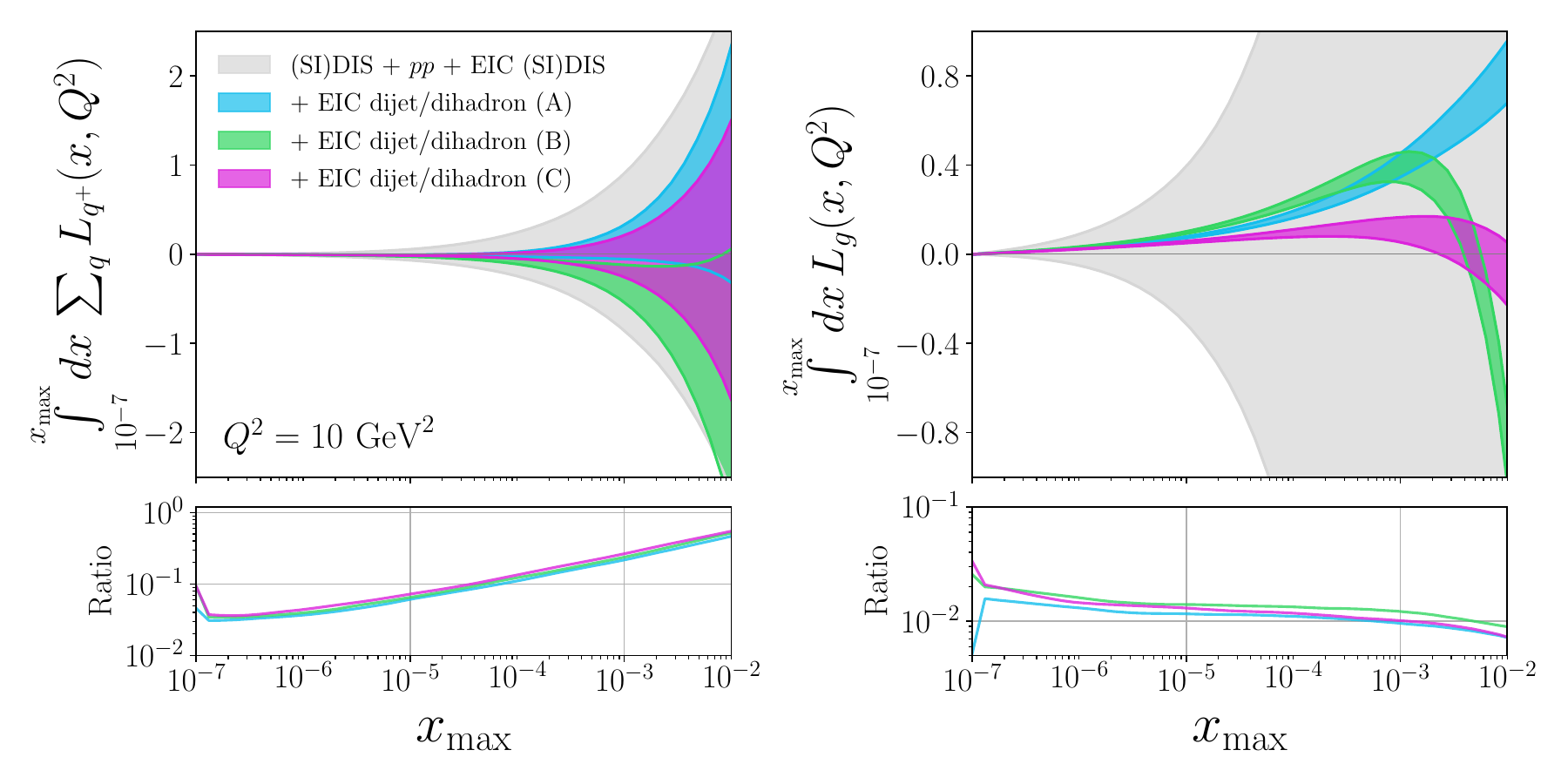}
    \vspace*{-0.3cm}
    \caption{Quark singlet $\sum_q L_{q^+}$ and gluon $L_g$ OAM distributions integrated over $x \in [10^{-7}, x_{\mathrm{max}}]$ at $Q^2 = 10\, \mathrm{GeV}^2$, where $x_{\mathrm{max}} \in [10^{-7}, 0.01]$. The legend is as in \fig{fig:xoam}. The bottom panels illustrate the ratios of the uncertainty bands for each projected extraction over those for the prior.} 
    \label{fig:int_oam}
\end{figure}

To quantify the small-$x$ OAM contribution to the Jaffe-Manohar sum rule, in \fig{fig:int_oam} we show the truncated quark singlet and gluon OAM integrals over $x \in [10^{-7}, x_{\mathrm{max}}]$ as functions of the upper integration limit, $x_{\mathrm{max}}$.\footnote{To be consistent with the JAMsmallx-25 analysis~\cite{JAMCollaborationSmall-xAnalysisGroup:2025tfa}, we have chosen the lower bound here to be $x=10^{-7}$. Below this value, at $Q^2 = 10$~GeV$^2$, saturation effects in the nucleon are expected to make the OAM distributions (and the hPDFs) negligibly small.} 
For both the quark and gluon OAM, the reduction in uncertainty compared with the prior when the dijet and dihadron pseudodata are included is substantial across the full kinematic range, and is again insensitive to the choice of OAM scenario. 
Since the $\widetilde{I}_q$ moment amplitude does not appear explicitly in the dijet/dihadron DSA expressions, and only couples through evolution to the moment amplitudes that enter those observables, the quark OAM is consequently less constrained than the gluon OAM.
In the bottom panels of \fig{fig:int_oam}, we illustrate this concretely with the ratios of the widths of the uncertainty bands to the width of the prior band. 
(Since all evolution equations employed in this work are linear, we expect the relative uncertainty reduction to be approximately independent of the chosen range of parameters in \eq{ic_ansatz}.)


For the full integral up to $x_{\mathrm{max}}=0.01$, dijet and dihadron pseudodata improve over the prior through an approximately 60\% reduction in the uncertainty for the quark OAM. 
For the gluon OAM, we observe a nearly 97\% reduction, reflecting the fact that $I_4$ and $I_5$, which determine the gluon OAM via \eq{gluon_OAM}, both appear explicitly in the dijet/dihadron DSA.
For the fit without $pp$ data, our conclusions are the same --- we observe an $\approx 60\%$ reduction on the prior uncertainty for the integrated quark OAM and a 95\% reduction for the integrated gluon OAM, consistent across all OAM scenarios. 
The stronger constraints in the gluon OAM sector may reflect the choice of the quark and antiquark dijets in the current fragmentation region as the observable. 
The DSA for elastic quark or gluon dijets, measured at mid-rapidity, away from the current fragmentation region of the longitudinally polarized $ep$ collisions, may couple directly to $\widetilde{L}_{q\,10}$ and hence to $\widetilde{I}_q$, thus allowing better constraint on the quark OAM distribution. 
More work is needed to test this possibility. 

\begin{figure}[t!]
\centering
    \includegraphics[scale=0.55]{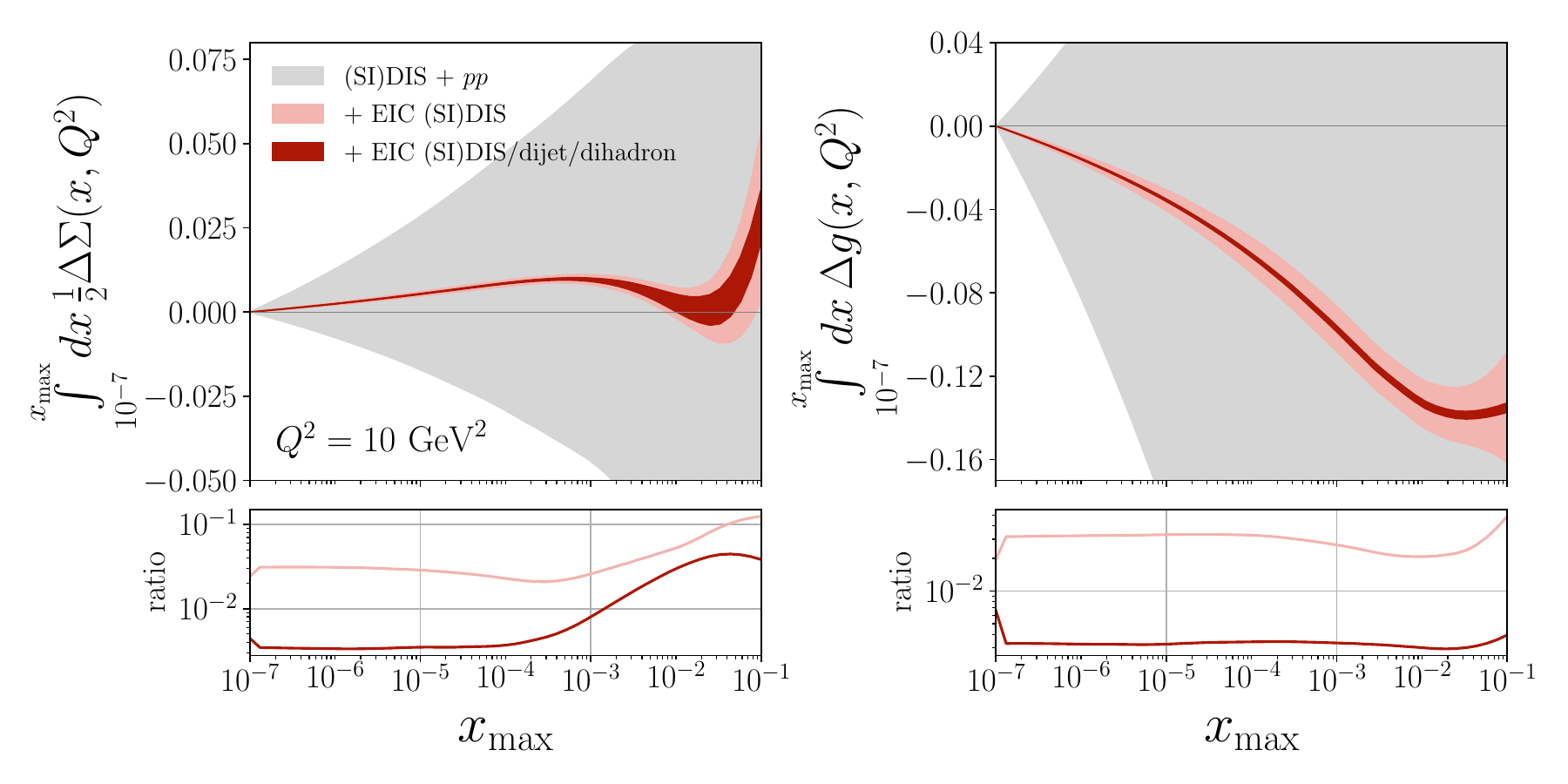}
    \vspace*{-0.3cm}
    \caption{Quark singlet (left) and gluon (right) hPDFs integrated over $x \in [10^{-7}, x_{\mathrm{max}}]$, for $x_{\mathrm{max}} \in [10^{-7}, 0.1]$ at $Q^2=10$~GeV$^2$. The baseline hPDFs from the fit to DIS, SIDIS, and $pp$ data from Sec.~\ref{sec:baseline} (gray bands) are compared with results including EIC pseudodata for DIS and SIDIS for the all $g^p_1$ (red) scenario, and including in addition EIC pseudodata for elastic dijet and dihadron production (dark red). The bands are given by the 68\% credible intervals of the extracted set of replicas. The bottom panels illustrate the ratios of the uncertainty bands for each projected extraction with the baseline fit.}
    \label{fig:int_hpdfs}
\end{figure}

Turning now to the helicity sector, \fig{fig:int_hpdfs} shows the truncated integrals of the quark singlet $\tfrac12\Delta\Sigma$ and gluon $\Delta g$ hPDFs in the range $x \in [10^{-7}, x_{\mathrm{max}}]$ for the ``all'' $g_1^p$ pseudodata scenario, compared with the baseline fit to the world DIS, SIDIS, and $pp$ data. 
In both the quark singlet and gluon channels there is a dramatic reduction in the uncertainties when EIC DIS and SIDIS pseudodata are included. 
(The reduction is broadly similar across all three $g_1^p$ scenarios.)
In particular, including EIC DIS and SIDIS pseudodata results in an $\approx 90\%$ improvement in uncertainty over the baseline fit in the quark sector and an $\approx 95\%$ improvement in the gluon sector. 
Adding the dijet and dihadron pseudodata further decreases the uncertainty in both channels relative to the EIC DIS+SIDIS-only fit, with the uncertainties on the integrated gluon and quark singlet helicities decreasing by an additional 70\% and 40\%, respectively.
The increased constraining power in the gluon sector is a direct reflection of the fact that $G_2$ appears explicitly in the dijet/dihadron DSA while $\widetilde{Q}_q$ only contributes via evolution.

\begin{figure}[t!]
\centering
    \includegraphics[scale=0.75]{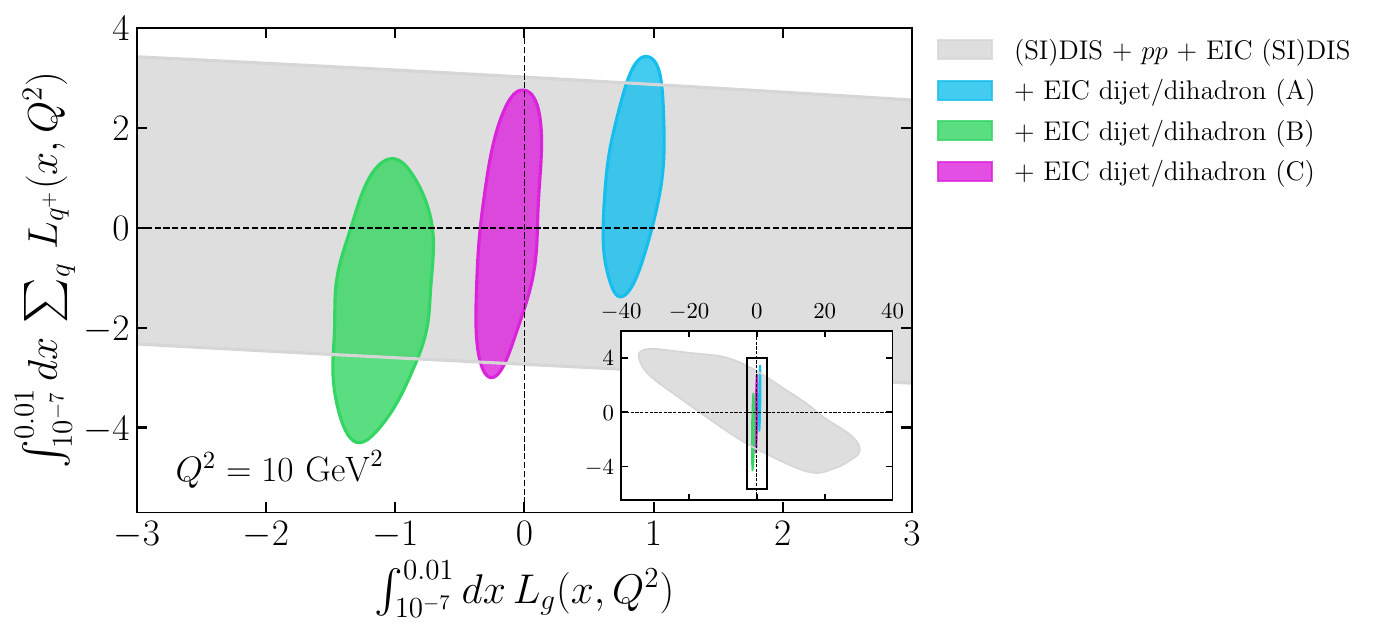}
    \caption{Small-$x$ contribution to the truncated moments of the quark singlet OAM (vertical axis) and gluon OAM (horizontal axis), integrated over $x \in [10^{-7}, 0.01]$ at $Q^2=10$~GeV$^2$. The 68\% credible interval results with DIS+SIDIS+$pp$ data along with EIC DIS+SIDIS pseudodata (gray contours) are compared with those including EIC dijet and dihadron pseudodata for the large-positive OAM (scenario A, blue), large-negative OAM (scenario B, green), and small OAM (scenario C, purple).
    The inset shows the plot zoomed-out on both axes.
   }
   \label{fig:contour_oam}
\end{figure}

\begin{figure}[b!]
    \centering
    \includegraphics[scale=0.75]{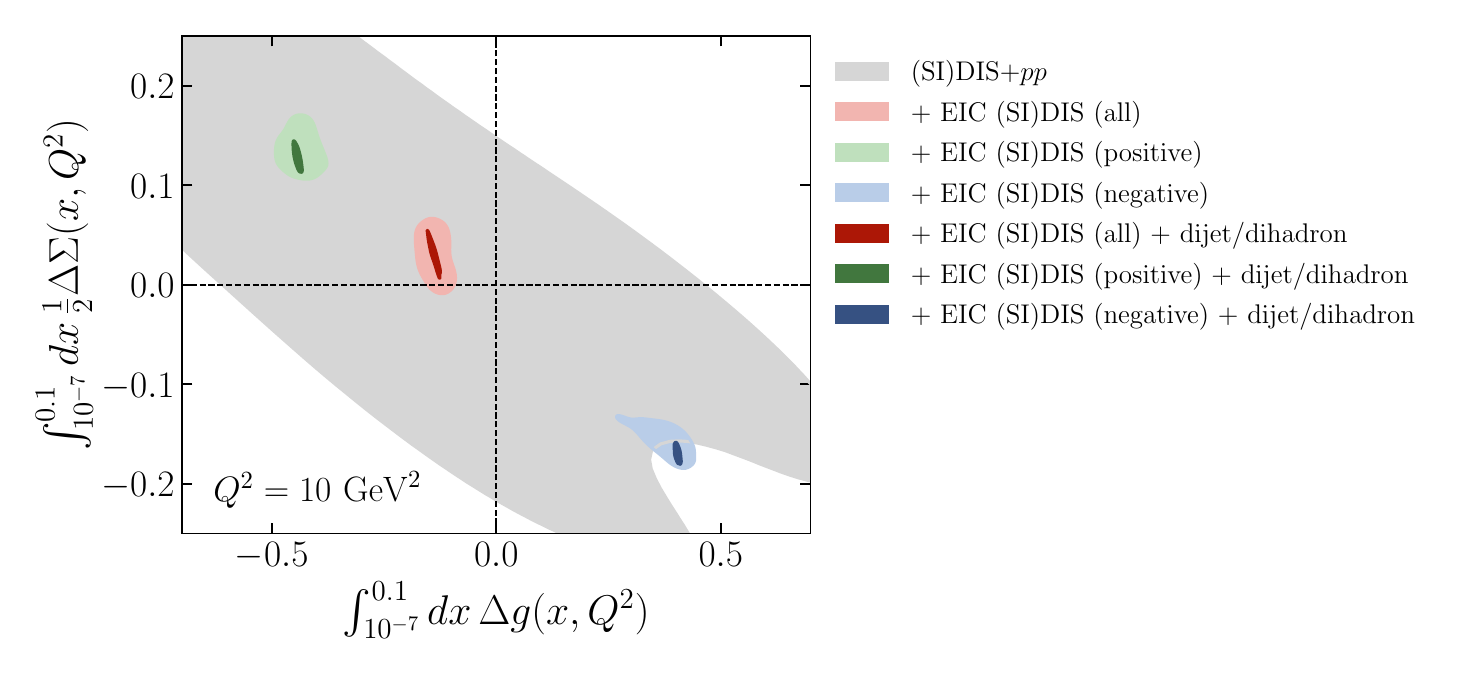}
   \caption{Small-$x$ contribution to the truncated moments of the quark singlet helicity (vertical axis) and gluon helicity (horizontal axis), integrated over $x \in [10^{-7}, 0.1]$ at $Q^2=10$~GeV$^2$. The 68\% credible interval results of the baseline fit to DIS+SIDIS+$pp$ data (gray contour) are compared with those including EIC DIS+SIDIS pseudodata for the all $g^p_1$ (red), positive $g_1^p$ (green), and negative $g_1^p$ (blue) scenarios, and including dijet and dihadron pseudodata for the same scenarios (dark red, dark green, and dark blue, respectively).
   }
   \label{fig:contour_hpdf}
\end{figure}

Finally, in \figs{fig:contour_oam} and \ref{fig:contour_hpdf} we show the 68\% credible intervals in the plane of the total small-$x$ quark and gluon OAM and helicity contributions, respectively, at $Q^2 = 10$~GeV$^2$, with the OAM integrated over $10^{-7} < x < 0.01$ and helicity integrated over $10^{-7} < x < 0.1$. 
The much greater range of the contours in \fig{fig:contour_oam} relative to \fig{fig:contour_hpdf} --- roughly two orders of magnitude in the 68\% credible contours
--- reflects the disparity in constraining power between the helicity and OAM sectors at the EIC. 
The small-$x$ helicity distributions are much more precise as compared to the OAM distributions mainly due to the fact that all the observables included in this study couple to the hPDFs, but only three of the five azimuthal harmonics of the dijet and dihadron DSAs couple to the OAM distributions. 
Although the uncertainty on the small-$x$ OAM contribution is sizable, it should be compared to the present situation, in which the small-$x$ OAM contribution is entirely unconstrained by data (corresponding to the gray bands in \fig{fig:contour_oam}). 
These results illustrate that elastic dijet and dihadron production measurements would provide the first direct experimental constraints on the small-$x$ OAM contributions to the Jaffe-Manohar sum rule.

Taken together, \figs{fig:xoam}--\ref{fig:contour_hpdf} establish two main results. 
First, the EIC dijet and dihadron measurements at $x < 0.01$ would, for the first time, provide a direct experimental constraint on the small-$x$ quark and gluon OAM distributions, with projected uncertainties essentially independent of the underlying truth value. 
Second, the same measurements would further sharpen the existing small-$x$ hPDF extractions from \cite{Adamiak:2023yhz, JAMCollaborationSmall-xAnalysisGroup:2025tfa}, beyond the improvement already provided by the EIC DIS and SIDIS pseudodata alone. 
Combined, these two extractions would allow all four terms of the Jaffe-Manohar sum rule to be confronted simultaneously against experimental data over the small-$x$ region, an outcome that is at present inaccessible to any other observable.

\clearpage
\section{Conclusions and outlook}
\label{sec:conclusions}

In this analysis we have performed the first feasibility study of extracting the small-$x$ quark and gluon OAM distributions from elastic dijet and dihadron production in longitudinally polarized $ep$ collisions at the future EIC. 
Building on earlier elastic dijet DSA calculations~\cite{Kovchegov:2024wjs}, we generalized the polarized partonic calculation to elastic dihadron production via convolution with collinear fragmentation functions (\Sec{sec:dihadrons}) and implemented the moment amplitude evolution equations in the large-$N_c\&N_f$ limit with running coupling on a discretized $(s_{10}, \eta)$ grid~(\Sec{sec:numerical_imp}). 
With these ingredients, we generated EIC dijet and dihadron pseudodata for proton, deuteron, and $^3\mathrm{He}$ beams across the $3 \times 3$ grid of $g_1^p$ and OAM scenarios spanning the range of currently allowed small-$x$ behaviors, and performed Bayesian Monte Carlo fits of these pseudodata together with the world polarized DIS, SIDIS, and $pp$ jet data and the projected EIC DIS and SIDIS pseudodata.

Two principal findings emerged from this analysis. 
First, the EIC elastic dijet and dihadron DSAs would provide the first direct constraint on the small-$x$ quark and gluon OAM distributions, with projected uncertainties that are essentially independent of the underlying truth value within the range of physically reasonable small-$x$ OAM behavior, albeit with larger relative uncertainties than for hPDFs. 
Second, the same measurements would further sharpen the small-$x$ hPDF extractions of the JAMsmallx-23~\cite{Adamiak:2023yhz} and JAMsmallx-25~\cite{JAMCollaborationSmall-xAnalysisGroup:2025tfa} analyses beyond the improvement already provided by the EIC DIS and SIDIS pseudodata alone, with the reduction broadly similar across the three $g_1^p$ scenarios considered.
Combined, these two extractions would, for the first time, allow all four terms of the Jaffe-Manohar sum rule to be confronted simultaneously against (projected) experimental data over the small-$x$ region --- an outcome inaccessible to any other currently proposed observable.

We should note several caveats to the present analysis.
The unpolarized elastic dijet and dihadron cross sections are computed at LLA in the rcBK framework with initial condition parameters fixed to the central values of Ref.~\cite{Albacete:2009fh}; the polarized DSA numerator is computed at sub-eikonal DLA in the large-$N_c\&N_f$ limit; the jet and fragmenting-jet functions are taken at their leading order expressions; only statistical experimental uncertainties are propagated, with systematic uncertainties from detector effects deferred to a dedicated simulation study. 
Each of these approximations is well motivated within the current state of the small-$x$ field, but each could be relaxed in a future expanded analysis.

Natural extensions of this work fall into three categories.
On the unpolarized side, the fixed-input rcBK calculation employed here could be replaced by a Bayesian posterior over the rcBK initial condition parameters \cite{Casuga:2023dcf}. The next-to-leading order BK equation \cite{Balitsky:2007feb} should also be eventually implemented, in a Bayesian framework. 
Similarly, the leading order expressions for the denominator could be replaced by next-to-leading order calculations of the unpolarized elastic dijet and dihadron cross sections~\cite{Caucal:2021ent, Caucal:2022ulg, Caucal:2023fsf, Caucal:2023nci, Caucal:2024bae, Caucal:2025xxh, Altinoluk:2022jkk, Altinoluk:2024zom, Agostini:2024xqs}. 
In the polarized sector, the leading single-logarithmic corrections beyond the DLA can be included as they become available.
Finally, additional OAM-sensitive observables could be added to the fit as they become phenomenologically tractable:~for example, elastic vector and axial-vector meson production~\cite{Bhattacharya:2026qnd, Kovchegov:2026lik} and elastic $\pi^0$ production~\cite{Bhattacharya:2023hbq} are all proposed probes of the OAM distributions (albeit at moderate $x$).

The present work establishes that the elastic dijet and dihadron DSAs at the future EIC are viable observables for a direct extraction of the small-$x$ OAM distributions. 
Together with the world polarized DIS, SIDIS, and $pp$ hadron and jet production data and the projected EIC DIS and SIDIS measurements, they would provide the first experimental test
of the full Jaffe-Manohar spin sum rule at small $x$, marking an important step toward a more complete understanding of the proton spin structure.

\section*{Acknowledgments}

The authors are grateful to Elke Aschenauer, Shohini Bhattacharya, Jeremy Borden, Luca Lavezzo, Brian Page, and Farid Salazar for informative discussions, and to Heikki Mäntysaari for generously providing the code to solve the rcBK equation. 
BM would like to thank Shohini Bhattacharya and the University of Connecticut for their hospitality during the intermediate stages of this work.
This work is supported by the U.S. Department of Energy, Office of Science, Office of Nuclear Physics under Contract No. DE-SC0004286 (YK and BM), DE-SC0020081 (AT), and 89243126CSC000213 (WM and NS), and by the National Science Foundation under Grants No.~PHY-2308567, PHY-2608477 (DP and BK), and PHY-2609761 (BM).
BM also gratefully acknowledges support from The Ohio State University via the Presidential Fellowship. 
DA is grateful for support from the Wu-Ki Tung Endowed Chair in particle physics.
YT is supported by the National Research Council of Thailand (NRCT) via the project number 220677 (contract number N42A690266), by the National Science, Research and Innovation Fund (NSRF) via the Program Management Unit for Human Resources \& Institutional Development, Research and Innovation (grant number B39G680009), and by grants for development of new faculty members, Ratchadaphiseksomphot Fund, Chulalongkorn University. 
This study was funded in part by the Coordination for the Improvement of Higher Education Personnel – Brazil (CAPES) – Finance Code 001. This work is also supported by the U.S. Department of Energy, Office of Science, Office of Nuclear Physics, within the framework of the Saturated Glue (SURGE) Topical Theory Collaboration.

\appendix
\section{The in-shock wave contribution to the $g_1$ structure function}
\label{sec:g1}

\begin{figure}[ht]
\centering
\includegraphics[width= 0.8 \textwidth]{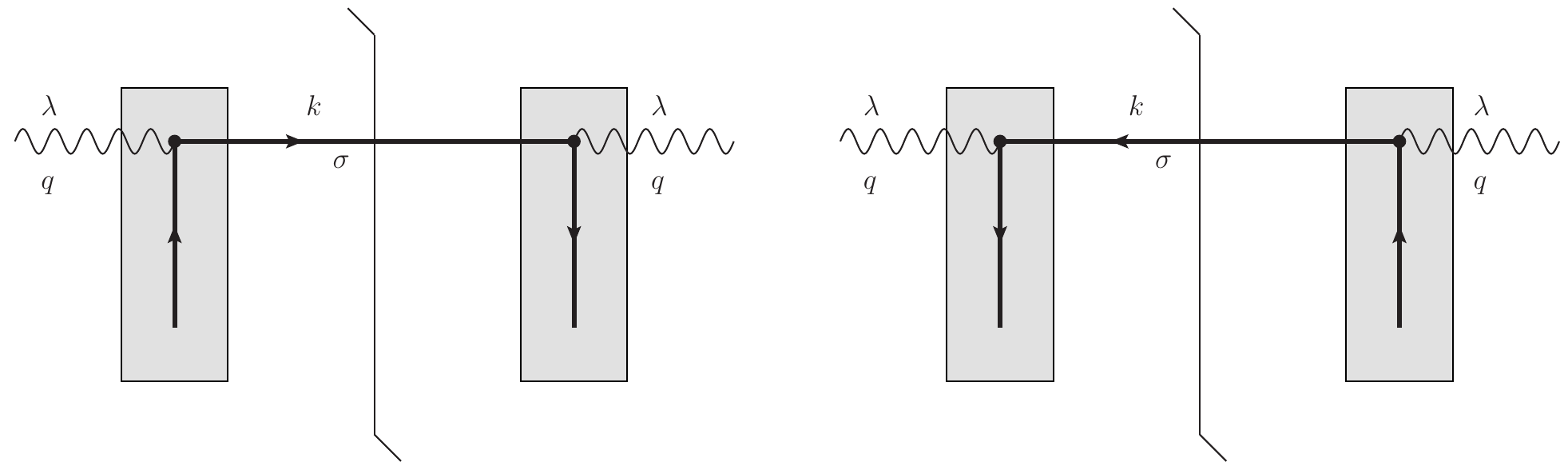} 
\caption{Diagrams contributing to the $g_1$ structure function with the electromagnetic currents (quark-photon interaction) inside the shock wave.}
\label{FIG:g1}
\end{figure}

In this appendix we use the shock-wave formalism to calculate the contribution to the $g_1$ structure function from the term where both electromagnetic currents are located inside the shock wave, illustrated in \fig{FIG:g1}. This term was neglected in earlier calculations \cite{Kovchegov:2015pbl, Cougoulic:2022gbk}, but was recently included by Chirilli in Refs.~\cite{Chirilli:2026vij, Chirilli:2026pkv}.

As shown by Cougoulic {\it et al.}~\cite{Cougoulic:2022gbk}, the $g_1$ structure function is related to the virtual photon--proton scattering cross section, $\sigma^{\gamma^* p}(\lambda, \Sigma)$,  via
\begin{align}\label{g1_xsect}
g_1 (x, Q^2) = - \frac{Q^2}{16 \pi^2 \, \alpha \, x} \, \left( \sigma^{\gamma^* p} (+,+) - \sigma^{\gamma^* p} (-, +) \right),
\end{align} 
where $\lambda = \pm 1$ is the polarization of the (transverse) virtual photon and $\Sigma = \pm 1$ is the proton helicity.
For the diagrams in \fig{FIG:g1}, the virtual photon--proton cross section can be readily shown to be
\begin{align}\label{LO_xsect}
\sigma^{\gamma^* p} (\lambda, \Sigma) 
= \int \frac{\dd^2 k_\perp}{(2 \pi)^2} \, \sum_\sigma \, \frac{1}{(2 s)^2} \, 
\Big\langle \big| M_{\lambda, \sigma, \Sigma} \big|^2 \Big\rangle ,
\end{align}
where $\sigma$ is the knocked-out quark polarization (as shown in \fig{FIG:g1}), while $M_{\lambda, \sigma, \Sigma}$ is the $\gamma^* + p \to \mathrm{quark} + X$ scattering amplitude. The brackets $\langle \ldots \rangle$ denote the small-$x$/saturation averaging. 
Using the approach of Borden {\it et al.}~\cite{Borden:2024bxa}, one can write the amplitude for the diagrams in \fig{FIG:g1} as,
\begin{align}\label{M}
\frac{1}{2s}
M_{\lambda, \sigma, \Sigma} 
= \int \dd^2 x_\perp \, e^{- i {\un k} \cdot {\un x}} \, 
\left( 
W^{q \gamma^*}_{\un x} [\infty, - \infty]  + W^{\bar q \gamma^*}_{\un x} [\infty, - \infty]  
\right),
\end{align}
where the $\gamma^* \to q$ and $\gamma^* \to \bar q$ transition operators can be found from Eqs.~(20) in Ref.~\cite{Borden:2024bxa} by replacing the gluon with the photon. 
This yields,
\begin{subequations}\label{Ws}
\begin{align}
& W_{\un x}^{q \gamma^*} [\infty, - \infty] 
= \frac{- i \, e \, e_q}{2 \, \sqrt{\sqrt{2} \, P^-}} \, \delta_{\lambda, \sigma} \, 
\int\limits_{-\infty}^\infty \dd x^- \, 
V_{\un x} [\infty, x^-]  \, 
\Big[ {\bar \rho}(+) + {\bar \rho}(-) - \sigma \, \big( {\bar \rho}(+) - {\bar \rho}(-) \big) 
\Big] \, {\psi} (x^-, {\un x} ) ,
\\
& W_{\un x}^{{\bar q} \gamma^*} [\infty, - \infty] 
= \frac{i \, e \, e_q}{2 \, \sqrt{\sqrt{2} \, P^-}} \, \delta_{\lambda, \sigma} \,
\int\limits_{-\infty}^\infty \dd x^- \, {\bar \psi} (x^-, {\un x}) \, 
\Big[ \rho(+) + \rho(-) + \sigma \, \big( \rho(+) - \rho(-) \big) 
\Big] \, V_{\un x} [x^-, \infty],
\end{align}
\end{subequations}
where $e$ is the magnitude of the electron's charge and $e_q$ is the fractional charge of the quark of flavor $q$.
The spinors $\rho (\sigma)$ are also defined in Ref.~\cite{Borden:2024bxa}. Substituting Eqs.~\eqref{Ws} into \eq{M} and using the result in Eqs.~\eqref{g1_xsect} and \eqref{LO_xsect} (while noting that the two terms in \eq{M} do not interfere in the cross section), after some algebra one arrives at 
\begin{align}\label{g1_in_shock}
g_1(x, Q^2)\Big|_{\mathrm{in} \,  \mathrm{shockwave}} = \half \sum_q \frac{e_q^2}{\pi^2 \, \as} \, {\widetilde Q}^{(0)}_q \!\left( x_{10}^2 = \frac{1}{Q^2}, s = \frac{Q^2}{x} \right),
\end{align}
in agreement with the first (inhomogeneous) term in \eq{dis_g1} in Sec.~\ref{sec:formalism}. The inhomogeneous term in \eq{sidis_g1} can also be derived using \eq{g1_in_shock}.

\section{Small-$x$ helicity evolution in the large-$N_c\&N_f$ limit}
\label{sec:AppendixB}

For completeness, we list here the evolution equations for the impact-parameter integrated amplitudes $Q_q, \widetilde{G}, G_2$, $\widetilde{Q}_q$, $Q^{\mathrm{NS}}_{q}$, and $\widetilde{Q}^{\mathrm{NS}}_q$ at large $N_c \& N_f$ and with running coupling \cite{Kovchegov:2021lvz}.
These are given by~\cite{Kovchegov:2015pbl, Kovchegov:2016zex, Kovchegov:2017lsr, Kovchegov:2018znm, Cougoulic:2022gbk, Borden:2024bxa},
\begin{subequations}\label{eq_LargeNcNf}
\begin{align}
& Q_q (x^2_{10},zs) = Q_q^{(0)}(x^2_{10},zs) + \frac{N_c}{2\pi} \int_{1/s x^2_{10}}^{z} \frac{\dd z'}{z'}   \int_{1/z's}^{x^2_{10}}  \frac{\dd x^2_{21}}{x_{21}^2} \, \as \Big( \frac{1}{x_{21}^2} \Big)   
\Big[ 2 \, {\widetilde G}(x^2_{21},z's) + 2 \, {\widetilde \Gamma}(x^2_{10},x^2_{21},z's)  
\notag \\
&\hspace*{5cm}
+ \; Q_q (x^2_{21},z's) -  \overline{\Gamma}_q (x^2_{10},x^2_{21},z's) + 2 \, \Gamma_2(x^2_{10},x^2_{21},z's) + 2 \, G_2(x^2_{21},z's)   
\Big] 
\notag \\
&\hspace*{3cm}+ \frac{N_c}{4\pi} \int_{\Lambda^2/s}^{z} \frac{\dd z'}{z'}   \int_{1/z's}^{\min \{ x^2_{10}z/z', 1/\Lambda^2 \}}  \frac{\dd x^2_{21}}{x_{21}^2} \, \as \Big( \frac{1}{x_{21}^2} \Big)  \left[Q_q (x^2_{21},z's) + 2 \, G_2(x^2_{21},z's) \right] ,    
\\ \notag 
&\overline{\Gamma}_q (x^2_{10},x^2_{21},z's) = Q_q^{(0)}(x^2_{10},z's) + \frac{N_c}{2\pi} \int_{1/s x^2_{10}}^{z'} \frac{\dd z''}{z''}   \int_{1/z''s}^{\min\{x^2_{10}, x^2_{21}z'/z''\}}  \frac{\dd x^2_{32}}{x_{32}^2} \, \as \Big( \frac{1}{x_{32}^2} \Big)
\left[ 2\, {\widetilde G} (x^2_{32},z''s)  \right. 
\\
&\hspace*{2.5cm}\left.+ \; 2\, {\widetilde \Gamma} (x^2_{10},x^2_{32},z''s) +  Q_q (x^2_{32},z''s) -  \overline{\Gamma}_q (x^2_{10},x^2_{32},z''s) + 2 \, \Gamma_2(x^2_{10},x^2_{32},z''s) + 2 \, G_2(x^2_{32},z''s) \right] \notag \\
&\hspace*{3cm}+ \frac{N_c}{4\pi} \int_{\Lambda^2/s}^{z'} \frac{\dd z''}{z''}   \int_{1/z''s}^{\min \{ x^2_{21}z'/z'', 1/\Lambda^2 \}}  \frac{\dd x^2_{32}}{x_{32}^2} \, \as \Big( \frac{1}{x_{32}^2} \Big)  \left[Q_q (x^2_{32},z''s) + 2 \, G_2(x^2_{32},z''s) \right] ,  
\\
\notag 
& {\widetilde G}(x^2_{10},zs) = {\widetilde G}^{(0)}(x^2_{10},zs) + \frac{N_c}{2\pi}\int_{1/s x^2_{10}}^z\frac{\dd z'}{z'}\int_{1/z's}^{x^2_{10}} \frac{\dd x^2_{21}}{x^2_{21}} \, \as \Big( \frac{1}{x_{21}^2} \Big)  \bigg[ 3 \, {\widetilde G}(x^2_{21},z's) + {\widetilde \Gamma}(x^2_{10},x^2_{21},z's) 
\\
&\hspace*{1.5cm}  + \; 2\,G_2(x^2_{21},z's)  +  \left(2 - \frac{N_f}{2N_c}\right) \Gamma_2(x^2_{10},x^2_{21},z's) - \frac{1}{4N_c}\, \sum_q \overline{\Gamma}_q (x^2_{10},x^2_{21},z's)  - \frac{1}{2 N_c} \, \sum_q {\widetilde Q}_q (x^2_{21},z's)   \bigg] 
\notag \\
&\hspace*{3cm}- \frac{1}{8\pi}   \sum_q \int_{\Lambda^2/s}^z \frac{\dd z'}{z'}\int_{\max\{x^2_{10},\,1/z's\}}^{\min \{ x^2_{10}z/z', 1/\Lambda^2 \}} \frac{\dd x^2_{21}}{x^2_{21}} \, \as \Big( \frac{1}{x_{21}^2} \Big)  \left[   Q_q (x^2_{21},z's) +     2 \, G_2(x^2_{21},z's)  \right] ,  
\\
& {\widetilde \Gamma} (x^2_{10},x^2_{21},z's) = {\widetilde G}^{(0)}(x^2_{10},z's) + \frac{N_c}{2\pi}\int_{1/s x^2_{10}}^{z'}\frac{\dd z''}{z''}\int_{1/z''s}^{\min\{x^2_{10},x^2_{21}z'/z''\}} \frac{\dd x^2_{32}}{x^2_{32}} \, \as \Big( \frac{1}{x_{32}^2} \Big)  \bigg[3 \, {\widetilde G} (x^2_{32},z''s) + \; {\widetilde \Gamma}(x^2_{10},x^2_{32},z''s) 
\notag \\
&\hspace*{1.2cm}   + 2 \, G_2(x^2_{32},z''s)  +  \left(2 - \frac{N_f}{2N_c}\right) \Gamma_2(x^2_{10},x^2_{32},z''s) - \frac{1}{4N_c} \, \sum_q \overline{\Gamma}_q (x^2_{10},x^2_{32},z''s)   - \frac{1}{2 N_c} \, \sum_q {\widetilde Q}_q (x^2_{32},z''s)  \bigg] 
\notag \\
&\hspace*{2cm}- \frac{1}{8\pi} \sum_q  \int_{\Lambda^2/s}^{z'x^2_{21}/x^2_{10}} \frac{\dd z''}{z''}\int_{\max\{x^2_{10},\,1/z''s\}}^{\min \{ x^2_{21}z'/z'', 1/\Lambda^2 \} } \frac{\dd x^2_{32}}{x^2_{32}} \, \as \Big( \frac{1}{x_{32}^2} \Big)  \left[   Q_q (x^2_{32},z''s) +  2  \,  G_2(x^2_{32},z''s)  \right] , 
\\
& G_2(x_{10}^2, z s)  =  G_2^{(0)} (x_{10}^2, z s) + \frac{N_c}{\pi} \, \int\limits_{\frac{\Lambda^2}{s}}^z \frac{\dd z'}{z'} \, \int\limits_{\max \left[ x_{10}^2 , \frac{1}{z' s} \right]}^{\min \{\frac{z}{z'} x_{10}^2, 1/\Lambda^2 \}} \frac{\dd x^2_{21}}{x_{21}^2} \, \as \Big( \frac{1}{x_{21}^2} \Big)  \left[ {\widetilde G} (x^2_{21} , z' s) + 2 \, G_2 (x_{21}^2, z' s)  \right] , 
\\
& \Gamma_2 (x_{10}^2, x_{21}^2, z' s)  =  G_2^{(0)} (x_{10}^2, z' s) + \frac{N_c}{\pi} \!\! \int\limits_{\frac{\Lambda^2}{s}}^{z' \frac{x_{21}^2}{x_{10}^2}} \frac{\dd z''}{z''}\!\!\!  \int\limits_{\max \left[ x_{10}^2 , \frac{1}{z'' s} \right]}^{\min \{ \frac{z'}{z''} x_{21}^2, 1/\Lambda^2 \}} \!\frac{\dd x^2_{32}}{x_{32}^2} \, \as \Big( \frac{1}{x_{32}^2} \Big)  \left[ {\widetilde G} (x^2_{32} , z'' s) + 2 \, G_2(x_{32}^2, z'' s)  \right]\!,
\\
&  {\widetilde Q}_q (x^2_{10}, z s) =  {\widetilde Q}_q^{(0)} (x^2_{10}, z s) - \frac{N_c}{2 \, \pi} \, \int\limits_\frac{\Lambda^2}{s}^{z} \frac{\dd z'}{z'} \, \int\limits_{\max \left[ x_{10}^2, \frac{1}{z' s} \right]}^{\min \{\frac{z}{z'} x_{10}^2, 1/\Lambda^2 \}} \frac{\dd x^2_{21}}{x^2_{21}} \, \as \Big( \frac{1}{x_{21}^2} \Big)  \left[ Q_q (x_{21}^2, z' s) + 2 \, G_2 (x_{21}^2, z' s) \right] , 
\label{Qtilde_evol} \\
&  Q^{\mathrm{NS}}_{q} (x^2_{10}, z s) =  Q^{\mathrm{NS} \, (0)}_{q} (x^2_{10}, z s) + \frac{N_c}{4 \, \pi} \, \int\limits_\frac{\Lambda^2}{s}^{z} \frac{\dd z'}{z'} \, \int\limits_{\frac{1}{z' s}}^{\min \{\frac{z}{z'} x_{10}^2, 1/\Lambda^2 \}} \frac{\dd x^2_{21}}{x^2_{21}} \, \as \Big( \frac{1}{x_{21}^2} \Big)  Q^{\mathrm{NS}}_{q} (x_{21}^2, z' s) , 
\\
& {\widetilde Q}^{\mathrm{NS}}_q (x^2_{10}, z s) =  {\widetilde Q}_q^{\mathrm{NS} \, (0)} (x^2_{10}, z s) + \frac{N_c}{2 \, \pi} \, \int\limits_\frac{\Lambda^2}{s}^{z} \frac{\dd z'}{z'} \, \int\limits_{\max \left[ x_{10}^2, \frac{1}{z' s} \right]}^{\min \{\frac{z}{z'} x_{10}^2, 1/\Lambda^2 \}} \frac{\dd x^2_{21}}{x^2_{21}} \, \as \Big( \frac{1}{x_{21}^2} \Big) \, Q_q^{\mathrm{NS}} (x_{21}^2, z' s) .
\end{align}
\end{subequations}
As before, $\overline{\Gamma}_q$, ${\widetilde \Gamma}$, and $\Gamma_2$ are the ``neighbor dipole" amplitudes for $Q_q, \widetilde{G}$, and $G_2$, respectively. Equations \eqref{eq_LargeNcNf} resum powers of $\as \ln^2(1/x)$ and $\as \ln(1/x)\ln(Q^2/\Lambda^2)$~\cite{Kovchegov:2015pbl,  Kovchegov:2016zex, Kovchegov:2017lsr, Kovchegov:2018znm, Cougoulic:2022gbk, Borden:2024bxa}: this resummation is our definition of DLA. Below, we will solve these equations (numerically), either as shown or in the approximation where we put $N_f =0$ and discard all the sums over quark flavors ($\sum_q$) \cite{JAMCollaborationSmall-xAnalysisGroup:2025tfa}. Following \cite{JAMCollaborationSmall-xAnalysisGroup:2025tfa}, the latter will be referred to as the large-$N_c^{+q}$ limit.

\section{Discretized form of the moment amplitude evolution equations}
\label{sec:AppendixC}

As discussed in Sec.~\ref{sec:formalism} above, \eqs{mom_eqn_rc} and \eqref{neigh_mom_eqn_rc} are not closed. 
They need to be solved simultaneously with the evolution equations \eqref{eq_LargeNcNf} for $Q_q, \widetilde{G}$, $G_2$, and $\widetilde{Q}_q$.
The discretized versions of these equations are given in Eqs.~(17) of Ref.~\cite{Adamiak:2023yhz}.\footnote{Eqs.~(17) from \cite{Adamiak:2023yhz} do not include the contributions from or evolution equation for $\widetilde{Q}_q$. However, these contributions can be readily inferred from \eqs{eq_LargeNcNf} of the current paper and in the companion paper~\cite{Becker:2026xcp}.} Applying the constraints and discretization procedure described above to \eqs{mom_eqn_rc}, we obtain\footnote{The initial condition for $I_E$ is zero to all orders in the quasi-classical approximation~\cite{Becker:2026xcp}.}
\begin{align} 
\notag 
    \begin{pmatrix}
        I_{3q} \\
        \tI_3\\ 
        I_4 \\
        I_5 \\
        \widetilde{I}_q \\ 
        I_E
    \end{pmatrix}[i,j]
    &= 
     \begin{pmatrix}
        I_{3q} \\
        \tI_3\\ 
        I_4 \\
        I_5 \\
        \widetilde{I}_q \\
        I_E
    \end{pmatrix}[i,j-1]
    +
    \begin{pmatrix}
        I_{3q}^{(0)} \\
        \tI_3^{(0)}  \\ 
        I_4^{(0)} \\
        I_5^{(0)} \\
        \widetilde{I}_q^{(0)} \\ 
        0
    \end{pmatrix} [i, j]
    - 
    \begin{pmatrix}
        I_{3q}^{(0)} \\
        \tI_3^{(0)}  \\ 
        I_4^{(0)} \\
        I_5^{(0)} \\
        \widetilde{I}_q^{(0)} \\
        0
    \end{pmatrix} [i, j-1]
    \\ \notag &
    + 
    \delta^2 \sum_{i' = i }^{j-2-j_0}
    \as[i']
    \begin{pmatrix}
        -1 & 2 & -2 & 1 & -1 \\
        - \frac{1}{4N_c} \sum_q   & 1 & \frac{N_f}{2N_c} -2  & 1- \frac{N_f}{4N_c} & \frac{N_f}{4N_c} - 1  \\
        0 & 0 & 0 & 0  &  0  \\ 
        0 & 0 & 0 & 0  &  0  \\ 
        0 & 0 & 0 & 0  &  0  \\
        0 & 0 & 0 & 0  &  0 
    \end{pmatrix}
    \begin{pmatrix}
        \Gamma_{3q} \\ 
        \tGm_3 \\
        \Gamma_4 \\
        \Gamma_5 \\
        \Gamma_2 
    \end{pmatrix}[i, i', j-1]
    \\ \notag & 
    + \frac{1}{2} \delta^2 
    \sum_{j' = j-1-i}^{j-2} 
    \as[i + j' - j + 1]
    \begin{pmatrix}
        0 & 4 & -4 & 2 & 0 & -2 \\
        -\frac{1}{2N_c} \sum_q  & 4 & \frac{N_f}{N_c} - 4 & 2 - \frac{N_f}{2N_c}   & \frac{1}{2N_c} \sum_q  & -2 \\
        0 & 0 & 4 & 2  & 0 & -1 \\
        0 & -2 & 2 & -1 & 0 & 1  \\
        -1 & 0 & 2 & -1 & 1 & 0 \\
        0 & -4 & 4 & -2 & 0 & 2
    \end{pmatrix}
    \begin{pmatrix}
        I_{3q} \\ 
        \tI_3 \\
        I_4 \\
        I_5 \\
        \widetilde{I}_q \\ 
        I_E
    \end{pmatrix}[i+j'-j+1, j']
        \\ & 
    + \frac{1}{2} \delta^2 
    \sum_{j' = j-1-i}^{j-2} 
    \as[i + j' - j + 1]
    \begin{pmatrix}
         0 & -4  & -6 & 0   \\
         0 & -4  & \frac{N_f}{2N_c} -6 & \frac{1}{4 N_c} \sum_q \\
         0 &  -2 & -3 & 0 \\
         0 & 4 & 7 & 0 \\
         1 & 0 & 3 & \thalf \\
         0 & 4 & 6 & 0
    \end{pmatrix}
    \begin{pmatrix}
        Q_q \\ 
        \tG \\
        G_2 \\
        \widetilde{Q}_q 
    \end{pmatrix}[i+j'-j+1, j'],
     \label{rc_mom_disc}
\end{align}
where $j_0 \equiv \lceil y_0/\delta \rceil$, and we have defined the discretized coupling $\as[i] = \alpha_s(s_{10} = i\delta)$. 
The running of the coupling in logarithmic variables is \cite{Adamiak:2021ppq, Adamiak:2023yhz, JAMCollaborationSmall-xAnalysisGroup:2025tfa}
\begin{align} \label{pol_rc}
    \as(s_{10}) = \sqrt{\frac{N_c}{2\pi}} \frac{12\pi}{(11\, N_c - 2 \, N_f)}\frac{1}{s_{10} + s_0}, \hspace{1cm} s_{0} \equiv \sqrt{\frac{N_c}{2\pi}} \ln \frac{4\, C^2 \Lambda^2}{\Lambda_{\mathrm{QCD}}^2}.
\end{align}
We take $\as(M_Z^2) = 0.1176$, for $N_f=3, N_c=3$ and $M_Z$ the mass of the $Z$ boson \cite{Albacete:2009fh}. 
Additionally, we set $C^2=1$ for the polarized evolution. Note our choice of $\Lambda$ is such that $\Lambda > \Lambda_{\mathrm{QCD}}$ (and hence $s_0>0$).

To speed up the computation, we have written \eq{rc_mom_disc} in a recursive form. Similarly, the discretized version of the neighbor amplitude evolution in \eqref{neigh_mom_eqn_rc}  
is 
\begin{align} 
 \notag 
    \begin{pmatrix}
        \Gamma_{3\,q} \\
        \tGm_3\\ 
        \Gamma_4 \\
        \Gamma_5 
    \end{pmatrix}[i, k,j] 
    &= 
    \begin{pmatrix}
        \Gamma_{3\,q} \\
        \tGm_3\\ 
        \Gamma_4 \\
        \Gamma_5 
    \end{pmatrix}[i, k-1 ,j-1]
    + 
    \begin{pmatrix}
        I_{3\,q}^{(0)} \\
        \tI_3^{(0)}  \\ 
        I_4^{(0)} \\
        I_5^{(0)} 
    \end{pmatrix} [i, j]
    - 
    \begin{pmatrix}
        I_{3\,q}^{(0)} \\
        \tI_3^{(0)}  \\ 
        I_4^{(0)} \\
        I_5^{(0)} 
    \end{pmatrix} [i, j-1]
    \\& 
    + \delta^2 \sum_{i' = k-1}^{j-2-j_0}
    \as[i']
    \begin{pmatrix}
        -1 & 2 & -2 & 1 & -1  \\
        - \frac{1}{4N_c} \sum_q   & 1 & \frac{N_f}{2N_c} -2  & 1- \frac{N_f}{4N_c} & \frac{N_f}{4N_c} - 1  \\
        0 & 0 & 0 & 0  &  0  \\ 
        0 & 0 & 0 & 0  &  0  
    \end{pmatrix}
    \begin{pmatrix}
        \Gamma_{3\,q} \\ 
        \tGm_3 \\
        \Gamma_4 \\
        \Gamma_5 \\
        \Gamma_2 
    \end{pmatrix}[i, i', j-1].
    \label{rc_neigh_disc}
\end{align}
After evolving the polarized dipole amplitudes and their moments, the quark and gluon OAM distributions along with the hPDFs can be computed via writing \eqs{hPDFs} and \eqref{OAMs} in terms of the logarithmic variables in \eqs{log_vars}~\cite{Kovchegov:2023yzd, Cougoulic:2022gbk, Becker:2026xcp} 
\begin{subequations} \label{logv_dists}
    \begin{align}
        \Delta q^+(x,Q^2)  &=  \frac{1}{\as (Q^2) \, \pi^2} \, {\widetilde Q}_q \!\left(s_{10} = \sqrt{\frac{N_c}{2\pi}} \ln \frac{Q^2}{\Lambda^2}, \,\,\eta = \sqrt{\frac{N_c}{2\pi}} \ln \frac{Q^2}{x\Lambda^2}\right) , \\  
            \Delta q^-(x,Q^2)  &=  \frac{1}{\as (Q^2) \, \pi^2} \, {\widetilde Q}^{\mathrm{NS}}_q \!\left(s_{10} = \sqrt{\frac{N_c}{2\pi}} \ln \frac{Q^2}{\Lambda^2}, \,\,\eta = \sqrt{\frac{N_c}{2\pi}} \ln \frac{Q^2}{x\Lambda^2}\right) , \\  
        \Delta g(x, Q^2) &= \frac{2\, N_c}{\as(Q^2) \pi^2 } \, G_2 \!\left(s_{10} = \sqrt{\frac{N_c}{2\pi}} \ln \frac{Q^2}{\Lambda^2}, \,\,\eta = \sqrt{\frac{N_c}{2\pi}} \ln \frac{Q^2}{x\Lambda^2}\right) , 
        \\ 
        L_{q^+}(x, Q^2) &= \frac{2}{\as (Q^2) \, \pi^2} \, {\widetilde I}_q \!\left(s_{10} = \sqrt{\frac{N_c}{2\pi}} \ln \frac{Q^2}{\Lambda^2}, \,\,\eta = \sqrt{\frac{N_c}{2\pi}} \ln \frac{Q^2}{x\Lambda^2}\right) , \\  
        L_g(x, Q^2) &= -\frac{2\, N_c}{\as(Q^2) \pi^2 } \Big[2\, I_4  + 3\, I_5 \Big ]\!\!\left(s_{10} = \sqrt{\frac{N_c}{2\pi}} \ln \frac{Q^2}{\Lambda^2}, \,\,\eta = \sqrt{\frac{N_c}{2\pi}} \ln \frac{Q^2}{x\Lambda^2}\right).
    \end{align}
\end{subequations}
Similarly, the Fourier-Bessel transforms in \eqs{correlations} are computed by expressing \eq{fb_transform} in terms of the logarithmic variables in \eqs{log_vars}.

\end{document}